# Towards the correct microscopic structure of aqueous CsCl solutions with a comparison of classical interatomic potential models

Ildikó Pethes

Wigner Research Centre for Physics, Konkoly Thege út 29-33., H-1121 Budapest, Hungary

E-mail address: pethes.ildiko@wigner.hu

Abstract

The structure of aqueous CsCl solutions was investigated by classical molecular dynamics simulations (MD) at three salt concentrations (1.5, 7.5, and 15 mol %). Thirty interatomic potential sets, based on the 12-6 Lennard-Jones model, parametrized for non-polarizable water solvent molecules were collected and tested. Some basic properties, such as density, static dielectric constant, and self-diffusion coefficients, predicted by the force fields (FF), were compared with available experimental data. The simulated particle configurations were used to calculate the partial radial distribution functions (PRDF) and the neutron and X-ray total scattering structure factors (TSSF). The TSSFs were compared with experimental data from the literature, to find the best FF models, which describe the structure correctly. It was found that, though several of the thirty models failed in the tests, some models are compatible with the measured data. Values of the structural parameters consistent with the experiments were determined (such as water-ion distances, the average number of water molecules around the ions, average number, and distance between anion-cation contact ion pairs, water-water hydrogen bonds). It was shown, that in addition to models in which the number of contact ion pairs is too high, models in which this number is too low are also unable to reproduce the experimental data.

# 1. Introduction

Simulation techniques play an essential role in studying the properties of molecular liquids. The same applies to describing the structure of aqueous salt solutions. Since at least four constituting elements (O and H from water, anion and cation from the salt) are present in such solutions, there are ten partial radial distribution functions (PRDF). There should be ten independent (diffraction) experimental data sets to determine all the ten PRDFs, which is impossible. Thus the use of simulation techniques is necessary to determine the structure.

Computer simulation techniques have been applied for investigating molecular liquids for nearly 70 years [1]. Initially, Metropolis Monte Carlo (MC), classical molecular dynamics (MD), and reverse Monte Carlo (RMC) methods yielded the most publications on this topic; but today, MD simulations using more complex potential models and *ab initio* molecular dynamics (AIMD) simulations have come to the fore. However, the classic MD simulations, using rigid, non-polarizable models, are still widely used nowadays due to their widespread acquaintance and lower computational costs. Moreover, the development of such interatomic potential parameters also continues (see e. g., Ref. [2]).

The success of the classical MD simulations depends fundamentally on the interatomic potential model (force field, FF) chosen. It was shown for aqueous NaCl solutions that there are crucial differences in the performance of the force fields, even in similar types of models [3]. Some of the tested potential models gave false results even for basic properties such as the solubility range: precipitation was observed in the simulated system at concentrations below the (experimental) solubility limit. It was found in a recent paper that pairwise additive MD models give poor agreement with experimental EXAFS data on aqueous CsCl solutions [4]. The authors of that paper concluded that it is necessary to include many-body effects in the simulations to get the correct description of the structure. Since the usability of the classical models highly depends on the FF parameters, the validity of such a statement should be tested against more FFs, rather than just the one used in Ref. [4].

Our previous study investigated the applicability of 29 different 12-6 type Lennard-Jones potential models for highly concentrated aqueous solutions of LiCl [5]. The comparison of several FFs with experimental diffraction data made it possible to describe the structure of aqueous LiCl solutions: the hydrate sphere of the ions (number and distance of water molecules around the ions), as well as the presence of contact and solvent separated ion pairs [6]. This technique has been completed with the analysis of the hydrogen-bonded (H-bond) network of water molecules and chloride ions [7], showing that considering the halide anions to be part of the H-bonded network makes a more meaningful interpretation of the structural details and is also helpful to separate better and worse performing potential models.

The structure of CsCl solutions and the hydrate sphere of $Cs^+$ and $Cl^-$ ions have been studied since 1967 [8] applying experimental and simulation techniques. Reviews of previous results are collected in Refs. [9 – 11]. Experimental techniques were applied in several papers, e. g.: X-ray diffraction (XRD) [8, 12, 13], neutron diffraction (ND) [12, 14, 15], inelastic neutron scattering [16], EXAFS [4, 17, 18], near infrared spectroscopy [19], pulse field gradient NMR [20]. Simulation techniques were applied in papers e.g.: classical MD [4, 12, 17, 21 – 26], QM/MM [22], RMC [12], EPSR [13], AIMD [4, 26, 27]. Some of the papers combined experimental data and simulation techniques, such as in Refs. [4, 12, 13, 17], which helps to obtain more realistic results.

According to the papers mentioned above, the hydrate sphere of $Cs^+$ ions consists of 6 – 10 water molecules, with a $Cs^+$-O distance of about 3 – 3.3 Å. Around the $Cl^-$ anions, 5 – 7 water molecules can be found with $Cl^-$-O distances of about 3.1 – 3.2 Å. The number of hydrating water molecules decreases while the formation of contact ion pairs increases with increasing salt concentration. The $Cs^+$-$Cl^-$ coordination number is around 2 – 3 in the most concentrated solutions, with a $Cs^+$-$Cl^-$ distance around 3.35 – 3.5 Å.

The values of the structural parameters found in the literature have a fairly broad distribution, either obtained by simulation or experimentally. Since the structural parameters cannot be measured directly, the interpretation of the measured data also influences the (so-called) experimental values. This also increases the uncertainty of these values. During the development of FFs, the appropriateness of targeted experimental data is crucial; hence it is necessary to provide more accurate target values.

Most FFs consist of complete parameter sets for all halogen and alkaline ions. The target properties of these FFs are different, but they are usually not the structural parameters of concentrated solutions. The ability of one model to describe the structure of one type of salt solution does not mean that the same FF will give correct values in the case of another solution. (It will be shown later, that FFs, which predict the structure of aqueous LiCl solution well, can completely fail for aqueous CsCl solution and vice versa).

30 different 12 – 6 Lennard-Jones type ionic interatomic potential models, developed for rigid, non-polarizable water models, are investigated in this study. They are used to simulate highly concentrated aqueous CsCl solutions. The appropriateness of the models is determined by comparing the simulated results with experimental data, concerning both the structure and some basic properties. PRDFs and total scattering structure factors (TSSF) are calculated from the obtained atomic configurations and are compared with diffraction data. The structural results of the best models are further investigated to describe the structure in more detail.

The purpose of this study is twofold: (1) to ascertain which FF parameters are appropriate for the description of aqueous CsCl solutions and (2) to determine the structural parameters (bond lengths, coordination numbers) more precisely to get values, which can be targets for future FF-development.

## 2. Methods

*2.1 Investigated concentrations, experimental ND and XRD diffraction data*

Aqueous cesium chloride solutions were investigated at three concentrations, namely at 1.5, 7.5, and 15 mol %. Neutron and X-ray diffraction data are known for all these concentrations [12]. The exact concentrations (number of ions per water molecules) and the corresponding number densities were taken from Ref [12] and are shown in Table 1. Simulations were performed in cubic simulation boxes, using periodic boundary conditions, the box sizes were calculated from the number densities. The number of atoms in the boxes was around 10000; the exact number of ions and water molecules are summarized in Table 1.

*2.2 Interatomic potentials (force field models)*

Pairwise additive, non-polarizable potential models were investigated, in which the intermolecular interactions between atoms *i* and *j* are described by the 12-6 Lennard-Jones (LJ) interaction and the Coulomb potential:

$$V_{ij}(r_{ij}) = \frac{1}{4\pi\varepsilon_0}\frac{q_i q_j}{r_{ij}} + 4\varepsilon_{ij}\left[\left(\frac{\sigma_{ij}}{r_{ij}}\right)^{12} - \left(\frac{\sigma_{ij}}{r_{ij}}\right)^{6}\right] \qquad (1)$$

Here $r_{ij}$ is the distance between two particles, *i* and *j*, $q_i$ and $q_j$ are the point charges of the two particles, and $\varepsilon_0$ is the vacuum permittivity. $\varepsilon_{ij}$ and $\sigma_{ij}$ represent the energy and distance parameters of the LJ potential. The ionic $\varepsilon_{ii}$, $\sigma_{ii}$, and $q_i$ values for the investigated models are collected in Table 2 (also shown in Fig. S1 in the Supplementary Material (SM)). The tested FFs apply one of the following water models: SPC/E [28], TIP3P [29], TIP4P [29], TIP4P-Ew [30] and TIP4P/2005 [31] (also shown in Table 2). All of these water models are rigid and non-polarizable, the first two are 3-site models (3 point-like atoms representing the oxygen and the two hydrogen atoms), the other three models use a fourth, virtual site also (the partial charge of the oxygen atom is located on the virtual site). The LJ parameters of the hydrogen atom are 0 for all of these water models; the other parameters are collected in Table 3.

The LJ parameters between unlike atoms (anion-cation and ion-oxygen) are calculated according to combination rules. Most tested models (except for one) use one of the following rules: a) the geometric combination rule (where both $\varepsilon_{ij}$ and $\sigma_{ij}$ are calculated as the geometric average of the homoatomic parameters) or b) the Lorentz-Berthelot combination rule (in which $\varepsilon_{ij}$ is calculated as geometric, while $\sigma_{ij}$ as the arithmetic average of the relevant parameters). The applied combination rules are shown in Table 2.

The tested interatomic potential sets are listed below. Several of the FFs studied here were also tested for aqueous LiCl solutions; a more detailed description of those FFs can be found in Ref. [5].

**Ion parameters with SPC/E water**:
- *Dang-Smith parameter set* (DS) [32, 33]: One of the most frequently used FFs, it was applied in Ref. [12] also.
- *Joung-Cheatham III set* (JC-S) [34]: Parameters were fitted for hydration free energies, lattice energies, and the lattice constants of salt crystals. The authors developed parameters for three different water models (all of them are tested here).
- *Horinek-Mamatkulov-Netz parameter sets* (6 models: HS-g, HS-LB, HM-g, HM-LB, HL-g, HL-LB) [35] Solvation free energies and solvation entropies were the targets of these models. The ion-oxygen parameters were determined, leaving the choice of the combination rule free. Therefore geometric and LB rules were also tested here. Three different parameter sets are given by the authors of Ref. [35]: with a small (S), intermediate (M), and large (L) LJ interaction strength.
- *Gee et all. FF* (Gee) [36]: This model was designed to reproduce Kirkwood-Buff integrals. It should be noted here that there is a typo in the original paper: the $\varepsilon_{CsCs}$ parameter is correctly 0.065 [37]. For the $\varepsilon_{CsO}$ parameter they used a modified geometric combination rule: $\varepsilon_{CsO} = 0.95(\varepsilon_{CsCs} \varepsilon_{OO})^{1/2}$.
- *Reif-Hünenberger force fields* (RH, RM, RL) [38]: Three different hydration free energy values have been used for the determination of these FF parameters (high (H), medium (M), and low (L)).
- *Fyta-Netz force fields* (FN6, FN9) [39, 40]: These parameters were determined to reproduce the experimental free energy of solvation for a given single ion, the ion-pair properties were taken into account through fitting osmotic coefficients. For the cesium ion two different LJ-parameters are given, which were similarly appropriate in the criteria mentioned above; the notation of Ref. [39] is followed here, the numbers 6 and 9 are used to distinguish the two sets.

- *Deublein-Vrabec-Hasse* (DVH) [41] *and Reiser-Deublein-Vrabec-Hasse* (RDVH) [42] *parameter sets*: The $\sigma$ parameters of these models were determined to obtain good agreement with the reduced density as a function of the solution salinity. The $\varepsilon$ parameters are fixed at a constant value for all ions: DVH and RDVH sets differ only in their $\varepsilon$ values.
- *Li-Song-Merz FF* (Li-HFE-S, Li-IOD-S) [43]: Two 12-6 type LJ parameter sets are available in Ref. [43], they were fitted to hydration free energies (Li-HFE) and ion-oxygen distance (Li-IOD). Both of them were optimized to all three water models (see later).

**Ion parameters using TIP3P water model:**
- *Joung-Cheatham III set* (JC-T3) [34]: Same as JC-S, adopted to TIP3P water.
- *Li-Song-Merz FF* (Li-HFE-T3) [43]: similar to Li-HFE-S.
- *Mamatkulov-Schwierz parameters* (MS) [44]: These parameters were developed to reproduce thermodynamic and kinetic properties simultaneously: namely the experimental solvation free energy, the activity derivative, and the characteristics of water exchange from the first hydration shell of the metal cations.

**Ion parameters for TIP4P water model**:
- *Jensen-Jorgensen set* (JJ) [45]: A model developed with the purpose to create a consistent parameter set for the frequently used OPLS-AA [46] force field. The authors tried to fit the parameters to liquid-phase data: the experimental free energies of hydration and the locations of the first maxima of the ion-oxygen radial distribution functions.

**Interatomic potential sets using the TIP4P-Ew water model**:
- *Joung-Cheatham III set* (JC-T) [34]: see above
- *Li-Song-Merz FF* (Li-HFE-T, Li-IOD-T) [43]: both the parameters for reproducing the hydration free energy (HFE) and the ion-oxygen distance (IOD) were calculated and tested with TIP4P-Ew water model.

**Solvent independent force field:**
- *Mao-Pappu parameter set* [47] (MP-S, MP-T): A solvent-independent approach was used to calibrate the parameters based on crystal lattice properties. It was tested here using the SPC/E (MP-S) and TIP4P-Ew (MP-T) water models.

**Scaled-ionic-charge models:**
- *Kann-Skinner models* (KS-RL, KS-Li-IOD-T) [48]: Kann and Skinner proposed a simple charge-scaling method to improve the diffusion trends of water. The method is based on the ratio of dielectric constants of the solvent calculated from simulation and experiment. The effect of charge-scaling was tested with two previously presented models: RL (with SPC/E water) and Li-IOD-T (with TIP4P-Ew water) models.

- *Madrid-2029 force field* (Madrid) [2]: This model uses the TIP4P/2005 water model. The charges of the monovalent ions are scaled to $\pm 0.85e$. The target property of this FF is the density, even at high salt concentration. The parameters were determined in a way to avoid the formation of precipitation, thus keeping the number of contact ion pairs relatively low. The authors also reported improvements in the viscosity and self-diffusion coefficient values obtained by simulations using this model. The model does not use combination rules to calculate the LJ parameters between unlike atoms, all $\varepsilon_{ij}$ and $\sigma_{ij}$ parameters are given one by one.

*2.3 Molecular dynamics simulations*

Classical molecular dynamics simulations were performed by the GROMACS software (version 2020.6) [49]. Ions and water molecules were initially placed into the boxes randomly. Water molecules were kept together rigidly by the SETTLE algorithm [50]. Van der Waals interactions were truncated at 10 Å, with added long-range corrections to energy and pressure [51]. Coulomb interactions were treated by the smoothed particle-mesh Ewald (SPME) method, using a 10 Å cutoff in the direct space [52, 53]. Energy minimization was carried out using the steepest descent method. After that, the leap-frog algorithm was used for integrating Newton's equations of motion, using a 2 fs time step. Equilibration was carried out at constant volume and temperature (NVT) at T = 300 K. In the first equilibration stage, which was 200 ps long, the Berendsen thermostat [54] was used with a time constant $\tau_T = 0.1$, after that the Nose-Hoover thermostat [55, 56] was applied to keep the temperature constant, with $\tau_T = 2.0$. After a 4 ns long second equilibration stage, configurations were saved in every 2 ps during a 20 ns long production run. For density determination, a 4 ns long NpT simulation was carried out. The pressure was kept at $p = 10^5$ Pa by Parrinello-Rahman barostat [57, 58], using a $\tau_p = 2.0$ coupling constant. In these simulations the Nose-Hoover thermostat was used with $\tau_T = 0.5$ coupling.

*2.4 Calculation of the partial radial distribution functions and the structure factors*

Every 10[th] (altogether 1000) configurations were used for calculating the partial radial distribution functions (PRDF, $g_{ij}(r)$) with the 'gmx rdf' program of the GROMACS package. The partial structure factors ($S_{ij}(Q)$) were calculated as

$$S_{ij}(Q) - 1 = \frac{4\pi\rho_0}{Q} \int_0^\infty r(g_{ij}(r) - 1)\sin(Qr)\mathrm{d}r. \tag{2}$$

Here $Q$ is the amplitude of the scattering vector, and $\rho_0$ is the average number density. The neutron and X-ray total scattering structure factors (TSSF) were obtained as

$$S(Q) = \sum_{i \leq j} w_{ij}^{X,N}(Q) S_{ij}(Q). \tag{3}$$

Here $w_{ij}^{X,N}$ are the scattering weights for X-ray ($w_{ij}^X$) and neutron ($w_{ij}^N$), they are given by equations (4 and 5):

$$w_{ij}^N = (2 - \delta_{ij}) \frac{c_i c_j b_i b_j}{\sum_{ij} c_i c_j b_i b_j}. \tag{4}$$

$$w_{ij}^X(Q) = (2 - \delta_{ij}) \frac{c_i c_j f_i(Q) f_j(Q)}{\sum_{ij} c_i c_j f_i(Q) f_j(Q)}, \tag{5}$$

where $\delta_{ij}$ is the Kronecker delta, $c_i$ denotes the atomic concentration, $b_i$ is the coherent neutron scattering length, and $f_i(Q)$ is the atomic form factor. The weighting factors for the investigated concentrations are shown in Fig. S2 (in the SM). In the experiments of Mile et al. [12], the hydrogen atoms of water molecules were isotopically substituted by deuterium (D), the weights are calculated accordingly. For simplicity, the related PRDFs (O-D, D-D, Cs$^+$-D, Cl$^-$-D) will be noted as O-H, H-H, Cs$^+$-H, Cl$^-$-H.

The comparison of measured and simulated TSSFs is performed by calculating the *R*-factors ('goodness-of-fit' values):

$$R = \frac{\sqrt{\sum_i (S_{\mathrm{mod}}(Q_i) - S_{\mathrm{exp}}(Q_i))^2}}{\sqrt{\sum_i (S_{\mathrm{exp}}(Q_i) - 1)^2}}. \tag{6}$$

Here $S_{\mathrm{mod}}$ and $S_{\mathrm{exp}}$ are the model and experimental structure factors, and the summation is over the $Q_i$ experimental points.

*2.5 Hydrogen bond calculations*

Every 100$^{\mathrm{th}}$ (altogether 101) configurations were used for hydrogen bond analysis. Hydrogen bonds (HBs) were identified using energetic criteria: two water molecules are identified as H-bonded if the intermolecular distance between an oxygen and a hydrogen atom is less than 2.5 Å, and the interaction energy is less than -12 kJ/mol. A similar definition is used in the case of the chloride ion using the same energy boundary and a cut-off distance (H...Cl$^-$), which is defined as the first minimum value of the $g_{\mathrm{ClH}}(r)$ curve (it depends on the FF). Determination of the H-bonded molecules and the calculations about the H-bonded network were performed by an in-house program based on the HBTOPOLOGY code [59].

## 3. Results and discussion

### 3.1 *Precipitation*

As was found for NaCl [3] and LiCl [5] solutions, if the parameters of a FF are not chosen appropriately, salt precipitation can appear in the models below the experimental solubility limit. It happened in the case of the CsCl solutions for the HS-LB FF: the presence of CsCl precipitate can be observed, even at the lowest investigated concentrations, see Fig. S3 (of the SM). This behavior is also apparent from the PRDFs of this model. The configuration, obtained by the HS-g model at 15 mol % concentration, is also precipitated. At the highest concentration, snapshots of some models (e. g., FN6, JC-S, MS) also show clustering of the ions; however, the other properties predicted by these FFs, do not indicate the presence of precipitation in these models so clearly.

### 3.2 *Density, static dielectric constant, self-diffusion coefficients*

These properties are calculated for each model; the results are shown in the Supplementary Material in more detail.

The experimental densities are well reproduced by the models (within 1-3 %); the differences are higher only in the case of the 15 mol % sample for some FFs: for the DS, RH, DVH, RDVH, and JJ models, where the simulated values are 5 – 7 % smaller than the experimental ones, see Fig. 1. Reducing the ionic charges cause lower density values.

The static dielectric constants are smaller than the experimental ones; the difference is higher at higher salt concentrations (10 – 38 % for the 1.5 mol % solution and 22 – 60 % for the 15 mol % sample), see Fig. 2. This difference is caused partly by the applied water models, which give values in the rather wide, 50 – 94 range for pure water (the experimental value is 78.5). Thus the reduced $\varepsilon/\varepsilon_{water}$ values are also calculated (Fig. S5 in the SM) to better evaluate the contribution of the ion parameter sets. The models which give results closer to the experimental values are HS-g, JC-S, RDVH, FN6, and DVH. The highest differences between experiential and simulated values are in models JC-T3, Li-HFE-T3, Madrid, MS, and RH. The scaled ionic charges result in higher static dielectric constant values.

The simulated self-diffusion coefficients are generally around the experimental values at the 1.5 mol % salt concentration, and they are significantly smaller than the experimental ones for higher concentrations. The FFs applying the TIP3P predict higher self-diffusion values than the other models.

The charge-scaled KS models perform slightly better than their full charge versions: all self-diffusion values (water, Cs, and Cl) increase with decreasing ionic charge. The ions in the HS-LB models do not move even at the 1.5 mol % salt concentration. The self-diffusion coefficients of ions and water are very low in RH and RM models. (See Figs. 3, S6, and S7 in the SM).

Another possibility offered by the detailed comparison of a multitude of FFs is to analyze subtle variations between members of model families in order to identify trends. Beside the effect of charge scaling, which was detailed previously, such example is the family of the Horinek models (HS, HM and HL). The analysis of these models show that the density and water self-diffusion values does not vary significantly between the HL-HM-HS models, but the static dielectric constant increases as the interaction strength decreases. At the same time the ionic mobility decreases and the tendency to precipitate increases.

3.3 *Neutron and X-ray total scattering structure factors*

Partial radial distribution functions were calculated from the obtained trajectories, they will be discussed in detail in the next section. Neutron and X-ray TSSFs were calculated from the PRDFs, and are compared with experimental results from Ref. [12]. Calculated $S^N(Q)$ and $S^X(Q)$ functions of some selected FFs are shown in Figs. 4 and 5. (The TSSFs obtained for the 15 mol % solution are presented in the SM (Figs. S8-S12 for all FFs.) Comparison with the experimental data is also made by calculating the *R*-factors, see (Eq. 6), which are shown for all FFs in Figs. 6 and 7.

In the case of the less concentrated composition (1.5 mol %), $S^N(Q)$ depends mainly on the applied water model: FFs using 4-site water models perform best, while the agreement with experimental data is the poorest for models using the TIP3P water. The *R*-factors for the FFs with 4-site water models are around 10 %, they are 16-18 % for the FFs with the SPC/E model, and more than 20 % for the models with the TIP3P water. At higher concentrations, the applied ionic parameters have more influence on the ND fits, and the *R*-factor has broader distribution: its value is between 10 % and 50 % for the 15 mol % sample. In general, the FFs with 4-site water models give better results. However, there are also significant differences between these FFs (see, for example, Li-HFE-T and Li-IOD-T models). Concerning the FFs with 3-site water models, the role of the water model seems to be less important: there are huge differences between the FFs: see e. g. RH ($R$ = 52 %) and RL ($R$ = 17 %) models. The most significant deviations are at around $Q = 1$ Å$^{-1}$ and $Q < 1$ Å$^{-1}$. In the experimental curve, a small pre-peak emerges around $Q = 1$ Å$^{-1}$, while in the simulated curves, this peak is missing for some FFs (e.g. FN9, DS, Gee) or has a much higher amplitude than the experimental value (e.g. RH, RM). For several simulated curves, $S^N(Q)$ increases as $Q \to 0$, e. g. DS, DVH, FN6, JC-S models. The best models, which give good results for all concentrations investigated are Li-IOD-T and KS-Li-IOD-T, followed by models MP-T and JC-T. The best 3-site models are: Li-IOD-S, HL-g, MP-S, FN9, HL-LB, RL, KS-RL.

The $R$-factor values of the X-ray TSSFs ($S^X(Q)$ functions) obtained by simulations using different FFs are between 6% and 20% in the case of the 1.5 mol % solution, except the HS-LB model, which has an $R$-factor as high as 45 %. In the 15 mol % sample, the $R$-factors are between 6% and 38%, except the HS-LB model with 80% and the HS-g model with 67%. Beside the HS-LB model, which is definitely precipitated even at the lowest concentration, the HS-g model also has salt precipitations at 15 mol %. Some other models show clustering of the ions at the highest concentration, e. g. JC-S or FN6 models, which FFs also produce high $R$-factors. The performance of the Madrid model is surprisingly poor, despite the absence of ion clusters in the configurations, which suggests that the lack of ion clusters is not sufficient, not even necessary to ensure a correct structural model for aqueous CsCl salt solutions.

In some simulated curves, there is a small pre-peak around $Q = 1$ Å$^{-1}$, similarly to the ND curve, however, in the experimental data, only a shoulder can be seen. In the experimental curve, the 1$^{st}$ (or main) peak is around $Q = 2.1$ Å$^{-1}$. A 2$^{nd}$ peak (for the 1.5 mol % solution) or a shoulder (for the 15 mol % solution) is at around $Q = 2.6 – 2.8$ Å$^{-1}$, followed by a 3$^{rd}$ peak at about $4 – 4.5$ Å$^{-1}$. Though these peaks are identifiable on the simulated curves, their positions and amplitudes vary from FF to FF. The ratio of the amplitudes of the 1$^{st}$ and 2$^{nd}$ peaks (at $Q = 2.1$ Å$^{-1}$ and 2.8 Å$^{-1}$) is also different. Concerning the 1.5 mol % sample, there are 12 models with similarly good $R$-factors. As the concentration increases, the $R$-factor of some models remains nearly constant (e. g. DS, Li-HFE-T), but for some FFs the $R$-factor increases (e. g. MP-T, RDVH). In some cases, the more concentrated solution has a better fit (mostly for the models with TIP3P water, see e. g. Li-HFE-T3). The lowest $R$-factors belong to the following models: KS-Li-IOD-T, KS-RL, Li-IOD-S, Li-IOD-T, Li-HFE-T, RL.

Since different partials play a significant role in the neutron and X-ray TSSFs, the most appropriate description of the structure is given by the models that have good results in both diffraction experiments. Thus the best models are KS-Li-IOD-T, Li-IOD-T, Li-IOD-S, KS-RL, and RL. It should be noted here that the JC-T model, which was one of the best in the case of aqueous LiCl solutions [6] and was also tested for experimental EXAFS data on CsCl solutions [4], is not among the best models concerning the structure of aqueous CsCl solutions. Thus the finding of ref. [4] that classical, non-polarizable MD models are not suitable for describing the structure of aqueous CsCl solutions is unfounded since there are classical models performing significantly better than the model they have chosen.

## 3.4. Structure determination via the comparison of FFs

### 3.4.1 Hydration shell of the anions (Cl⁻)

The $g_{ClH}(r)$ and $g_{ClO}(r)$ curves for some FFs are shown in Figs. 8 and 9. They have a well-defined 1$^{st}$ peak, followed by a relatively deep minimum for all investigated FFs. This peak originates from the water molecules hydrating the anion. The positions of the first peaks ($r_{ClH}$ and $r_{ClO}$) do not depend on salt concentration in the investigated concentration range, but they depend strongly on the applied ionic potential model, see Table 4 and Fig. 10. The position of the first minimum is well defined and concentration-independent for the Cl⁻-H pairs. In the case of the Cl⁻-O pairs, the first minimum is not as deep, it is indistinct, and slightly depends on concentration (see Tables S6 and S7).

The coordination numbers are calculated from the PRDFs as:

$$N_{ij} = 4\pi\rho c_j \int_0^R g_{ij} r^2 dr, \qquad (7)$$

where $c_j$ is the concentration of the $j^{th}$ species, and $R$ is the upper limit of the coordination sphere. Usually, the upper limit of the integration is defined as the first minimum after the first peak. The $N_{ClH}$, $N_{ClO}$ coordination numbers (calculated up to the first minima) are also collected in Table S6 and S7, and shown in Fig. 11.

The $N_{ClH}$ and $N_{ClO}$ coordination numbers are close to each other (within 10 %), although $N_{ClO}$ is slightly higher than $N_{ClH}$. This shows that water molecules connect to Cl⁻ ions with only one of their H atoms. The water molecules are oriented with one of their H atoms towards the anion; the other H atoms of these first shell water molecules produce the 2$^{nd}$ peak of the $g_{ClH}(r)$ curve. This orientation is also verified by the Cl⁻–H–O angle, which is nearly 180°, as shown for some models in Fig. S13. The slightly higher $N_{ClO}$ coordination number can be an artifact, caused by the not-so-well defined minimum in $g_{ClO}(r)$ and the slight overlap of the 1$^{st}$ and 2$^{nd}$ Cl⁻-O peaks (hydration shells). The 2$^{nd}$ peak of the $g_{ClO}(r)$ curves can originate from Cl⁻-O distances in Cl⁻-H-O-H...O motifs or Cl⁻-Cs⁺-O connections.

For most models 5.7 – 7 water molecules can be found on the 1$^{st}$ hydration shell of Cl⁻ ions at 1.5 mol %, which number decreases to 4.8 – 6.5 for the 7.5 mol % solution and 3.7 – 5.9 in the 15 mol % sample. (For those models, in which precipitation appeared, the $N_{ClH}$ coordination number is, of course, much lower.) The best model, the KS-Li-IOD-T, gives 6.15, 5.38, and 4.58 for 1.5, 7.5, and 15 mol %, respectively.

The Cl⁻-H bond lengths are distributed between 1.96 Å and 2.44 Å. Taking into account the performance of the FFs in reproducing the experimental total structure factors, it can be concluded that for the most appropriate FFs, the $r_{ClH}$ distance is in the 2.11 – 2.23 Å range (for the KS-Li-IOD-T model this number is 2.23 Å). The $r_{ClO}$ distances are 3.1 – 3.18 Å for these models (3.17 Å for KS-Li-IOD-T). This finding is in good agreement with the experimental EXAFS results in Ref. [18], where $r_{ClO}$ = 3.15 Å was obtained.

3.4.2 *First hydration shell of the cations (Cs$^+$)*

The first hydration shell of the cesium ions can be studied through the $g_{CsO}(r)$ curves; for some FFs, see Fig. 12. The upper limit of the first shell is not well defined, the minimum following the first maxima is not deep. At 1.5 mol %, in the case of some FFs, a broad and small 2$^{nd}$ peak represents the 2$^{nd}$ coordination shell, but for several FFs and at higher concentrations, the 2$^{nd}$ shell cannot be determined at all. The position of the 1$^{st}$ peak depends on the FF and is between 2.80 and 3.50 Å. The $r_{CsO}$ distance is in the 3.05 – 3.16 Å range in the most appropriate models (3.16 Å for KS-Li-IOD-T), which agrees well with the value 3.08 Å obtained in EXAFS measurement [4]. Because of the lack of a well-defined upper limit of the shell, the uncertainty of the $N_{CsO}$ coordination number is also high. The bond length and coordination numbers are shown in Tables 4 and S8, and Figs. 10 and 11. For most FFs, the $N_{CsO}$ coordination numbers are between 6.6 – 9, 6 – 8.5, and 5 – 8 for the 1.5, 7.5, and 15 mol % solutions, respectively. Some FFs give lower hydration numbers; these are the models with precipitations (HS-LB, HS-g). In contrast, there are two models with significantly higher $N_{CsO}$ values: the RH and RM models, which also have the longest bond distance; in these models, the 1$^{st}$ peak is strongly asymmetric with an elongated decrease, the limit of the 1$^{st}$ hydration shell cannot be reliably determined. For the best model, the KS-Li-IOD-T, $N_{CsO}$ is 8.56, 7.58, and 7.4 for 1.5, 7.5, and 15 mol % solutions, respectively.

3.4.3 *Contact ion-pairing*

The first peak of the $g_{CsCl}(r)$ curve (around 3.5 Å) originates from the Cs$^+$ – Cl$^-$ contact ion pairs (Fig. 13). It is a sharp, well-defined peak, followed by (for most models) a rather deep minimum (except Madrid FF, see below). The 2$^{nd}$ peak is small and broad; it is produced by those Cs$^+$ – Cl$^-$ distances, for which the ions are connected through a common water molecule (Cs$^+$ - O-H – Cl$^-$ motifs). The position of the 1$^{st}$ peak ($r_{CsCl}$) depends on the FFs, its value is in the 3.08 – 3.80 Å range, see Table 4 or Fig. 10. The most appropriate FFs give results between 3.45 – 3.60 Å (it is 3.55 Å for the KS-Li-IOD-T), which are close to the value obtained in EXAFS measurements (3.5 Å) [18]. The smallest values belong to the HS-LB, HS-g, JC-S, and FN6 models, showing that the ion-pairing and thus the tendency of ion clustering is strong in these models. At the other end of this range are the RH, RM, JJ, and RDVH models that all give $r_{CsCl} > 3.6$ Å.

The $N_{CsCl}$ coordination number (the number of contact ion pairs) is usually in the range of 0.16 – 0.48, 0.68 – 1.65, and 1.76 – 3 in the 1.5 mol %, 7.5 mol % and 15 mol % solutions, respectively, see Table S9 or Fig. 11. (The exceptions are HS-LB, which has $N_{CsCl} > 4$ in all solutions, HS-g in the 15 mol % sample, and Madrid, which is discussed later in detail.) In the most appropriate models, these numbers are 0.28 – 0.38, 0.99 – 1.35 and 1.86 – 2.42, KS-Li-IOD-T gives 0.38, 1.35 and 2.42 for the 1.5, 7.5 and 15 mol % solutions, respectively.

As the concentration increases the number of contact ion pairs increases, while the number of hydrating water molecules decreases. Around Cl⁻ ions the total number of water molecules and contact ion pairs together ($N_{Cl} = N_{ClH} + N_{ClCs}$) is similar in all concentrations, the deviation from their average is below 4 %. ($N_{Cl}$ either slightly increases with the concentration or does not change monotonically, depending on the FF). The $N_{Cl}$ total coordination number depends on the FF and is usually $6.5 < N_{Cl} < 7.5$ (except Madrid and HS-LB, for which FFs $N_{Cl} = 5.8$). For the best models $N_{Cl}$ is between 6.75 and 7.2 (for KS-Li-IOD-T $N_{Cl} = 6.75$).

A similar rule holds for the Cs⁺ ions: $N_{Cs}=N_{CsCl}+N_{CsO}$ is more or less constant: the difference is mostly below 5 %. The total coordination number of Cs⁺ ions obtained by different FFs has a broad distribution, however, between 7 and 9.5. For the most appropriate models, it is $8.5 < N_{Cs} < 9.5$ (for the KS-Li-IOD-T, $N_{Cs}$ is 9.23).

The Madrid model is unique in this comparison because it was developed specifically to produce an extra-low number of ion pairs. The shape of the $g_{CsCl}(r)$ curve obtained by this FF is different from that of the other models (see Fig. 13). There is no sharp first peak in the 3 – 4 Å region. The first visible peak is around 5 Å, with a shoulder on the left side. The peak around 5 Å originates from the solvent-separated ion pairs, while the shoulder represents the contact ion pairs. Since the contributions of contact and solvent-separated ion pairs overlap, the number of contact ion pairs cannot be determined precisely. The upper limit of the integration was chosen to be the slight minimum between the shoulder and the peak. The coordination numbers obtained that way are in line with the values of Ref. [2]; they are 0.06, 0.25, and 0.5 for the 1.5, 7.5, and 15 mol % solutions, respectively. Compared to other FFs, the number of ion pairs only slightly increases, and the number of hydrating water molecules around the ions only slightly decreases as the salt concentration increases. The total coordination numbers of ions are also smaller than for the other FFs ($N_{Cl} = 5.8$ and $N_{Cs} = 6.7$). For aqueous LiCl solutions, keeping the number of contact ion pairs as low as possible proved to be a promising strategy to generate appropriate models [6]. However, in the case of aqueous CsCl solutions, models with too high and too low number of contact ion pairs both result in poor agreement with diffraction data.

3.4.4 *Neighborhood of water molecules*

The environment of the water molecules can be deduced from the O-H (O-O), Cl$^-$-H, and Cs$^+$-O PRDFs. The ion-water nearest neighbors were discussed in the previous sections. The water-water, hydrogen-bonded connections can be seen as the first intermolecular peak in the $g_{OH}(r)$, which is located around 1.8 Å (1.76 – 1.84 Å); the position of the peak depends only on the applied water model (Fig. 14). The $N_{OH}$ coordination numbers (which are uniformly determined up to 2.4 Å for all FFs) are collected in Table S10 and shown in Fig. 11. As the salt concentration increases, the number of water-ion connections increases and the number of H-bonded water pairs decreases. The water molecules connect with their H atoms to O atoms of other water molecules (H-bond) or to Cl$^-$ ions. The $N_H = N_{HO} + N_{HCl}$ coordination number is around 0.9 – 1 for all investigated FFs; the higher numbers belong to the higher concentrations, which means that nearly all of the H atoms of water molecules participate in H-bond or water – Cl$^-$ connections.

In pure water, the oxygen atoms of water molecules link to two water molecules through H-bonds. In the salt solution, they lose one (or two) of their H-bonded water pairs to form water – Cs$^+$ pairs instead. Since the upper limit of the hydration shell of Cs$^+$ ions and thus the $N_{OCs}$ coordination number has high uncertainty, the $N_O = N_{OH} + N_{OCs}$ total coordination of O atoms is also quite uncertain. Its value is more or less around 2 for the 1.5 and 7.5 mol % solutions, but it is mostly higher than 2 in the 15 mol % solution. However, the lack of a well-defined minimum in the $g_{CsO}(r)$ can cause the overestimation of the $N_{CsO}$ coordination number.

The first peak in the $g_{OO}(r)$ PRDF (around 2.75 Å) originates mainly from the H-bonded water molecules (Fig. 15). However, the first minimum of this peak is not well-defined, and for the FFs applying the TIP3P water model, there are no minima at all. Thus the $N_{OO}$ coordination number was determined uniformly up to 3.2 Å for all FFs (Table S11). O – O distances up to this value still contain water pairs, which are not connected via H-bonds; this can be seen in the difference of the $N_{OO}$ and 2*$N_{OH}$ coordination numbers, which are around 0.2 – 0.5.

In the 1.5 mol % solution, the 2$^{nd}$ peak in the $g_{OO}(r)$ is around 4.5 Å; these distances are the 2$^{nd}$ neighbors, the O – O pairs, which have a common H-bonded water molecule. In the 15 mol % solution, a small 2$^{nd}$ peak can be seen in the O-O PRDF at shorter distances, around 3.5 Å, while the pair distances around 4.5 Å are significantly less frequent. Distances of 3.5 Å may result from water pairs, which are paired to a common ion. (Water molecules, which are in the hydrate sphere of the same ion).

3.5 *Structural origin of special features in the total scattering structure factors*

The shapes of the total scattering structure factors depend on the applied potential model. The TSSFs are calculated from the PRDFs, so dissimilarities in TSSFs result from differences in PRDFs. In this section, two $Q$ regions are investigated to find the reason for the difference in TSSFs obtained by various FFs. Only the results and the main steps of these calculations are presented here; a detailed description can be found in the SM.

Two significant differences can be observed between the potential models in the $S^N(Q)$ curves of the 15 mol % solution at low $Q$ values: around $Q = 1$ Å$^{-1}$ and below that value. At $Q = 1$ Å$^{-1}$, there is a small peak in the experimental curve, it is usually called "pre-peak", or "first sharp diffraction peak" (FSDP). The presence of this peak is often associated with the medium range order in the system. The existence and height of this peak in the simulated curves highly depend on the applied potential model (see Fig. 4). Below the pre-peak region, as $Q \to 0$, some of the calculated $S^N(Q)$ functions begin to increase, which is in contrast to the behavior of the experimental curve (e. g. for the MS, Li-HFE-S, Li-HFE-T, Li-HFE-T3, DVH, RDVH, JC-S, and FN6 models). This behavior was observed previously (see Ref. [12]), in that study, the DS FF parameters were used in the MD simulation.

By comparing the structural predictions and calculated TSSFs of several different FFs, it is possible to determine the structure of the solutions more accurately [6]. Here this method is used to reveal the structural origin of the discrepancies in the $S^N(Q)$ curves at $Q = 1$ Å$^{-1}$ and $Q = 0.5$ Å$^{-1}$.

The amplitude of $S^N(Q)$ at a given $Q$ value is determined as a weighted sum of the partial $S_{ij}(Q)$ curves ($w_{ij}^N * S_{ij}(Q)$), see Eq. 3. The comparison of the weighted partial structure factors at the pre-peak position ($Q = 1$ Å$^{-1}$) shows that the main differences are in the Cl$^-$-H, H-H, and O-H partial structure factors, while at $Q = 0.5$ Å$^{-1}$, the O-H and H-H partials cause the main difference. The $S_{ij}(Q)$ curves are determined as the integral of the $r*(g_{ij}(r)-1)*\sin(Qr)$ function (Eq. 2), thus with calculating the latter at $Q = 0.5$ Å$^{-1}$ and $Q = 1$ Å$^{-1}$, one can find the $r$ regions, from which the main differences between the FFs originate. (These calculations for some selected FFs are shown in the SM.) It turned out that several $r$-regions, between 2 and 8 Å, have significant contributions to the differences (see more details below).

Concerning the $Q = 0.5$ Å$^{-1}$ curves: a significant part of the discrepancies can be derived to come from the 2 – 4 Å region of the H-H and O-H PRDFs. It is caused mainly by the different numbers of H-bonded water molecules in the different FFs. In Fig. 16a the $S^N(Q)$ values of the FFs at $Q = 0.5$ Å$^{-1}$ are shown as a function of the $N_{OH}$ coordination number predicted by the models. The FFs which give higher $S^N(Q=0.5$ Å$^{-1})$ values are the same as the models that give a higher number of H-bonded water molecules. A similar relation can be observed between the $S^N(Q = 0.5$ Å$^{-1})$ and the $N_{CsCl}$ coordination number (Fig. 16b). The clustering of the ions goes together with the decreasing number of ion-water pairs: the hydration shells of clustered ions are connected, they create a common hydrate shell. At the same time, the smaller number of ion-water pairs eventuates that the water molecules can establish more H-bonds between themselves. This phenomenon causes the higher values observed in the structure factor at low Q values.

The origin of the pre-peak cannot be assigned to one $r$ region or one structural motif: there are 3 – 4 regions between 2 and 8 Å where significant differences can be seen in the $r*(g_{ij}(r)-1)*\sin(Qr)$ functions of different FFs (in the $Cl^-$-H, O-H, H-H partials). Some of the shortest distances can be associated with one structural motif, such as the differences of the FFs in $Cl^-$-H curves around 2 Å are caused by the different number and distance of the water molecules around the chloride ions. In most cases, however, these distances cannot be attributed to a single structural motif; e. g. even the first intermolecular peak of the H-H PRDFs can be the result, not just the H-H distances in H-bonded water molecules, but also the H-H pairs of water molecules hydrating the same chloride ion.

The values of $S^N(Q = 1$ Å$^{-1})$, as a function of the short range order parameters ($Cl^-$-H, $Cs^+$-O, and $Cs^+$-$Cl^-$ bond lengths, and the $Cl^-$-H, $Cs^+$-O, $Cl^-$-$Cs^+$, and O-H coordination numbers) predicted by the models, were examined. Significant correlation was detectable only for one parameter: the $S^N(Q = 1$ Å$^{-1})$ highly depends on the $r_{CsO}$ bond-length predicted by the FF, see Fig. 17 and S24. This result is quite surprising since the contribution of $S_{CsO}$ to $S^N(Q)$ at $Q = 1$ Å$^{-1}$ is not significant. However, the $Cs^+$-O distance affects other quantities, such as the number of $Cs^+$-O pairs, the H-bonded water molecules around $Cs^+$ ions, the number of contact ion pairs, etc. The $Cs^+$-O distance changes the other partial $S_{ij}(Q)$ functions through these quantities, including those with significant weight in the TSSF. This result also shows that many structural motifs cause the pre-peak of the S(Q) curve together; the presence of the FSDP cannot be assigned to one special motif in the solution.

3.6 *H-bonded network*

Traditionally, hydrogen bonding (H-bond) means the interaction between water molecules; however, an identical mechanism is responsible for the connection between the chloride ion and the hydrating water molecule. The halide ion can be considered a constituent of the hydrogen-bonded network, as was discussed in a previous paper [7]. Here the H-bonded network structure of aqueous CsCl solutions is investigated for some selected models.

The H-bonded water-water and water-chloride ion pairs were specified via an energetic definition of H-bonding, see section 2.5. The size distribution of H-bonded clusters for the water-anion system and the water subsystem (taking into account only the water-water connections) are shown in Figs. 18 and 19. Considering the water-anion system, the size of the most extended clusters are in the order of the number of water molecules plus $Cl^-$ ions for all concentrations; only a slight decrease can be observed as the salt concentration increases. There are small differences between the FFs: the size of the longest clusters are slightly bigger for models in which the ion clustering is less frequent. The network of water molecules is percolating for the 1.5 and 7.5 mol % solutions; however, the size of the clusters decreases significantly as the salt concentration increases. There is a huge difference between the FFs in this regard: those models, in which there are more ion clusters, contain more water-water connections as well, and the water H-bonded network is percolating even for 15 mol % salt content. However, in models, which has a weak ion-paring tendency, the size of the water clusters is significantly smaller than the total number of water molecules.

H-bonded chains whose first and last molecules are H-bonded together are called cyclic entities or cycles. The size of a cyclic entity can be defined as the shortest path through the H-bonded molecules in the cycle. A cyclic entity that cannot be decomposed into smaller cycles is called a primitive ring (or shortly, ring). The size distribution of the rings is shown in Fig. 20. A behavior similar to that of the cluster distribution can be seen: the main differences between the models are in the number of pure water rings in the 15 mol % solution. The most frequent ring size is the 6-membered ring in the 1.5 mol % solution, while at 15 mol % salt content, the 5-membered rings are the most frequent.

The compositions of rings (the number of water molecules and $Cl^-$ ions in the ring) were also calculated for some FFs, see Figs. 21, S25, and S26 (in the SM). The purely water rings are the most frequent at the lowest investigated salt concentration in all tested models. As the salt concentration increases, the frequency of the rings containing 1 or 2 $Cl^-$ ions increases. At 15 mol % salt content, there is a significant difference in the ring type distribution of different FFs. For FFs which give the best S(Q) fits (e. g. KS-Li-IOD-T model, Fig. 21), (5 and 6 membered) rings containing 1 or 2 $Cl^-$ ions are nearly most frequent, their frequencies are nearly equal. In the models where ion clusters are present (such as the JC-S model, Fig. S25), rings with 1 $Cl^-$ ion are the most common. However, in the models where the number of the ion pairs is smaller (e. g. the RH model, Fig. S26), the rings with 2 $Cl^-$ ions are the most frequent types (the number of rings with 2 $Cl^-$ ions and 3 water molecules is particularly high).

4. Conclusions

Aqueous CsCl solutions were investigated by classical MD simulations. Thirty interaction models, all of them using 12-6 LJ potentials and rigid, non-polarizable water molecules, were examined. The calculations were made at three different concentrations from 1.5 mol % to 15 mol %. Density, static dielectric constant, and self-diffusion predictions of simulations were compared to experimental values from the literature. The partial radial distribution functions were calculated for all FFs, together with neutron and X-ray total scattering structure factors that were also compared with previous experimental results.

The simulated density values are close to the experimental values, the static dielectric constant values are smaller than the experimental ones. The self-diffusion coefficient values are close to the experimental ones at the lowest concentration; however, they are significantly smaller at the highest investigated concentration. Two FFs, namely HS-LB and HS-g, precipitate below the solubility limit.

The structure parameters predicted by the potential models were investigated in detail. There are huge differences between the models; however, comparing the calculated structure factors with those obtained in neutron and X-ray diffraction measurements made it possible to find the best performing models. Special features that appear in the structure factors calculated from different models also helped to classify the FFs and narrow down the possible values of bond distances and coordination numbers in the system.

It was found that there are models among the tested FFs, which appropriately describe the structure of aqueous CsCl solutions. The best models fit the diffraction data and even reproduce well the ion-oxygen distances obtained by EXAFS measurements. The best performing model is the charge scaled KS-Li-IOD-T model, closely followed by the Li-IOD-T, Li-IOD-S, KS-RL, and RL models.

It was found that the best models are those in which the $Cl^-$ - O, $Cs^+$ - O, and the $Cl^-$ - $Cs^+$ bond lengths are $2.17 \pm 0.06$ Å, $3.11 \pm 0.05$ Å, and $3.52 \pm 0.07$ Å, respectively.

Contact ion pairing is significant in aqueous CsCl solutions; however, it was found that the number of ion pairs should be limited ($N_{CsCl}$ is around 0.4 for 1.5 mol % and 2.4 for 15 mol % solution). Potential models that result in a too high or too low number of contact ion pairs failed in the tests equally.

The number of water molecules and contact ion pairs together in the vicinity of a chloride ion is about 7, while around a cesium ion it is about 9, in all investigated concentrations – according to the best models. The water molecules are participating in 4 connections (two by their oxygen atoms and 1-1 with the two hydrogen atoms): H-bonding with other water molecules and ion hydration. As the number of ions increases, the number of H-bonded water molecules decreases. The H-bonded network of the water molecules transforms into a network of $Cl^-$ ions and water molecules as the salt concentration increases. The H-bonded network formed by the water molecules and $Cl^-$ ions together is percolating, even at the highest salt concentration; while the water subsystem contains somewhat smaller clusters in the 15 mol % solution.

Some additional comments about the performance of the models:

(i) The charge scaling is successful concerning both the physical properties and structural behavior. The density decreases, the static dielectric constant and the self-diffusion coefficients increase as the charge of the ions decreases.

(ii) The suitability of the models for one type of salt does not automatically mean their good performance for another salt.

(iii) Though the number of contact ion pairs should not be too high to avoid precipitation below the solubility limit, the experimental data clearly shows that some amount of such pairs is present in the CsCl solutions.

(iv) Structure parameters (bond length, coordination numbers) should also be taken into account during the development of force field models. Thus it is very important to determine them (using experimental and simulation techniques together) as accurately as possible.

(v) Despite the serious limitations of simple, non-polarizable, point charge classical potential models, they can be suitable to describe the structure of aqueous salt solutions, but it is necessary to use experimental diffraction data to find (or create) the best possible parameter sets for that purpose.

Acknowledgments

The author is grateful to the National Research, Development and Innovation Office (NKFIH) of Hungary for financial support through Grant No 124885. Financial support from the Eötvös Loránd Research Network ('ELKH', Hungary), via their special fund, Grant No. SA-89/2021., is also acknowledged. The author would like to thank Zoltán Németh, László Temleitner and László Pusztai for helpful discussions. The author thanks Imre Bakó for sharing the computer code used for hydrogen bond analysis.

References

[1] N. Metropolis, A.W. Rosenbluth, M.N. Rosenbluth, A.H. Teller, E. Teller, Equation of state calculations by fast computing machines, J. Chem. Phys. 21 (1953) 1087–1092. doi:10.1063/1.1699114.

[2] S. Blazquez, M.M. Conde, J.L.F. Abascal, C. Vega, The Madrid-2019 force field for electrolytes in water using TIP4P/2005 and scaled charges: Extension to the ions $F^-$, $Br^-$, $I^-$, $Rb^+$, and $Cs^+$, J. Chem. Phys. 156 (2022) 044505. doi:10.1063/5.0077716.

[3]  F. Moučka, I. Nezbeda, W.R. Smith, Molecular force fields for aqueous electrolytes: SPC/E-compatible charged LJ sphere models and their limitations, J. Chem. Phys. 138 (2013) 154102. doi:10.1063/1.4801322.

[4]  D. Zhuang, M. Riera, G.K. Schenter, J.L. Fulton, F. Paesani, Many-body effects determine the local hydration structure of $Cs^+$ in solution, J. Phys. Chem. Lett. 10 (2019) 406–412. doi:10.1021/acs.jpclett.8b03829.

[5]  I. Pethes, A comparison of classical interatomic potentials applied to highly concentrated aqueous lithium chloride solutions, J. Mol. Liq. 242 (2017) 845–858. doi:10.1016/j.molliq.2017.07.076.

[6]  I. Pethes, The structure of aqueous lithium chloride solutions at high concentrations as revealed by a comparison of classical interatomic potential models, J. Mol. Liq. 264 (2018) 179–197. doi:10.1016/j.molliq.2018.05.044.

[7]  I. Pethes, I. Bakó, L. Pusztai, Chloride ions as integral parts of hydrogen bonded networks in aqueous salt solutions: the appearance of solvent separated anion pairs, Phys. Chem. Chem. Phys. 22 (2020) 11038–11044. doi:10.1039/D0CP01806F.

[8]  R.M. Lawrence, R.F. Kruh, X-ray diffraction studies of aqueous alkali-metal halide Solutions, J. Chem. Phys. 47 (1967) 4758–4765. doi:10.1063/1.1701694.

[9]  P.R. Smirnov, V.N. Trostin, Structures of the nearest surroundings of the $K^+$, $Rb^+$, and $Cs^+$ ions in aqueous solutions of their salts, Russ. J. Gen. Chem. 77 (2007) 2101–2107. doi:10.1134/S1070363207120043.

[10] P.R. Smirnov, Structural parameters of the nearest surrounding of halide ions in the aqueous electrolyte solutions, Russ. J. Gen. Chem. 83 (2013) 1469–1481. doi:10.1134/S107036321308001X.

[11] P.R. Smirnov, Structure of the nearest environment of $Na^+$, $K^+$, $Rb^+$, and $Cs^+$ ions in oxygen-containing solvents, Russ. J. Gen. Chem. 90 (2020) 1693–1702. doi:10.1134/S1070363220090169.

[12] V. Mile, L. Pusztai, H. Dominguez, O. Pizio, Understanding the structure of aqueous cesium chloride solutions by combining diffraction experiments, molecular dynamics simulations, and reverse Monte Carlo modeling, J. Phys. Chem. B 113 (2009) 10760–10769. doi:10.1021/jp900092g.

[13] Y. Zhou, Y. Soga, T. Yamaguchi, Y. Fang, C. Fang, Structure of aqueous RbCl and CsCl solutions using X-ray scattering and empirical potential structure refinement modelling, Acta Physico-Chimica Sin. 34 (2018) 483–491. doi:10.3866/PKU.WHXB201709111.

[14] N. Ohtomo, K. Arakawa, Neutron diffraction study of aqueous ionic solutions. I. Aqueous solutions of lithium chloride and caesium chloride, Bull. Chem. Soc. Jpn. 52 (1979) 2755–2759. doi:10.1246/bcsj.52.2755.

[15] T. Yamaguchi, H. Ohzono, M. Yamagami, K. Yamanaka, K. Yoshida, H. Wakita, Ion hydration in aqueous solutions of lithium chloride, nickel chloride, and caesium chloride in ambient to supercritical water, J. Mol. Liq. 153 (2010) 2–8. doi:10.1016/j.molliq.2009.10.012.

[16] A.G. Novikov, M.N. Rodnikova, V.V. Savostin, O.V. Sobolev, The study of hydration effects in aqueous solutions of LiCl and CsCl by inelastic neutron scattering, J. Mol. Liq. 82 (1999) 83–104. doi:10.1016/S0167-7322(99)00044-6.

[17] D.Z. Caralampio, J.M. Martínez, R.R. Pappalardo, E.S. Marcos, The hydration structure of the heavy-alkalines $Rb^+$ and $Cs^+$ through molecular dynamics and X-ray absorption spectroscopy: surface clusters and eccentricity, Phys. Chem. Chem. Phys. 19 (2017) 28993–29004. doi:10.1039/C7CP05346K.

[18] V.-T. Pham, J.L. Fulton, Contact ion-pair structure in concentrated cesium chloride aqueous solutions: An extended X-ray absorption fine structure study, J. Electron Spectros. Relat. Phenomena 229 (2018) 20–25. doi:10.1016/j.elspec.2018.09.004.

[19] A.G. Davidian, A.G. Kudrev, L.A. Myund, O.S. Khlynova, M.K. Khripun, Structure of aqueous electrolyte solutions estimated by near infrared spectroscopy and chemometric analysis of spectral data, Russ. J. Gen. Chem. 84 (2014) 1877–1887. doi:10.1134/S1070363214100028.

[20] K. Hayamizu, Y. Chiba, T. Haishi, Dynamic ionic radius of alkali metal ions in aqueous solution: a pulsed-field gradient NMR study, RSC Adv. 11 (2021) 20252–20257. doi:10.1039/D1RA02301B.

[21] S. Koneshan, J.C. Rasaiah, R.M. Lynden-Bell, S.H. Lee, Solvent structure, dynamics, and ion mobility in aqueous solutions at 25 °C, J. Phys. Chem. B 102 (1998) 4193–4204. doi:10.1021/jp980642x.

[22] C.F. Schwenk, T.S. Hofer, B.M. Rode, "Structure breaking" effect of hydrated $Cs^+$, J. Phys. Chem. A 108 (2004) 1509–1514. doi:10.1021/jp037179v.

[23] I. Kalcher, J. Dzubiella, Structure-thermodynamics relation of electrolyte solutions, J. Chem. Phys. 130 (2009) 134507. doi:10.1063/1.3097530.

[24] C.J. Fennell, A. Bizjak, V. Vlachy, K. a. Dill, Ion pairing in molecular simulations of aqueous alkali halide solutions, J. Phys. Chem. B 113 (2009) 6782–6791. doi:10.1021/jp809782z.

[25] R. Hartkamp, B. Coasne, Structure and transport of aqueous electrolytes: From simple halides to radionuclide ions, J. Chem. Phys. 141 (2014) 124508. doi:10.1063/1.4896380.

[26] S. Roy, V.S. Bryantsev, Finding order in the disordered hydration shell of rapidly exchanging water molecules around the heaviest alkali $Cs^+$ and $Fr^+$, J. Phys. Chem. B 122 (2018) 12067–12076. doi:10.1021/acs.jpcb.8b08414.

[27] T. Ikeda, M. Boero, Communication: Hydration structure and polarization of heavy alkali ions: A first principles molecular dynamics study of $Rb^+$ and $Cs^+$, J. Chem. Phys. 137 (2012) 041101. doi:10.1063/1.4742151.

[28] H.J.C. Berendsen, J.R. Grigera, T.P. Straatsma, The missing term in effective pair potentials, J. Phys. Chem. 91 (1987) 6269–6271. doi:10.1021/j100308a038.

[29] W.L. Jorgensen, J. Chandrasekhar, J.D. Madura, R.W. Impey, M.L. Klein, Comparison of simple potential functions for simulating liquid water, J. Chem. Phys. 79 (1983) 926–935. doi:10.1063/1.445869.

[30] H.W. Horn, W.C. Swope, J.W. Pitera, J.D. Madura, T.J. Dick, G.L. Hura, et al., Development of an improved four-site water model for biomolecular simulations: TIP4P-Ew, J. Chem. Phys. 120 (2004) 9665–9678. doi:10.1063/1.1683075.

[31] J.L.F. Abascal, C. Vega, A general purpose model for the condensed phases of water: TIP4P/2005, J. Chem. Phys. 123 (2005) 234505. doi:10.1063/1.2121687.

[32] L.X. Dang, Free energies for association of $Cs^+$ to 18-crown-6 in water. A molecular dynamics study including counter ions, Chem. Phys. Lett. 227 (1994) 211–214. doi:10.1016/0009-2614(94)00810-8.

[33] D.E. Smith, L.X. Dang, Computer simulations of NaCl association in polarizable water, J. Chem. Phys. 100 (1994) 3757–3766. doi:10.1063/1.466363.

[34] I.S. Joung, T.E. Cheatham, Determination of alkali and halide monovalent ion parameters for use in explicitly solvated biomolecular simulations, J. Phys. Chem. B 112 (2008) 9020–9041. doi:10.1021/jp8001614.

[35] D. Horinek, S.I. Mamatkulov, R.R. Netz, Rational design of ion force fields based on thermodynamic solvation properties, J. Chem. Phys. 130 (2009) 124507. doi:10.1063/1.3081142.

[36] M.B. Gee, N.R. Cox, Y. Jiao, N. Bentenitis, S. Weerasinghe, P.E. Smith, A Kirkwood-Buff derived force field for aqueous alkali halides, J. Chem. Theory Comput. 7 (2011) 1369–1380. doi:10.1021/ct100517z.

[37] personal communication with one of the authors of Ref. [36], P.E.Smith.

[38] M.M. Reif, P.H. Hünenberger, Computation of methodology-independent single-ion solvation properties from molecular simulations. IV. Optimized Lennard-Jones interaction parameter sets for the alkali and halide ions in water, J. Chem. Phys. 134 (2011) 144104. doi:10.1063/1.3567022.

[39] M. Fyta, I. Kalcher, J. Dzubiella, L. Vrbka, R.R. Netz, Ionic force field optimization based on single-ion and ion-pair solvation properties, J. Chem. Phys. 132 (2010) 024911. doi:10.1063/1.3292575.

[40] M. Fyta, R.R. Netz, Ionic force field optimization based on single-ion and ion-pair solvation properties: Going beyond standard mixing rules, J. Chem. Phys. 136 (2012) 124103. doi:10.1063/1.3693330.

[41] S. Deublein, J. Vrabec, H. Hasse, A set of molecular models for alkali and halide ions in aqueous solution, J. Chem. Phys. 136 (2012) 084501. doi:10.1063/1.3687238.

[42] S. Reiser, S. Deublein, J. Vrabec, H. Hasse, Molecular dispersion energy parameters for alkali and halide ions in aqueous solution, J. Chem. Phys. 140 (2014) 044504. doi:10.1063/1.4858392.

[43] P. Li, L.F. Song, K.M. Merz, Systematic Parameterization of monovalent ions employing the nonbonded model, J. Chem. Theory Comput. 11 (2015) 1645–1657. doi:10.1021/ct500918t.

[44] S. Mamatkulov, N. Schwierz, Force fields for monovalent and divalent metal cations in TIP3P water based on thermodynamic and kinetic properties, J. Chem. Phys. 148 (2018) 074504. doi:10.1063/1.5017694.

[45] K.P. Jensen, W.L. Jorgensen, Halide, ammonium, and alkali metal ion parameters for modeling aqueous solutions, J. Chem. Theory Comput. 2 (2006) 1499–1509. doi:10.1021/ct600252r.

[46] W.L. Jorgensen, J.D. Madura, C.J. Swenson, Optimized intermolecular potential functions for liquid hydrocarbons, J. Am. Chem. Soc. 106 (1984) 6638–6646. doi:10.1021/ja00334a030.

[47] A.H. Mao, R. V Pappu, Crystal lattice properties fully determine short-range interaction parameters for alkali and halide ions, J. Chem. Phys. 137 (2012) 064104. doi:10.1063/1.4742068.

[48] Z.R. Kann, J.L. Skinner, A scaled-ionic-charge simulation model that reproduces enhanced and suppressed water diffusion in aqueous salt solutions, J. Chem. Phys. 141 (2014) 104507. doi:10.1063/1.4894500.

[49] M.J. Abraham, T. Murtola, R. Schulz, S. Páll, J.C. Smith, B. Hess, et al., GROMACS: High performance molecular simulations through multi-level parallelism from laptops to supercomputers, SoftwareX 1–2 (2015) 19–25. doi:10.1016/j.softx.2015.06.001.

[50] S. Miyamoto, P.A. Kollman, SETTLE: An analytical version of the SHAKE and RATTLE algorithm for rigid water models, J. Comput. Chem. 13 (1992) 952–962. doi:10.1002/jcc.540130805.

[51] M.P. Allen, D.J. Tildesley, Computer Simulation of Liquids, Oxford University Press, Oxford, 1987. doi:10.1093/oso/9780198803195.001.0001.

[52] T. Darden, D. York, L. Pedersen, Particle mesh Ewald: An N·log(N) method for Ewald sums in large systems, J. Chem. Phys. 98 (1993) 10089–10092. doi:10.1063/1.464397.

[53] U. Essmann, L. Perera, M.L. Berkowitz, T. Darden, H. Lee, L.G. Pedersen, A smooth particle mesh Ewald method, J. Chem. Phys. 103 (1995) 8577–8593. doi:10.1063/1.470117.

[54] H.J.C. Berendsen, J.P.M. Postma, W.F. van Gunsteren, A. DiNola, J.R. Haak, Molecular dynamics with coupling to an external bath, J. Chem. Phys. 81 (1984) 3684–3690. doi:10.1063/1.448118.

[55] S. Nosé, A molecular dynamics method for simulations in the canonical ensemble, Mol. Phys. 52 (1984) 255–268. doi:10.1080/00268978400101201.

[56] W.G. Hoover, Canonical dynamics: Equilibrium phase-space distributions, Phys. Rev. A 31 (1985) 1695–1697. doi:10.1103/PhysRevA.31.1695.

[57] M. Parrinello, A. Rahman, Polymorphic transitions in single crystals: A new molecular dynamics method, J. Appl. Phys. 52 (1981) 7182–7190. doi:10.1063/1.328693.

[58] S. Nosé, M.L. Klein, Constant pressure molecular dynamics for molecular systems, Mol. Phys. 50 (1983) 1055–1076. doi:10.1080/00268978300102851.

[59] V. Chihaia, S. Adams, W.F. Kuhs, Molecular dynamics simulations of properties of a (001) methane clathrate hydrate surface, Chem. Phys. 317 (2005) 208–225. doi:10.1016/j.chemphys.2005.05.024.

[60] E.W. Washburn, C.J. West, eds., International Critical Tables of Numerical Data, Physics, Chemistry and Technology, McGraw-Hill Book Company, New York, volume III (1928), p97 doi:10.17226/20230.

[61] P. Novotny, O. Sohnel, Densities of binary aqueous solutions of 306 inorganic substances, J. Chem. Eng. Data 33 (1988) 49–55. doi:10.1021/je00051a018.

[62] A.C. Tikanen, W.R. Fawcett, Application of the mean spherical approximation and ion association to describe the activity coefficients of aqueous 1:1 electrolytes, J. Electroanal. Chem. 439 (1997) 107–113. doi:10.1016/S0022-0728(97)00376-8.

[63] K.J. Müller, H.G. Hertz, A Parameter as an indicator for water−water association in solutions of strong electrolytes, J. Phys. Chem. 100 (1996) 1256–1265. doi:10.1021/jp951303w.

**Table 1** Investigated CsCl-water solutions. The number of ion pairs and water molecules and the respective densities are taken from Ref. [12]

| short name | 1.5 mol % | 7.5 mol % | 15 mol % |
| --- | --- | --- | --- |
| Concentration [mol %] | 1.49 | 7.52 | 15.07 |
| $N_{Cs} = N_{Cl}$ | 50 | 257 | 529 |
| $N_{water}$ | 3300 | 3162 | 2981 |
| Number density [Å$^{-3}$] | 0.0978 | 0.0877 | 0.08071 |
| Box length [nm] | 4.676 | 4.849 | 4.98547 |

**Table 2** Parameters of the investigated potential models. (The ion charges are + for Cs$^+$ and – for Cl$^-$ ions, only the numbers are shown.) Combination rules: g: geometric, LB: Lorentz-Berthelot, mg: modified geometric (see text).

| Model | $\sigma_{CsCs}$ [nm] | $\varepsilon_{CsCs}$ [kJ/mol] | $\sigma_{ClCl}$ [nm] | $\varepsilon_{ClCl}$ [kJ/mol] | $q_{ion}$ [e] | Comb. rule | water model | Ref. |
|---|---|---|---|---|---|---|---|---|
| DS | 0.3884 | 0.4184 | 0.4401 | 0.4184 | 1 | LB | SPC/E | [32,33] |
| JC-S | 0.3601 | 0.3760 | 0.4830 | 0.0535 | 1 | LB | SPC/E | [34] |
| HL-g | 0.3340 | 1.5400 | 0.4520 | 0.4200 | 1 | g | SPC/E | [35] |
| HL-LB | 0.3330 | 1.5400 | 0.4400 | 0.4200 | 1 | LB | SPC/E | [35] |
| HM-g | 0.3440 | 0.6500 | 0.4520 | 0.4200 | 1 | g | SPC/E | [35] |
| HM-LB | 0.3430 | 0.6500 | 0.4400 | 0.4200 | 1 | LB | SPC/E | [35] |
| HS-g | 0.5490 | 0.0006 | 0.4520 | 0.4200 | 1 | g | SPC/E | [35] |
| HS-LB | 0.5170 | 0.0006 | 0.4400 | 0.4200 | 1 | LB | SPC/E | [35] |
| Gee | 0.4130 | 0.0650 | 0.4400 | 0.4700 | 1 | mg | SPC/E | [36] |
| RH | 0.5371 | 0.0827 | 0.3493 | 1.7625 | 1 | g | SPC/E | [38] |
| RM | 0.4548 | 0.2242 | 0.3771 | 1.1137 | 1 | g | SPC/E | [38] |
| RL | 0.3904 | 0.5608 | 0.4096 | 0.6785 | 1 | g | SPC/E | [38] |
| KS-RL | 0.3904 | 0.5608 | 0.4096 | 0.6785 | 0.9326 | g | SPC/E | [38,48] |
| FN6 | 0.3490 | 0.3250 | 0.4400 | 0.4168 | 1 | LB | SPC/E | [39,40] |
| FN9 | 0.3331 | 1.5400 | 0.4400 | 0.4168 | 1 | LB | SPC/E | [39,40] |
| MP-S | 0.3440 | 2.0974 | 0.4612 | 0.1047 | 1 | LB | SPC/E | [47] |
| DVH | 0.3580 | 0.8314 | 0.4410 | 0.8314 | 1 | LB | SPC/E | [41] |
| RDVH | 0.3580 | 1.6629 | 0.4410 | 1.6629 | 1 | LB | SPC/E | [42] |
| Li-HFE-S | 0.3480 | 1.4507 | 0.4112 | 2.6931 | 1 | LB | SPC/E | [43] |
| Li-IOD-S | 0.3564 | 1.6294 | 0.3852 | 2.2240 | 1 | LB | SPC/E | [43] |
| JC-T3 | 0.3521 | 1.7010 | 0.4478 | 0.1489 | 1 | LB | TIP3P | [34] |
| Li-HFE-T3 | 0.3542 | 1.5838 | 0.4013 | 2.5227 | 1 | LB | TIP3P | [43] |
| MS | 0.3549 | 0.6040 | 0.4410 | 0.2840 | 1 | LB | TIP3P | [44] |
| JJ | 0.6200 | 0.0021 | 0.4020 | 2.9706 | 1 | g | TIP4P | [45] |
| JC-T | 0.3364 | 1.6503 | 0.4918 | 0.0488 | 1 | LB | TIP4P-Ew | [34] |
| MP-T | 0.3440 | 2.0974 | 0.4612 | 0.1047 | 1 | LB | TIP4P-Ew | [47] |
| Li-HFE-T | 0.3450 | 1.3863 | 0.4136 | 2.7309 | 1 | LB | TIP4P-Ew | [43] |
| Li-IOD-T | 0.3564 | 1.6294 | 0.3852 | 2.2240 | 1 | LB | TIP4P-Ew | [43] |
| KS-Li-IOD-T | 0.3564 | 1.6294 | 0.3852 | 2.2240 | 0.9235 | LB | TIP4P-Ew | [43,48] |
| Madrid | 0.3521 | 0.3760 | 0.4699 | 0.0769 | 0.85 | – [a] | TIP4P/2005 | [2] |

[a]The Madrid FF do not use combination rule, every LJ parameters are given separately: $\sigma_{CsO}$=0.3663, $\varepsilon_{CsO}$= 0.1, $\sigma_{CsCl}$0.4319, $\varepsilon_{CsCl}$=0.1616, $\sigma_{ClO}$=0.4239, $\varepsilon_{ClO}$=0.062.

**Table 3** Water potential models. For the 3 site models, the 3 point charges are in the positions of the atoms of the molecule (the positive charges in the H atoms, the negative charge in the O atom). In the case of the TIP4P, TIP4P-Ew, and TIP4P/2005 models, a fourth (virtual) site (M) is situated along the bisector of the H-O-H angle, coplanar with the oxygen and hydrogen atoms. The negative charge is placed in M.

|  | $\sigma_{OO}$ [nm] | $\varepsilon_{OO}$ [kJ/mol] | $q_H$ [e] | $d_{O-H}$ [nm] | $\theta_{H-O-H}$ [deg] | $d_{O-M}$ [nm] | Ref. |
|---|---|---|---|---|---|---|---|
| SPC/E | 0.3166 | 0.6502 | +0.4238 | 0.1 | 109.47 | - | [28] |
| TIP3P | 0.3151 | 0.6364 | +0.417 | 0.09572 | 104.52 | - | [29] |
| TIP4P | 0.3154 | 0.6485 | +0.52 | 0.09572 | 104.52 | 0.015 | [29] |
| TIP4P-Ew | 0.3164 | 0.6809 | +0.52422 | 0.09572 | 104.52 | 0.0125 | [30] |
| TIP4P/2005 | 0.3159 | 0.7749 | +0.5564 | 0.09572 | 104.52 | 0.01546 | [31] |

**Table 4** Position of the first maximum in the Cl$^-$-H, Cl$^-$-O, Cs$^+$-O, and Cs$^+$-Cl$^-$ partial radial distribution functions obtained in MD simulations (in Å). (The best models are in bold.)

|            | Cl$^-$-H | Cl$^-$-O | Cs$^+$-O | Cs$^+$-Cl$^-$ |
|------------|------|------|------|------|
| DS         | 2.24 | 3.21 | 3.12 | 3.50 |
| JC-S       | 2.14 | 3.13 | 2.96 | 3.25 |
| HL-g       | 2.23 | 3.21 | 3.00 | 3.45 |
| HL-LB      | 2.23 | 3.21 | 3.00 | 3.43 |
| HM-g       | 2.23 | 3.22 | 2.94 | 3.39 |
| HM-LB      | 2.23 | 3.21 | 2.94 | 3.35 |
| HS-g       | 2.24 | 3.22 | 2.81 | 3.21 |
| HS-LB      | 2.25 | 3.22 | 2.83 | 3.08 |
| Gee        | 2.20 | 3.17 | 2.94 | 3.36 |
| RH         | 1.95 | 2.93 | 3.49 | 3.79 |
| RM         | 2.03 | 3.01 | 3.32 | 3.62 |
| **RL**     | **2.11** | **3.09** | **3.15** | **3.51** |
| **KS-RL**  | **2.13** | **3.11** | **3.18** | **3.55** |
| FN6        | 2.24 | 3.22 | 2.88 | 3.27 |
| FN9        | 2.24 | 3.21 | 3.00 | 3.42 |
| MP-S       | 2.13 | 3.11 | 3.09 | 3.45 |
| DVH        | 2.34 | 3.31 | 3.05 | 3.55 |
| RDVH       | 2.43 | 3.41 | 3.14 | 3.75 |
| Li-HFE-S   | 2.35 | 3.31 | 3.07 | 3.59 |
| **Li-IOD-S** | **2.18** | **3.15** | **3.11** | **3.51** |
| JC-T3      | 2.18 | 3.12 | 3.09 | 3.43 |
| Li-HFE-T3  | 2.35 | 3.27 | 3.10 | 3.59 |
| MS         | 2.24 | 3.18 | 2.98 | 3.33 |
| JJ         | 2.32 | 3.24 | 3.19 | 3.79 |
| JC-T       | 2.21 | 3.15 | 3.02 | 3.41 |
| MP-T       | 2.17 | 3.11 | 3.10 | 3.45 |
| Li-HFE-T   | 2.40 | 3.32 | 3.05 | 3.59 |
| **Li-IOD-T** | **2.21** | **3.15** | **3.14** | **3.53** |
| **KS-Li-IOD-T** | **2.23** | **3.17** | **3.16** | **3.55** |
| Madrid     | 2.1  | 3.04 | 2.85 | 3.8? |

**Figures**

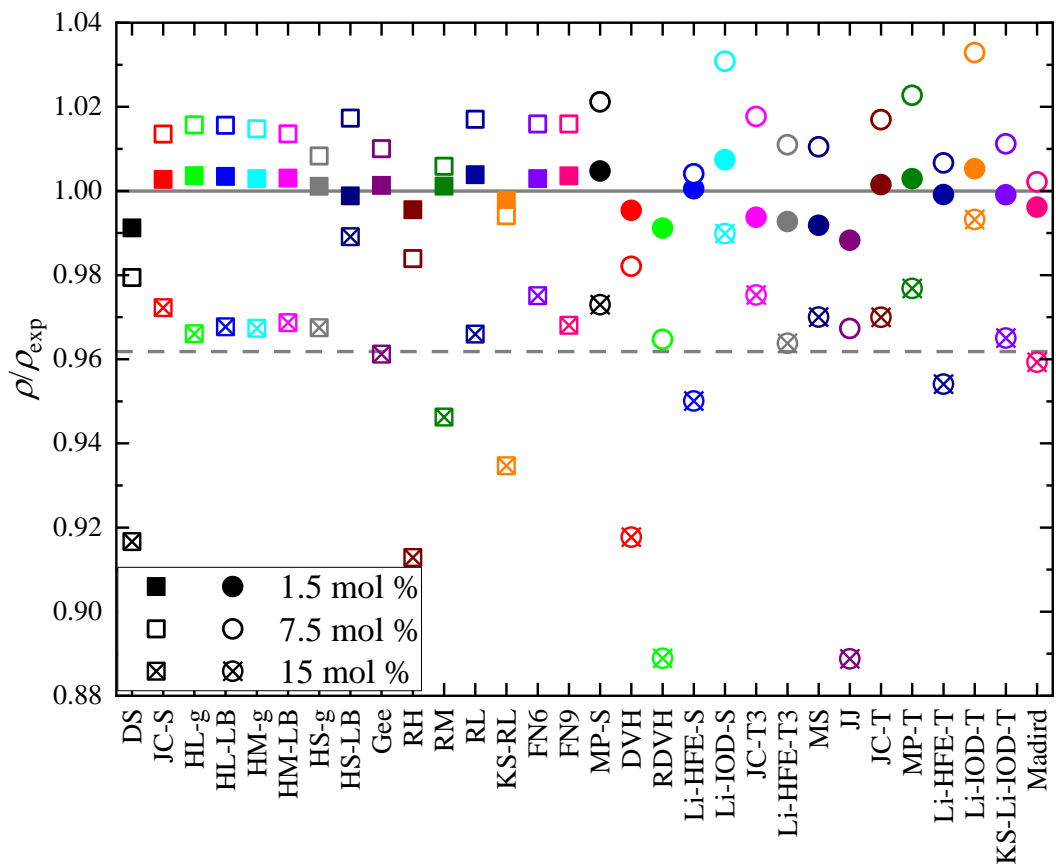

**Figure 1.** Ratio of the simulated and experimental density values of the CsCl solutions. The experimental densities of 1.5 mol %, 7.5 mol % and 15mol % solutions are 1102, 1460 and 1913 kg/m$^3$, respectively [12]. It should be noted that the experimental density of the 15 mol % sample in Ref [12] is somewhat higher than the value from other references (1840 kg/m$^3$ [60], 1870 kg/m$^3$ [61]). For the 15 mol % solution, the 1840 kg/m$^3$ value [60] is also shown for reference, as dashed line.

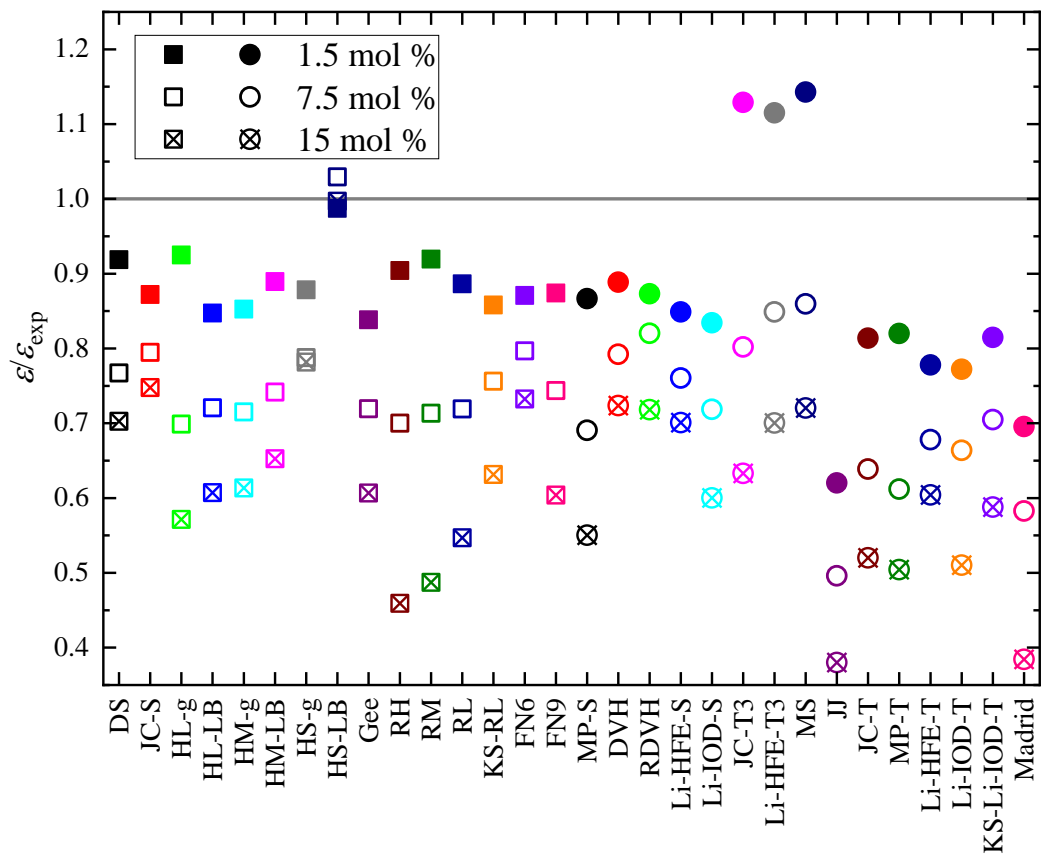

**Figure 2.** Ratio of simulated and experimental static dielectric constants. The experimental values at the given concentration were calculated using the formula of Ref. [62].

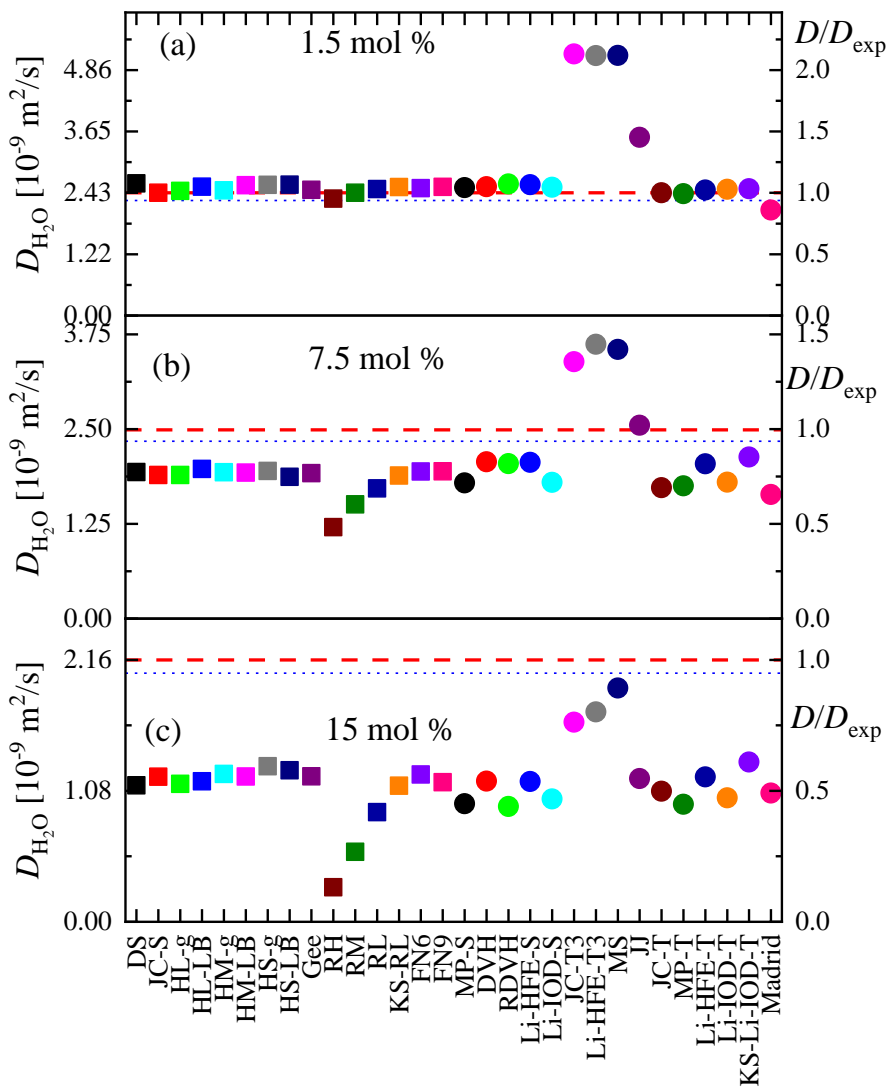

**Figure 3.** Self-diffusion coefficients of water molecules in aqueous CsCl solutions obtained in MD simulations. The experimental values are marked by dashed lines. The experimental $D_{H2O}$ values were estimated using the data of Ref. [63] (2.42, 2.49, and 2.16 × $10^{-9}$ m$^2$/s for 1.5 mol %, 7.5 mol %, and 15 mol % solutions, respectively). The experimental value corrected by the finite size effect (see detailed in the SM) is represented by dotted lines.

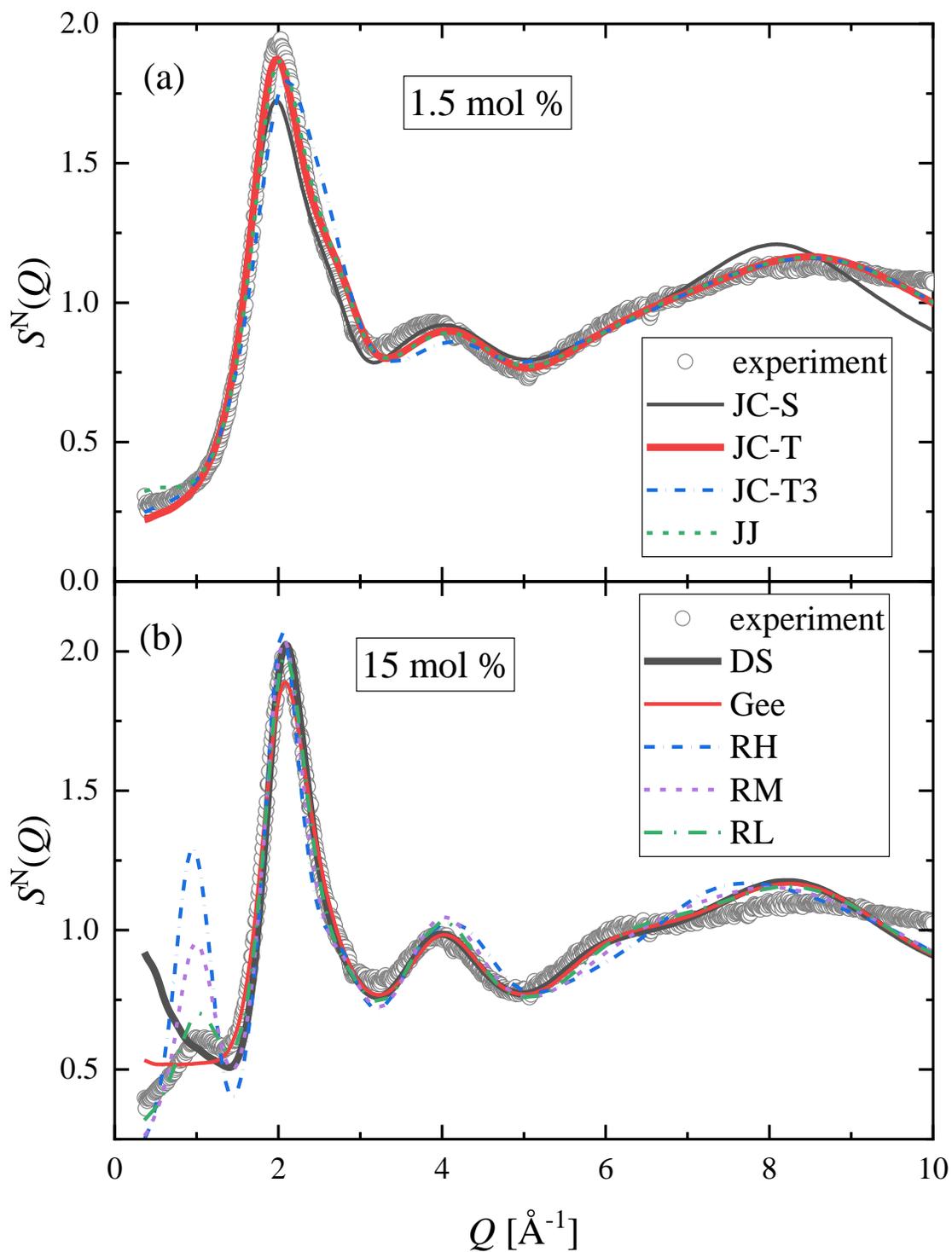

**Figure 4** Neutron total structure factors of the (a) 1.5 mol % and (b) 15 mol % solutions obtained by simulations (lines) compared with the experimental curve (symbols) from Ref. [12].

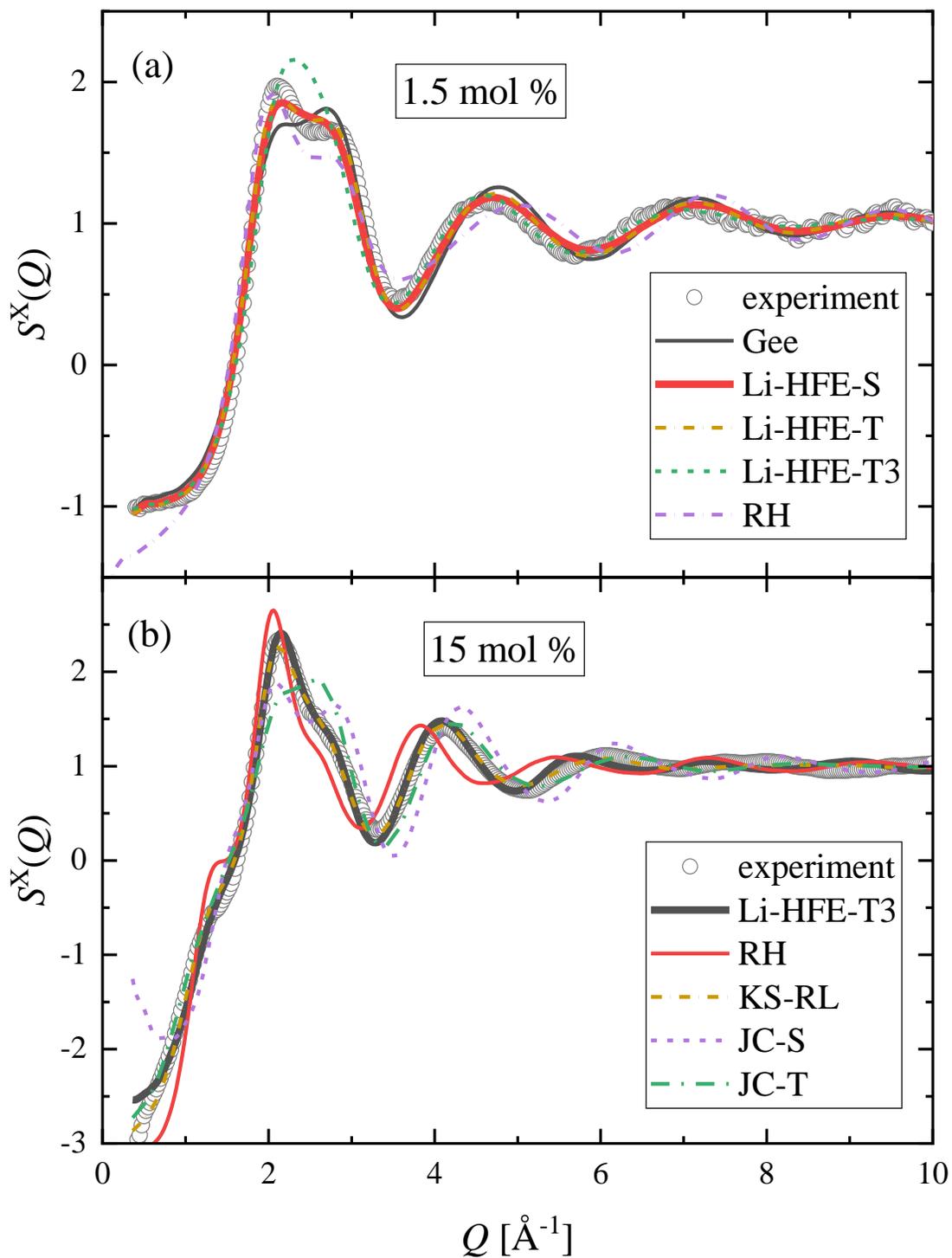

**Figure 5** X-ray total structure factors of the (a) 1.5 mol % and (b) 15 mol % solutions obtained by simulations (lines) compared with the experimental curve (symbols) from Ref. [12].

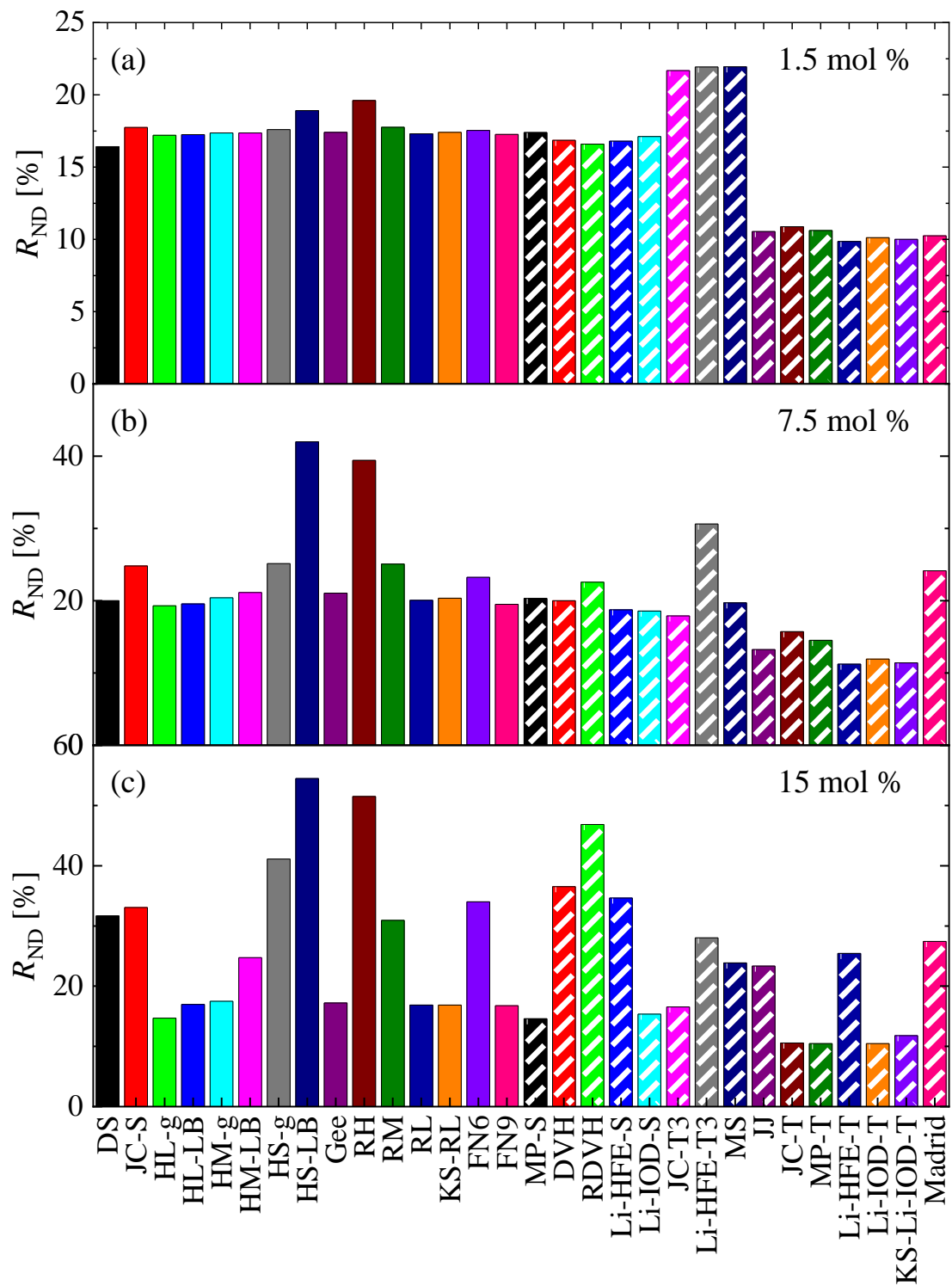

**Figure 6** *R*-factors of the simulated neutron structure factors. (For the definition of *R*-factor, see text.)

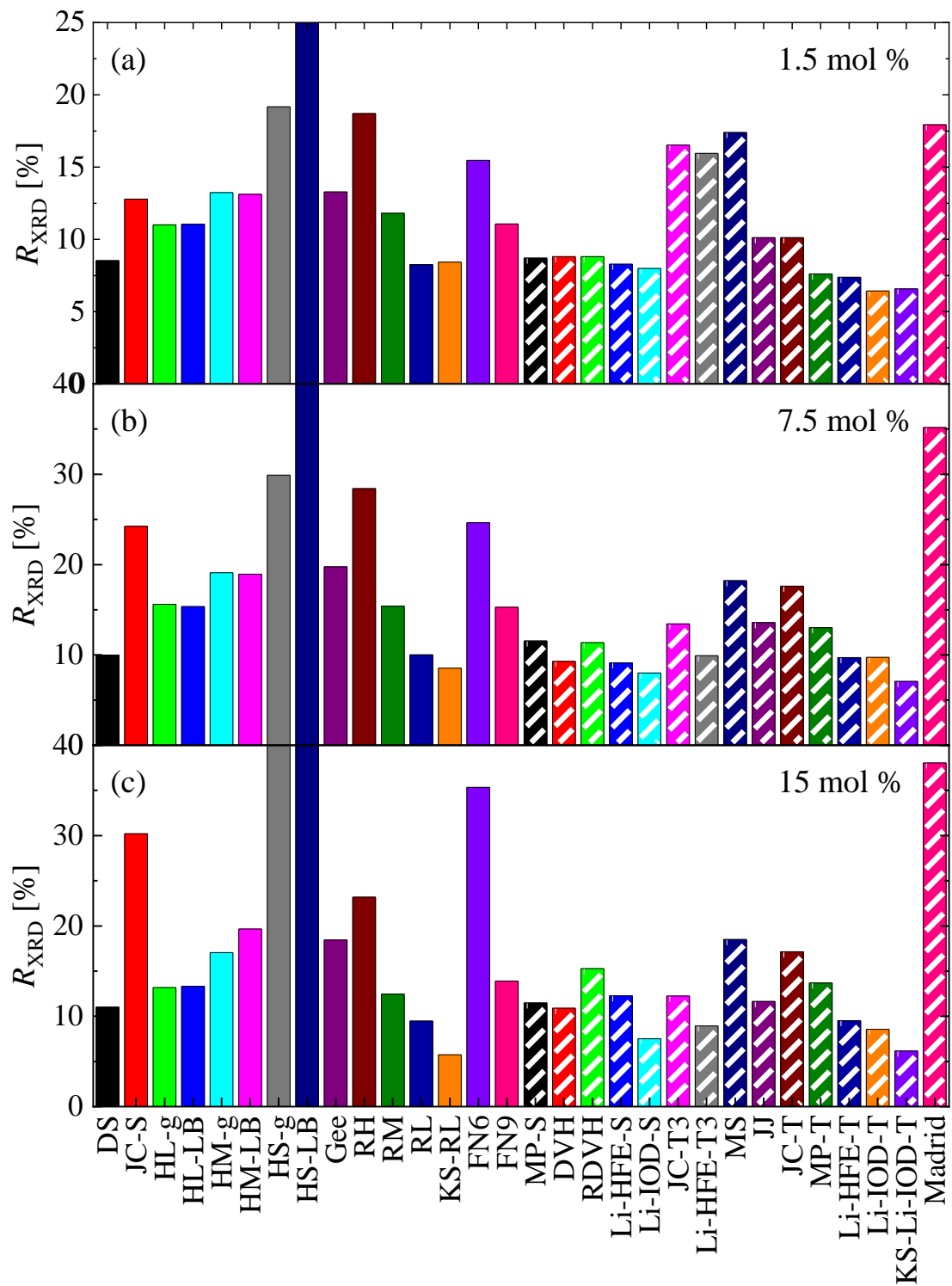

**Figure 7** $R$-factors of the simulated X-ray structure factors. (For the definition of $R$-factor, see text.)

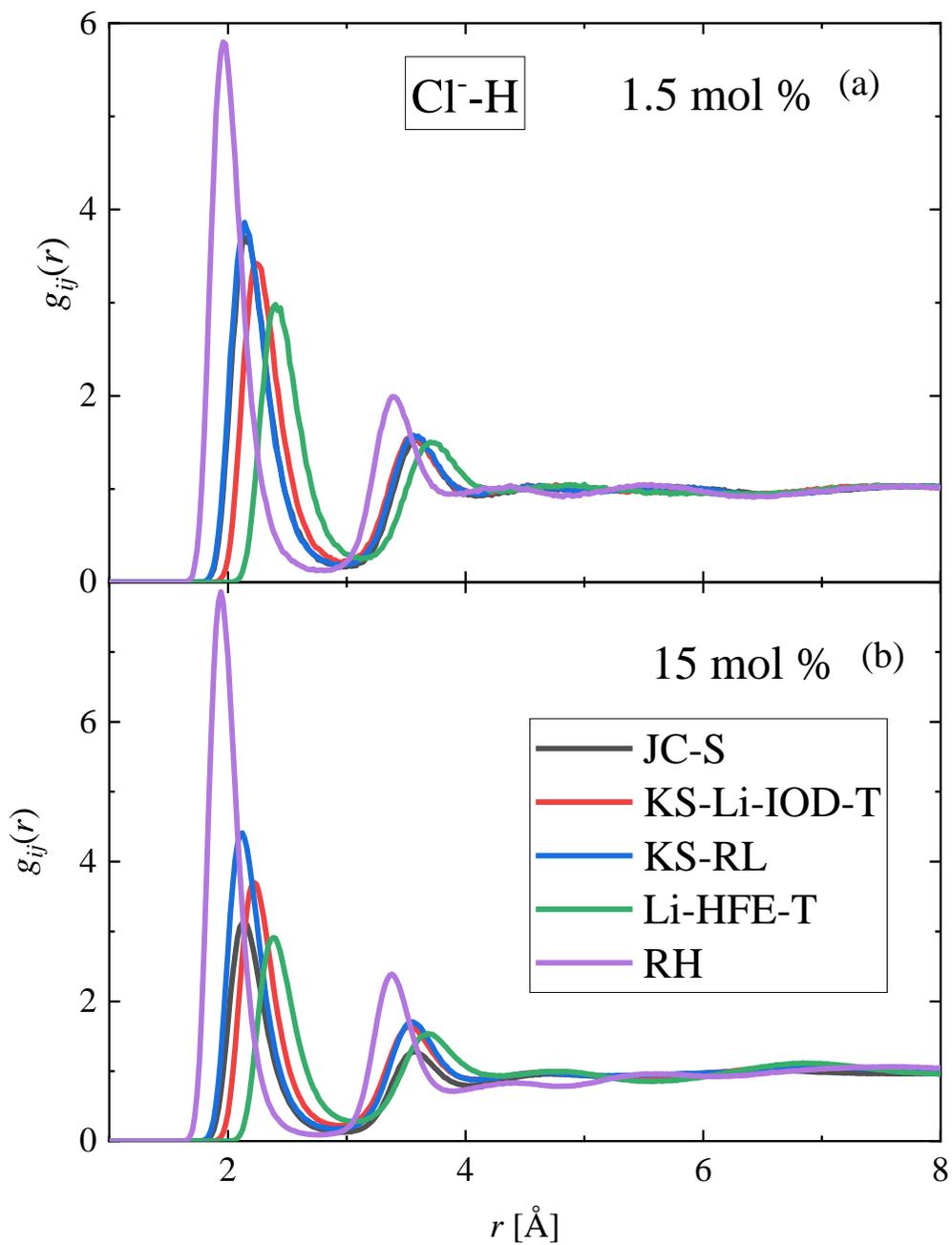

**Figure 8** Cl⁻ - H partial radial distribution functions, as calculated from MD simulations with different interatomic potential models. The curves are shown for (a) 1.5 mol % and (b) 15 mol % solutions.

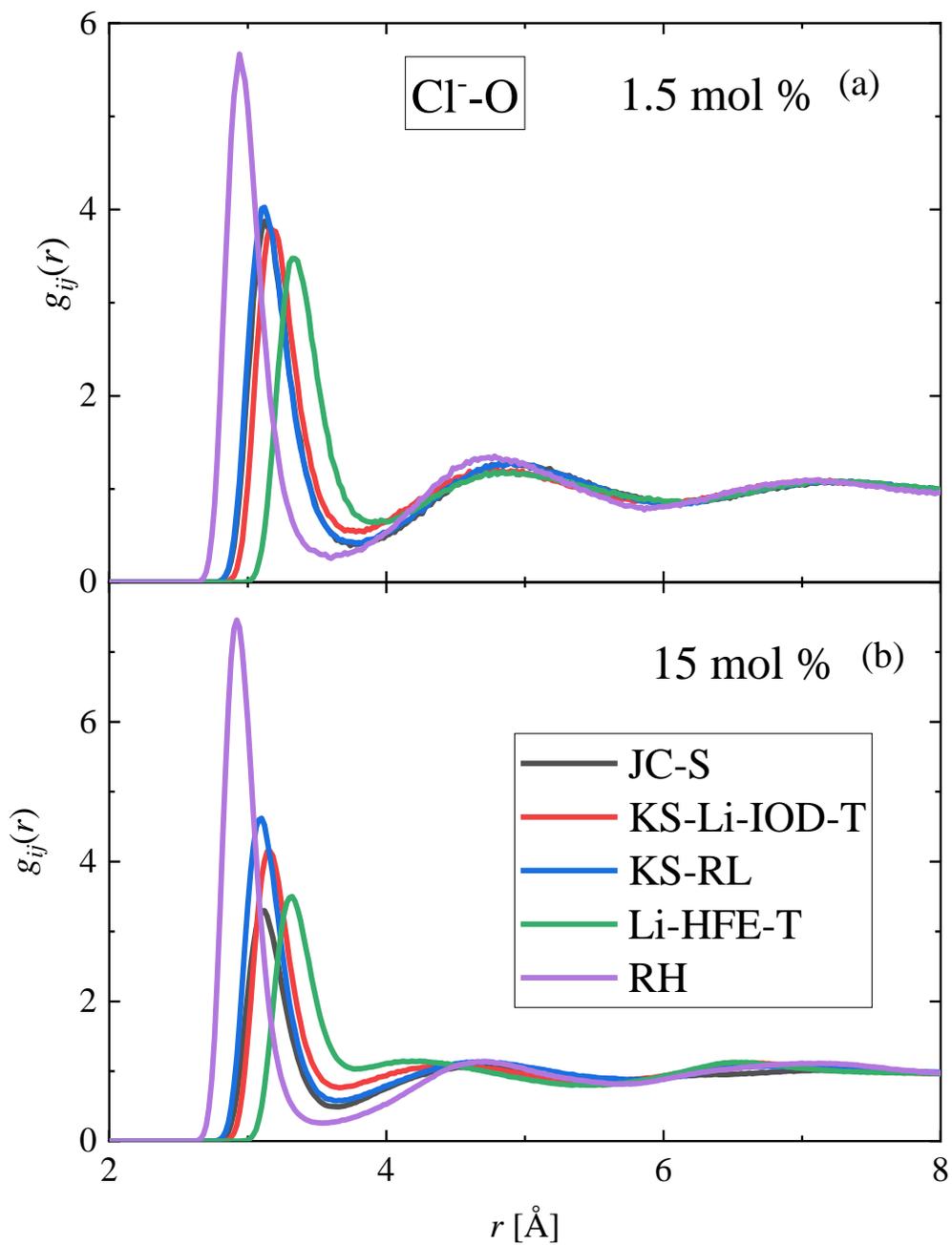

**Figure 9** Cl⁻ - O partial radial distribution functions, as calculated from MD simulations with different interatomic potential models. The curves are shown for (a) 1.5 mol % and (b) 15 mol % solutions.

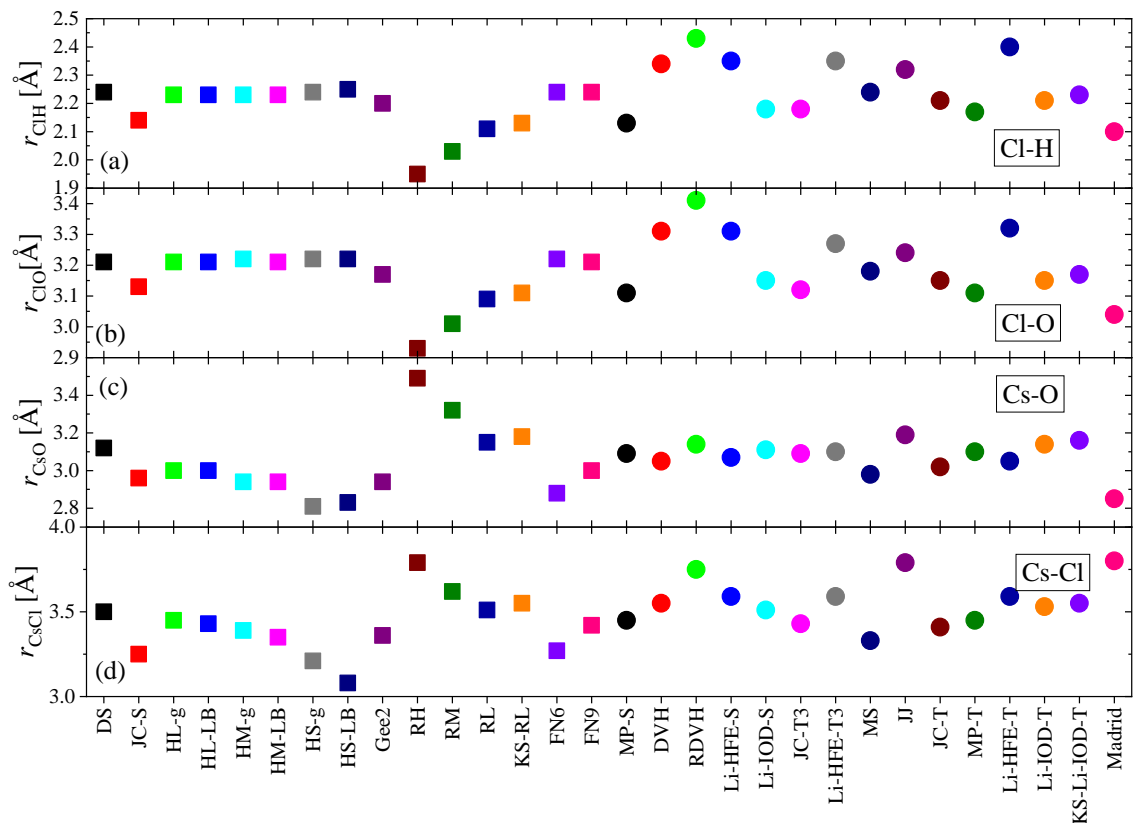

**Figure 10** (a) Cl$^-$-H, (b) Cl$^-$-O, (c) Cs$^+$-O, and (d) Cs$^+$-Cl$^-$ bond lengths (first maxima of the respective $g_{ij}(r)$ curve) obtained in MD simulations with different interatomic potential models. (The positions of these first peaks were almost the same at all concentrations tested, and their averaged values are shown.)

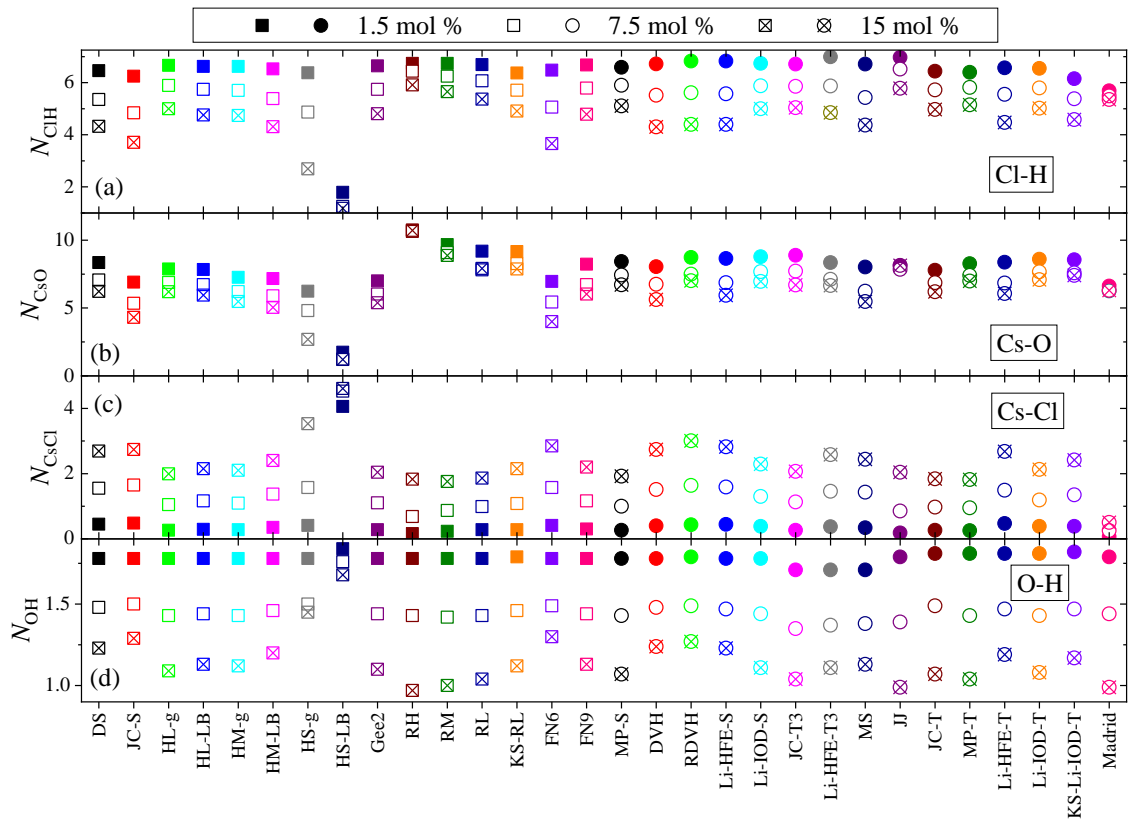

**Figure 11** (a) $Cl^-$-H, (b) $Cs^+$-O, (c) $Cs^+$-$Cl^-$, and (d) O-H coordination numbers obtained in MD simulations with different interatomic potential models.

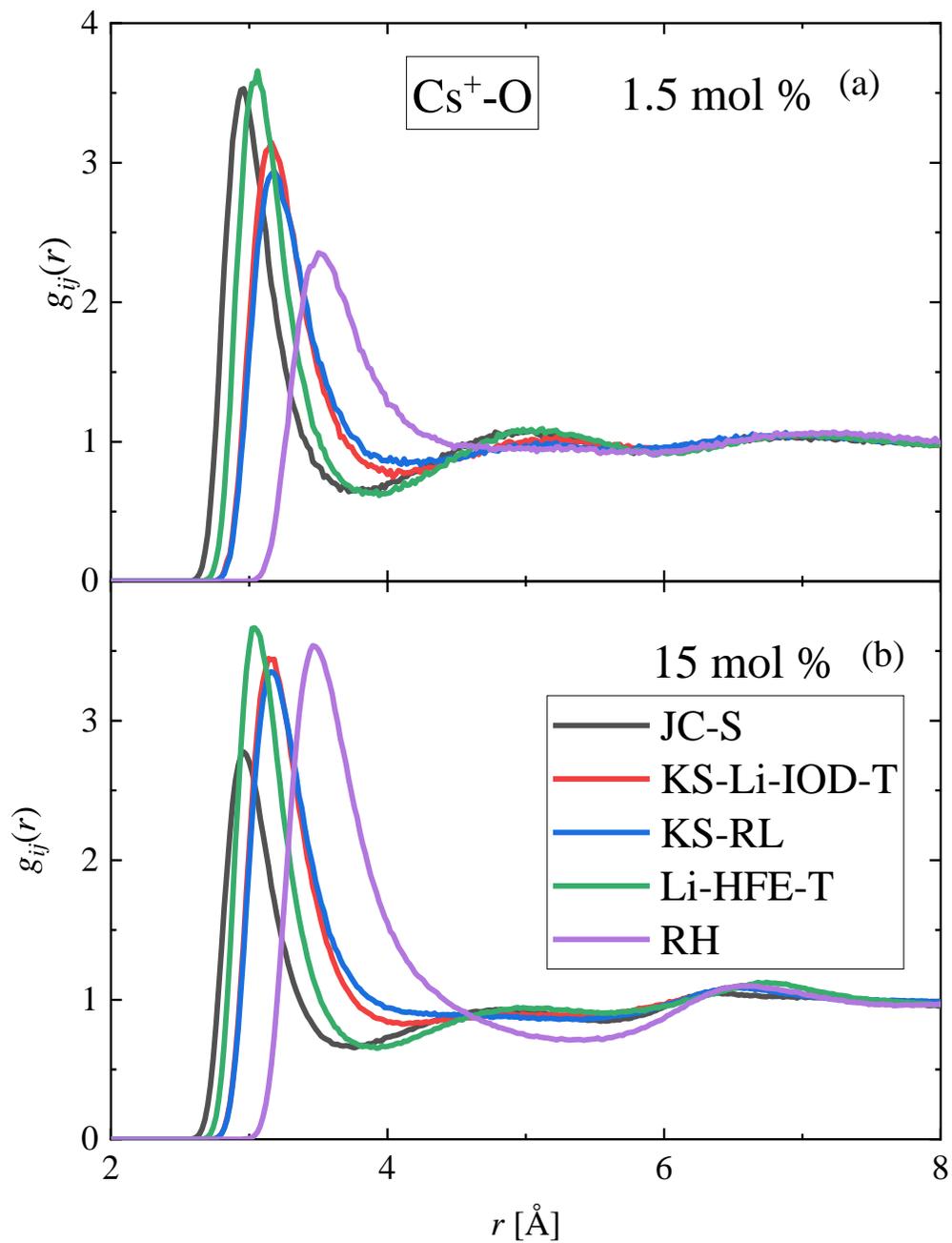

**Figure 12** $Cs^+$ - O partial radial distribution functions, as calculated from MD simulations with different interatomic potential models. The curves are shown for (a) 1.5 mol % and (b) 15 mol % solutions.

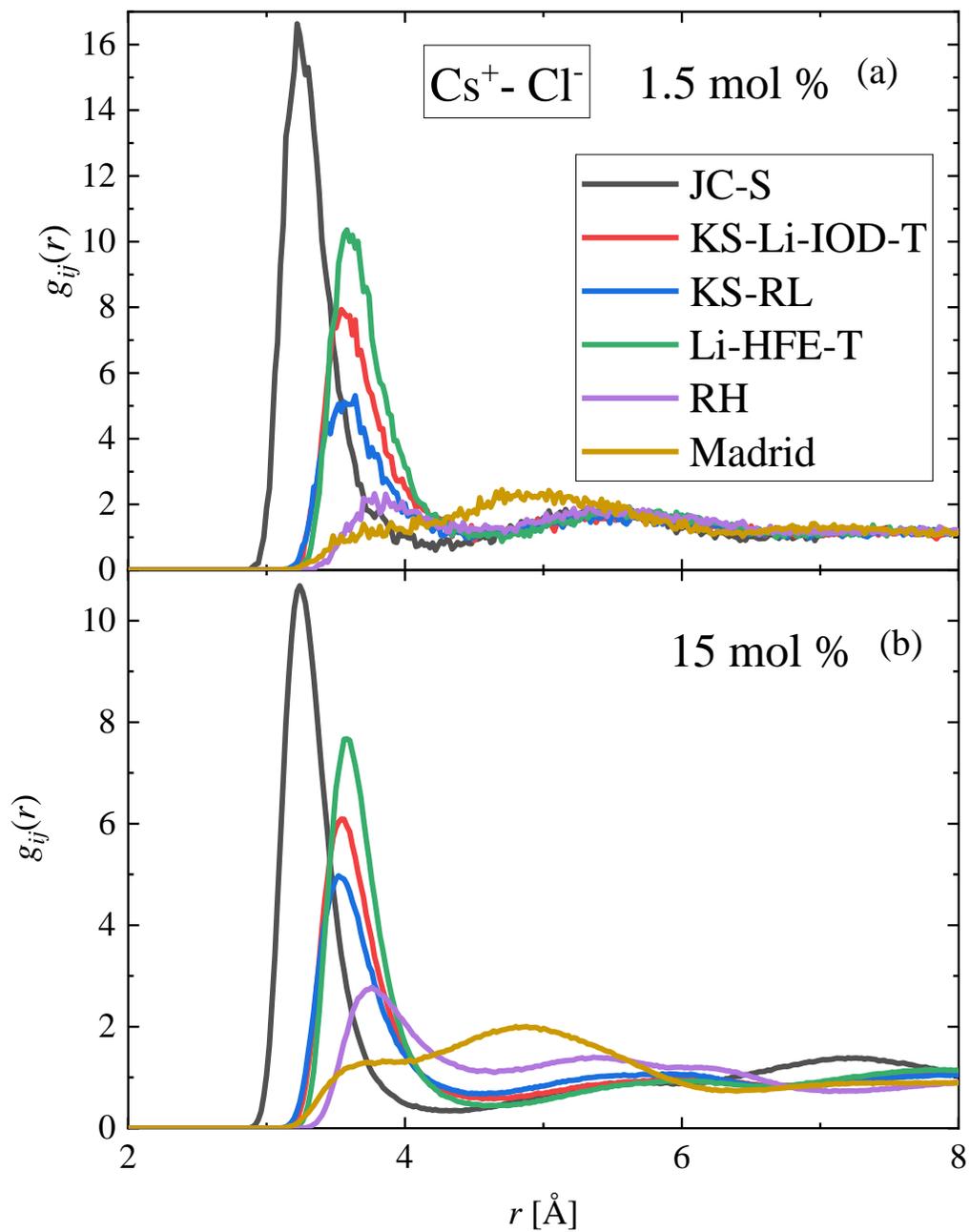

**Figure 13** $Cs^+$ - $Cl^-$ partial radial distribution functions, as calculated from MD simulations with different interatomic potential models. The curves are shown for (a) 1.5 mol % and (b) 15 mol % solutions.

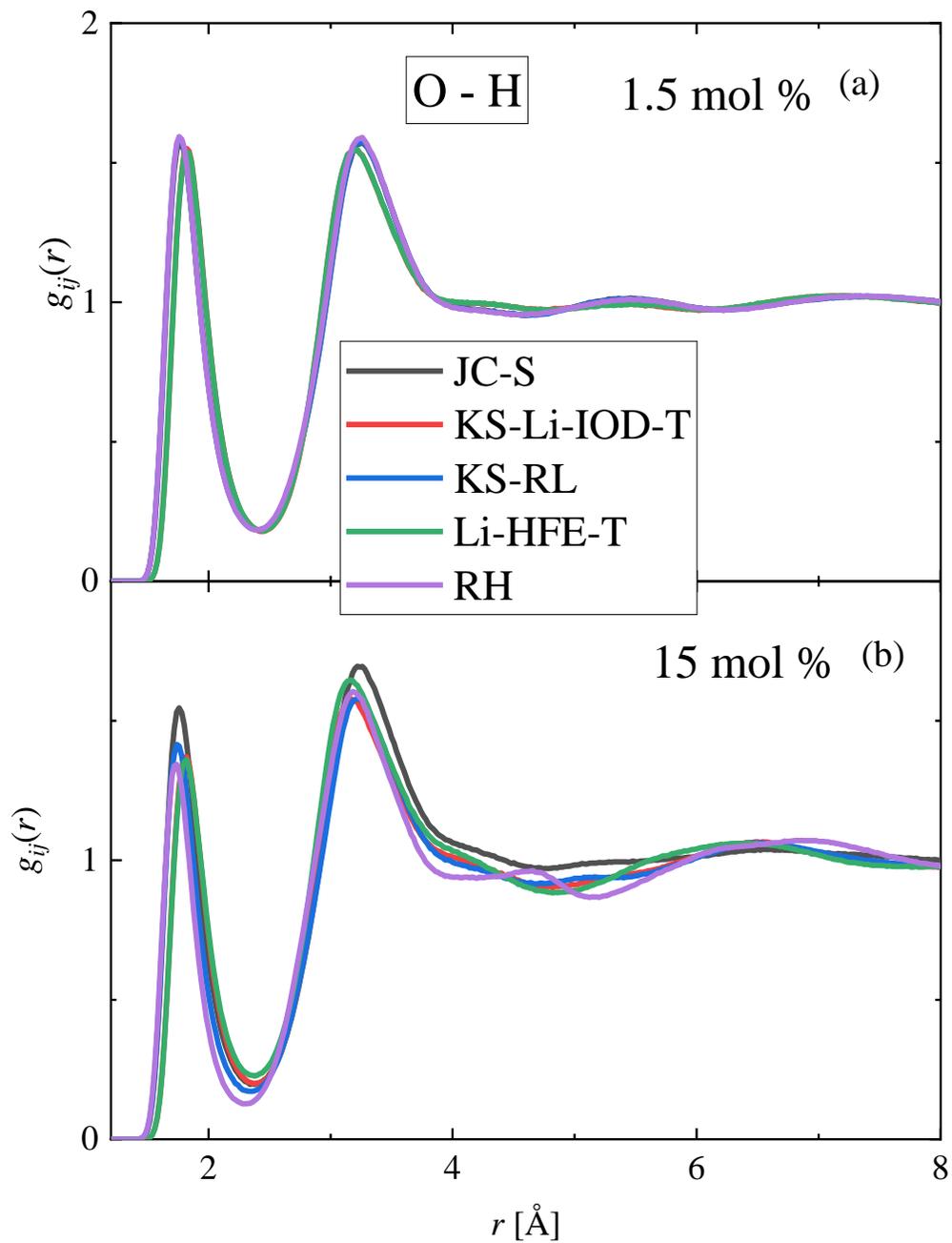

**Figure 14** O – H partial radial distribution functions, as calculated from MD simulations with different interatomic potential models. The curves are shown for (a) 1.5 mol % and (b) 15 mol % solutions.

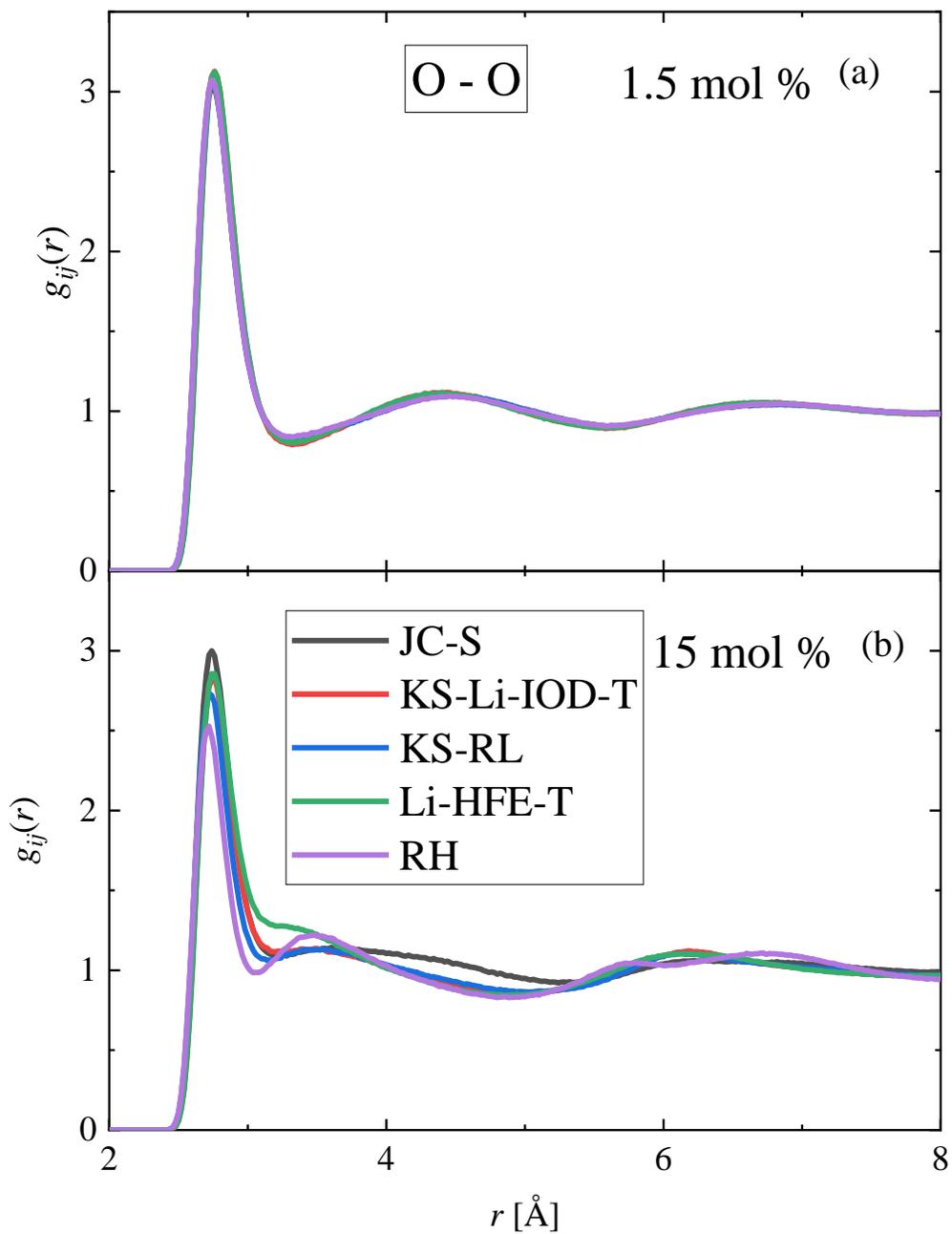

**Figure 15** O – O partial radial distribution functions, as calculated from MD simulations with different interatomic potential models. The curves are shown for (a) 1.5 mol % and (b) 15 mol % solutions.

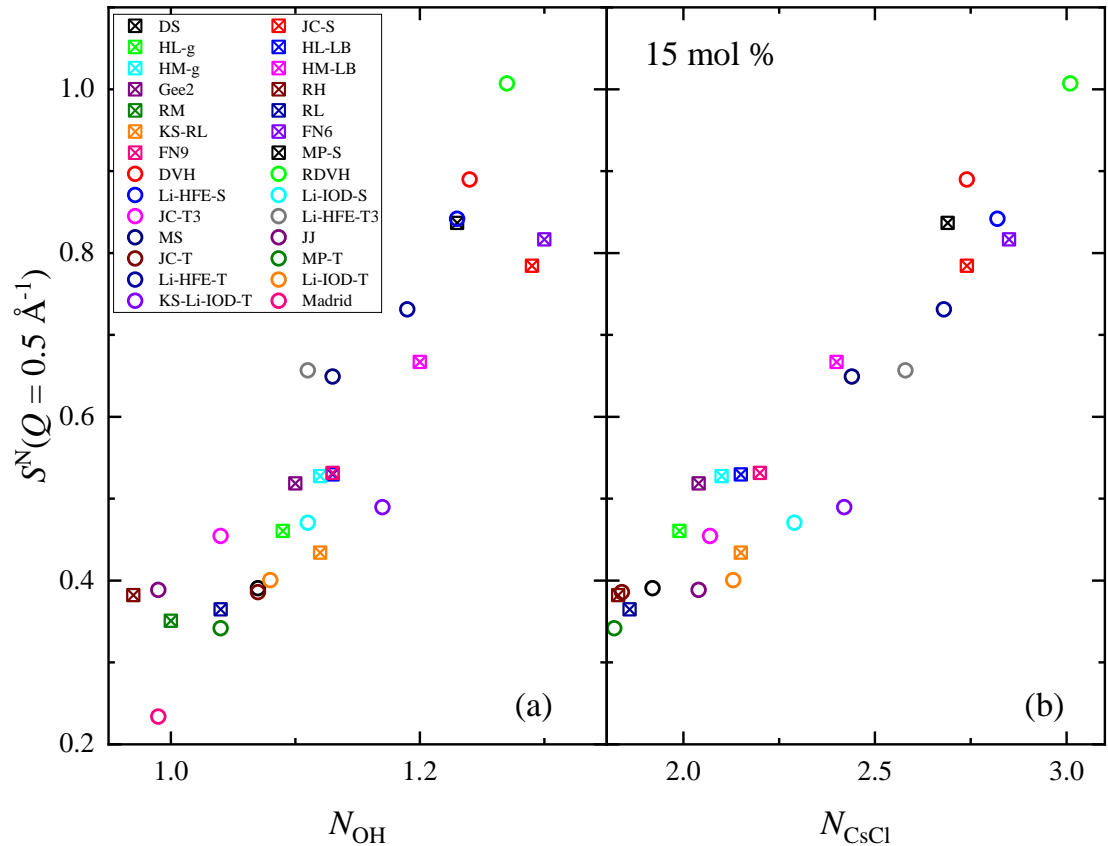

**Figure 16** Neutron total structure factors ($S^N(Q)$) values at $Q = 0.5$ Å$^{-1}$ as a function of the (a) $N_{OH}$ and (b) $N_{CsCl}$ coordination numbers obtained in simulations using different force fields. (Note: models, where ion precipitation occurred under the solubility limit (HS-g and HS-LB), are not shown.)

**Figure 17** Neutron total structure factors ($S^N(Q)$) values at $Q = 1$ Å$^{-1}$ as a function of the $r_{CsO}$ bond distance obtained in simulations using different force fields. (Note: models, where ion precipitation occurred under the solubility limit (HS-g and HS-LB), are not shown.)

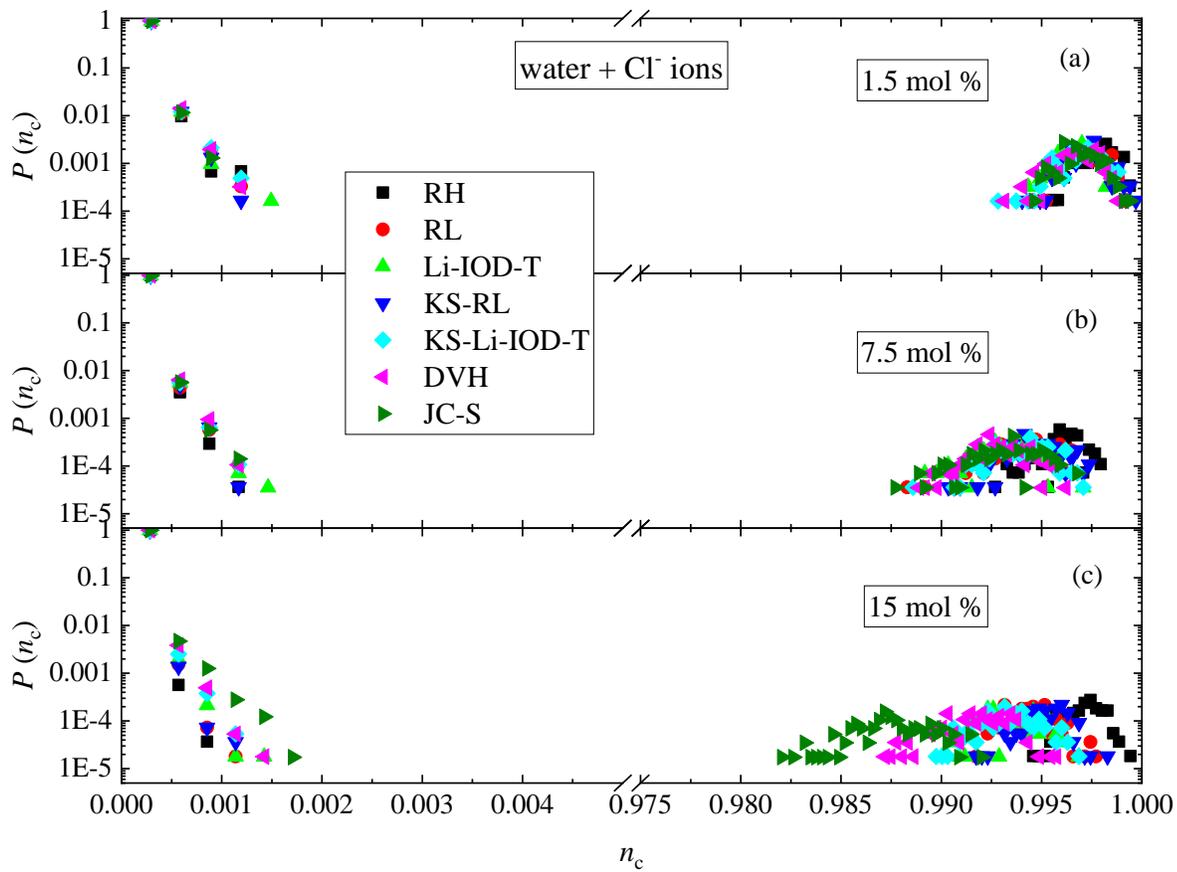

**Figure 18** Cluster size distributions calculated for the atomic configurations obtained from some selected models at different salt concentrations. Both water molecules and chloride ions are taken into account. The *x*-axes are normalized with the number of water molecules and Cl⁻ ions in the configurations.

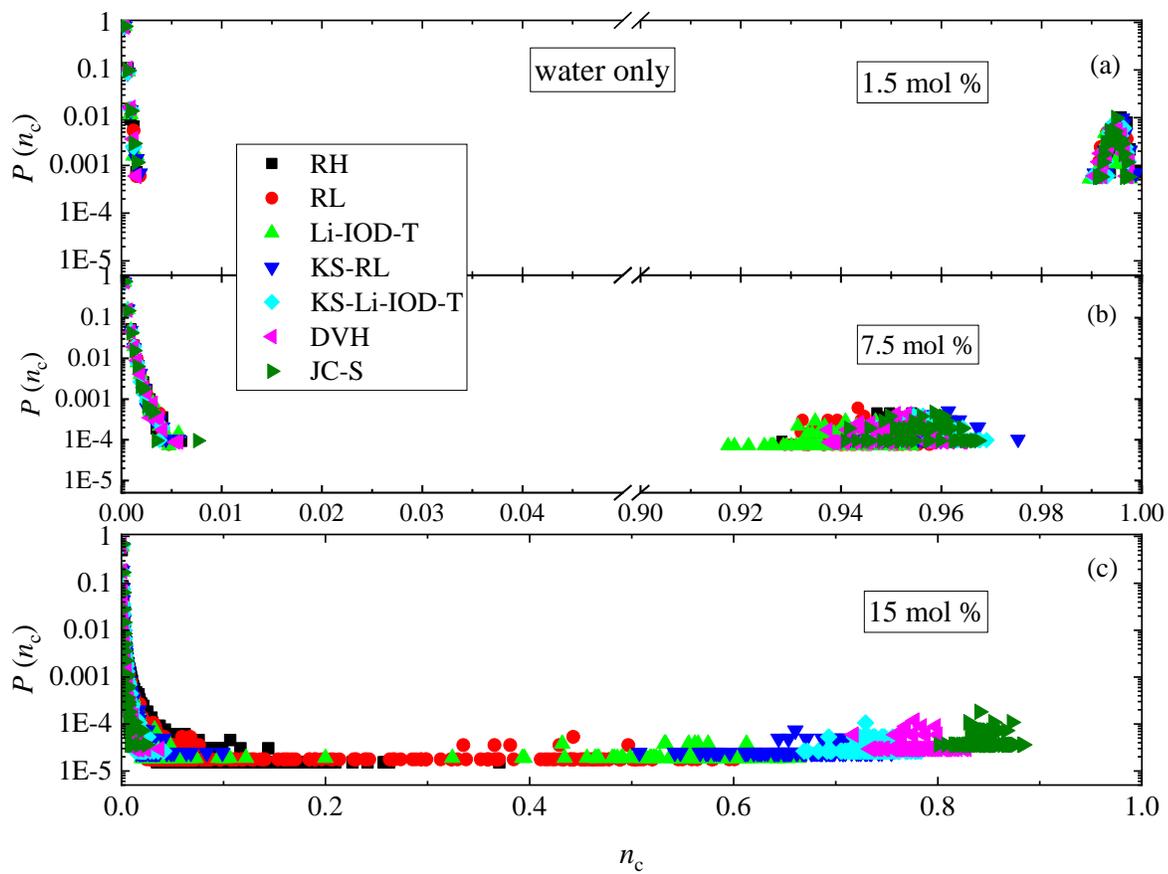

**Figure 19** Cluster size distributions calculated for the atomic configurations obtained from some selected models at different salt concentrations. Only water molecules are taken into account. The *x*-axes are normalized with the number of water in the configurations. Note: the *x*-scales of (a) and (b) are different from that of (c)!

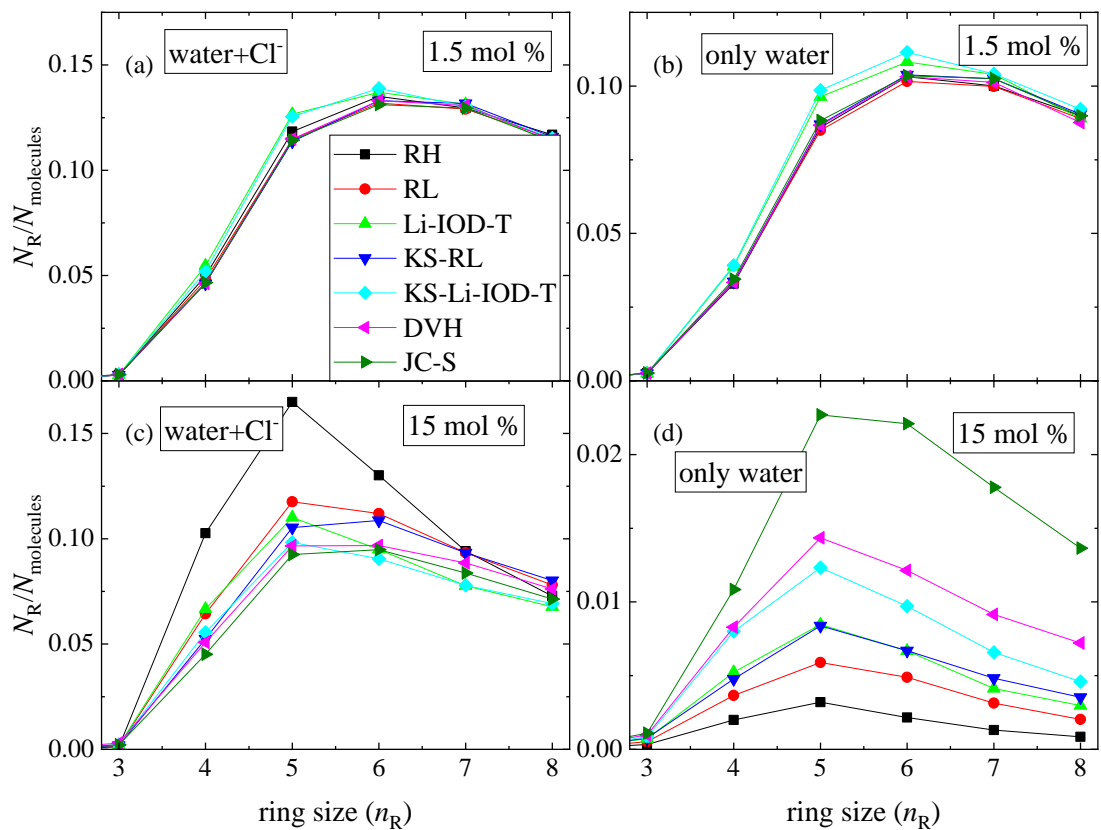

**Figure 20** Size distribution of cyclic entities calculated for the atomic configurations obtained from some selected models at salt concentrations (a and b) 1.5 mol % and (c and d) 15 mol %. During the calculations (b and d) only water molecules and (a and c) water molecules and chloride ions together were taken into account.

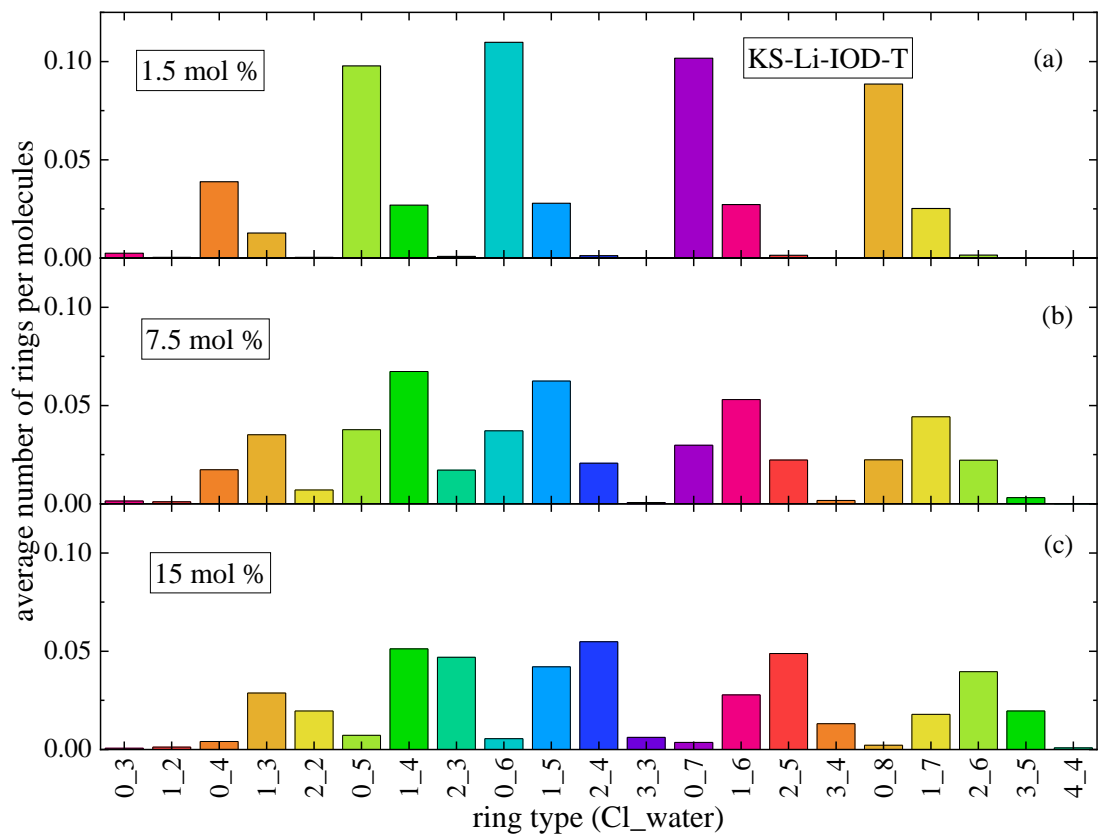

**Figure 21** Distribution of different types of rings (rings contain water molecules and Cl$^-$ ions) normalized by the number of molecules (water + Cl$^-$ ions) in the configurations at different salt concentrations obtained from the KS-Li-IOD-T model.

Supplementary Material

for

Towards the correct microscopic structure of aqueous CsCl solutions with a comparison of classical interatomic potential models

Ildikó Pethes

Wigner Research Centre for Physics, Konkoly Thege út 29-33., H-1121 Budapest, Hungary

E-mail address: pethes.ildiko@wigner.hu

# 1. Interatomic potential parameters

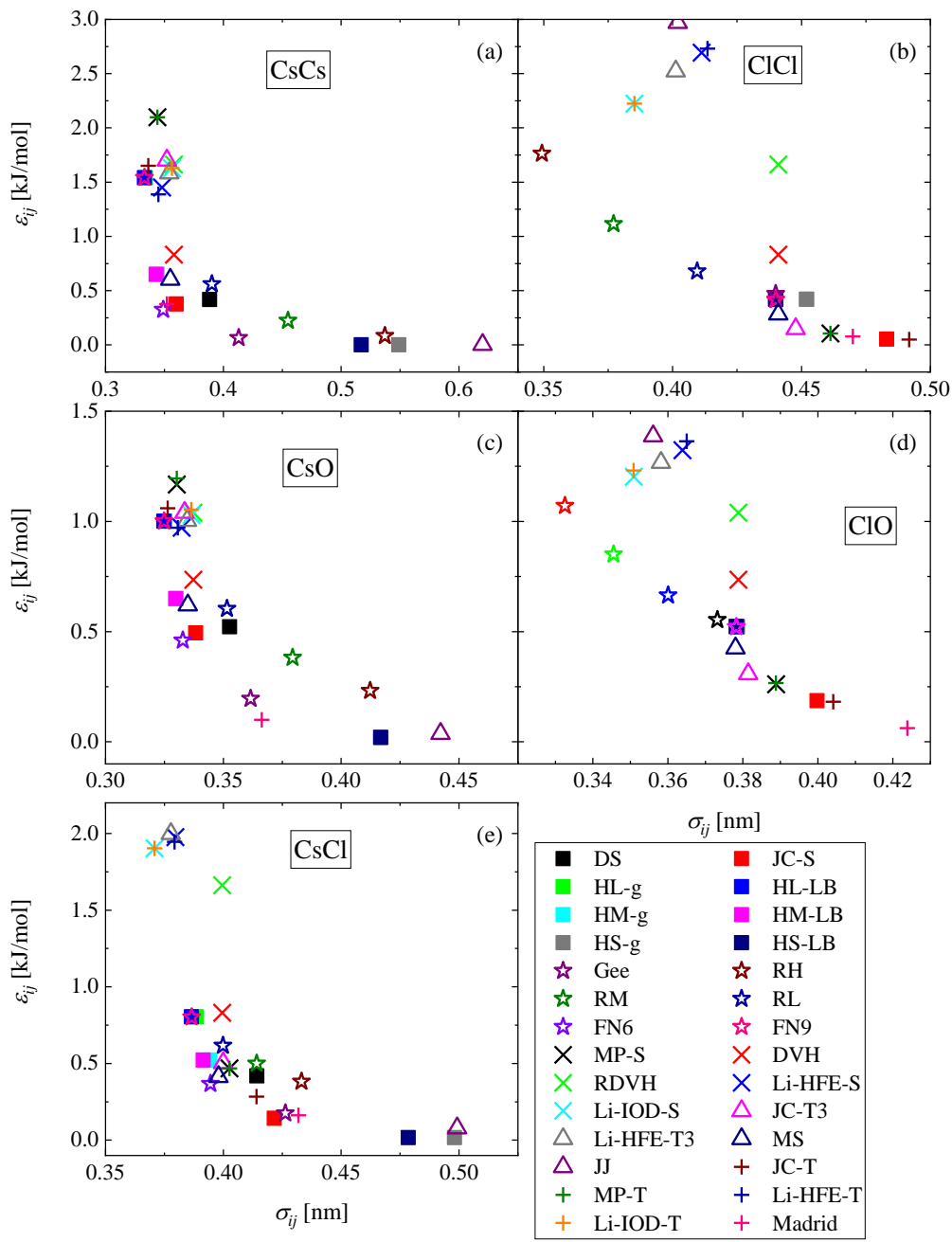

**Figure S1.** Graphical representation of the Lennard-Jones ($\varepsilon_{ij}$ and $\sigma_{ij}$) parameters of the investigated models. (Some of the points overlap; see the text for the values!) (The $\varepsilon_{ij}$ and $\sigma_{ij}$ parameters of the KS charge scaled models are the same as the original models.)

## 2. Contribution of partial structure factors to the neutron and X-ray weighted total scattering structure factors

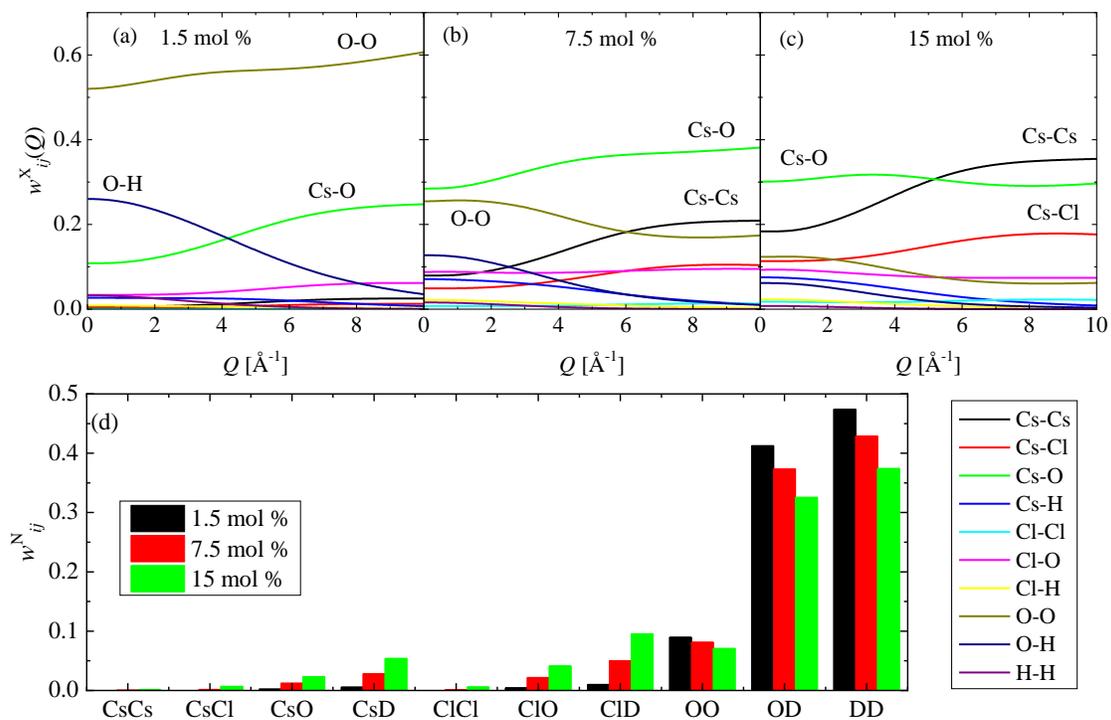

**Figure S2.** (a-c) X-ray and (d) neutron scattering weights used for the calculations of the X-ray and neutron total structure factors.

## 3. Snapshots of atomic configurations obtained in molecular dynamics simulations

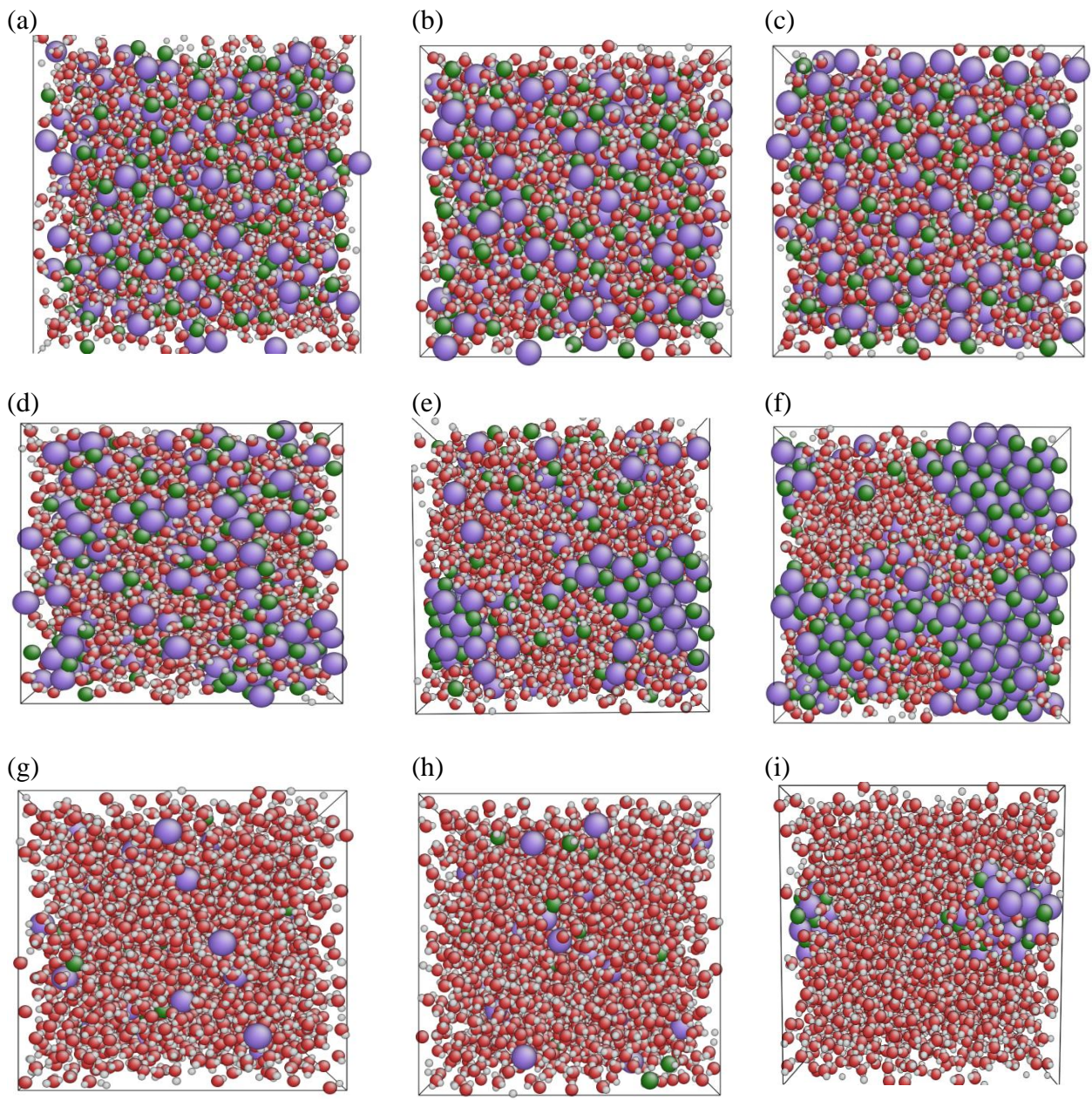

**Figure S3.** Snapshots of typical configurations obtained with different models at the concentrations (a-f) 15 mol %, and (g-i) 1.5 mol %. Red, gray, purple, and green balls represent oxygen, hydrogen, cesium, and chlorine atoms, respectively. (a) KS-Li-IOD-T 15 mol %, (b) Madrid 15 mol %, (c) RH 15 mol %, (d) JC-S 15 mol %, (e) FN6 15 mol %, (f) HS-g 15 mol %, (g) Li-IOD-T 1.5 mol %, (h) RH 1.5 mol %, and (i) HS-LB 1.5 mol %.

## 4. Calculation of basic properties: density, static dielectric constant, and self-diffusion coefficients

*4.1 Density*

A 4 ns long NpT simulation was carried out to determine the density predicted by the tested model. The pressure was kept at p = $10^5$ Pa by Parrinello-Rahman barostat [S1,S2], using a $\tau_p$ = 2.0 coupling constant. The Nose-Hoover thermostat was used in these simulations, with $\tau_T$ = 0.5 coupling. Densities were calculated every 2 ps from 2 to 4 ns, and the average of the 1000 values was determined.

Density values obtained by MD simulations are collected in Table S1. In Fig. 1 in the main article, the simulated/experimental density ratios are shown, where the experimental values are taken from Ref. [S3] ($\rho$ values in Ref. [S3] refer to heavy water, the reported number densities were used here to calculate $\rho$ values for $H_2O$.) For the lowest concentration, all models perform excellently: the simulated density values agree with the experimental ones within 1 %. For the 7.5 mol % solution, the discrepancies are in the 1 – 3 % range; the simulated values are, in most cases, higher than the experimental ones. In the case of the 15 mol % sample, the deviations are below 4 % for several models, the simulated values are lower than the experimental ones, and the highest difference is about 10 % (for JJ and RDVH models). It should be noted that the experimental density of the 15 mol % sample in Ref [S3] (1.91 g/cm$^3$) is somewhat higher than the value from other references (1.84 g/cm$^3$ [S4], 1.87 g/cm$^3$ [S5]).

**Table S1** Densities of the investigated solutions obtained in MD simulations (in kg/m$^3$). Experimental values are also presented (in bold) for comparison.

|  | 1.5 mol % | 7.5 mol % | 15 mol % |
|---|---|---|---|
| **experimental[S3]** | **1102** | **1464** | **1913** |
| DS | 1092 | 1430 | 1754 |
| JC-S | 1105 | 1480 | 1860 |
| HL-g | 1106 | 1483 | 1848 |
| HL-LB | 1106 | 1483 | 1851 |
| HM-g | 1105 | 1482 | 1851 |
| HM-LB | 1105 | 1480 | 1853 |
| HS-g | 1103 | 1472 | 1851 |
| HS-LB | 1101 | 1485 | 1892 |
| Gee | 1103 | 1475 | 1839 |
| RH | 1097 | 1436 | 1746 |
| RM | 1103 | 1469 | 1810 |
| RL | 1106 | 1485 | 1848 |
| KS-RL | 1099 | 1451 | 1788 |
| FN6 | 1105 | 1483 | 1865 |
| FN9 | 1106 | 1483 | 1852 |
| MP-S | 1107 | 1491 | 1861 |
| DVH | 1097 | 1434 | 1756 |
| RDVH | 1092 | 1408 | 1700 |
| Li-HFE-S | 1103 | 1466 | 1817 |
| Li-IOD-S | 1110 | 1505 | 1894 |
| JC-T3 | 1095 | 1486 | 1866 |
| Li-HFE-T3 | 1094 | 1476 | 1844 |
| MS | 1093 | 1475 | 1856 |
| JJ | 1089 | 1412 | 1700 |
| JC-T | 1104 | 1485 | 1856 |
| MP-T | 1105 | 1493 | 1869 |
| Li-HFE-T | 1101 | 1470 | 1825 |
| Li-IOD-T | 1108 | 1508 | 1900 |
| KS-Li-IOD-T | 1101 | 1476 | 1846 |
| Madrid | 1098 | 1463 | 1835 |

*4.2 Static dielectric constant*

The static dielectric constant was calculated according to Eq. S1:

$$\varepsilon = 1 + \frac{\langle M^2 \rangle - \langle M \rangle^2}{3\varepsilon_0 V k_B T};  \qquad (Eq.\ S1)$$

where $M$ is the total dipole moment, $V$ is the volume of the system, $k_B$ is the Boltzmann-constant, $\varepsilon_0$ is the vacuum permittivity, and $T$ is the temperature. All saved configurations were used as an input for the calculations performed with the 'gmx dipole' program of the GROMACS software. The average of the $\varepsilon$ values in the last 10 ns part of the $\varepsilon(t)$ curves (10 ns – 20 ns) was determined and taken as the final result.

Some typical $\varepsilon(t)$ curves are shown in Fig. S4. Long trajectories are necessary until the curves converge to the final $\varepsilon$ values, as was previously demonstrated for pure water in Ref [S6]. A 6 ns long trajectory is enough for pure water, as was found in Ref [S6]. However, in concentrated salt solutions, $\varepsilon(t)$ curves are not saturated even after 8 ns [S7]. Thus 20 ns long trajectories were used in this work; and the last 10 – 20 ns of the curves were averaged to determine the $\varepsilon$ value. The 20 ns was sufficient for $\varepsilon(t)$ to saturate at all concentrations tested, except for two FFs: HS-LB and HS-g.

$\varepsilon$ values obtained from the simulations with different FFs are shown in Fig. 2 of the main text, and Table S2. The experimental values at the given concentration were calculated using the formula of Ref. [S8]. The simulated values are mostly smaller than the experimental ones. An exception is the HS-LB model, where the $\varepsilon(t)$ curve for the 15 mol % concentration does not converge. The static dielectric constant values obtained by using the 3 FFs using the TIP3P water model (JC-T3, Li-HFE-T3, and MS) at 1.5 mol % concentration are higher than the experimental ones. It is not surprising considering that the TIP3P water model gives a simulated $\varepsilon$ value significantly higher than the experimental one even for pure water (94 compared to 78.5) [S9]. For the remaining 26 FFs, the deviation from experimental value is 10 – 38 %, 18 – 50 %, and 22 – 60 % for the 1.5, 7.5, and 15 mol % solutions, respectively. The smaller values correspond to HS-g, JC-S, FN6, DVH, and RDVH models, while the worst-performing models are the RH, RM models and the FFs with the TIP4P and the TIP4P-Ew water model.

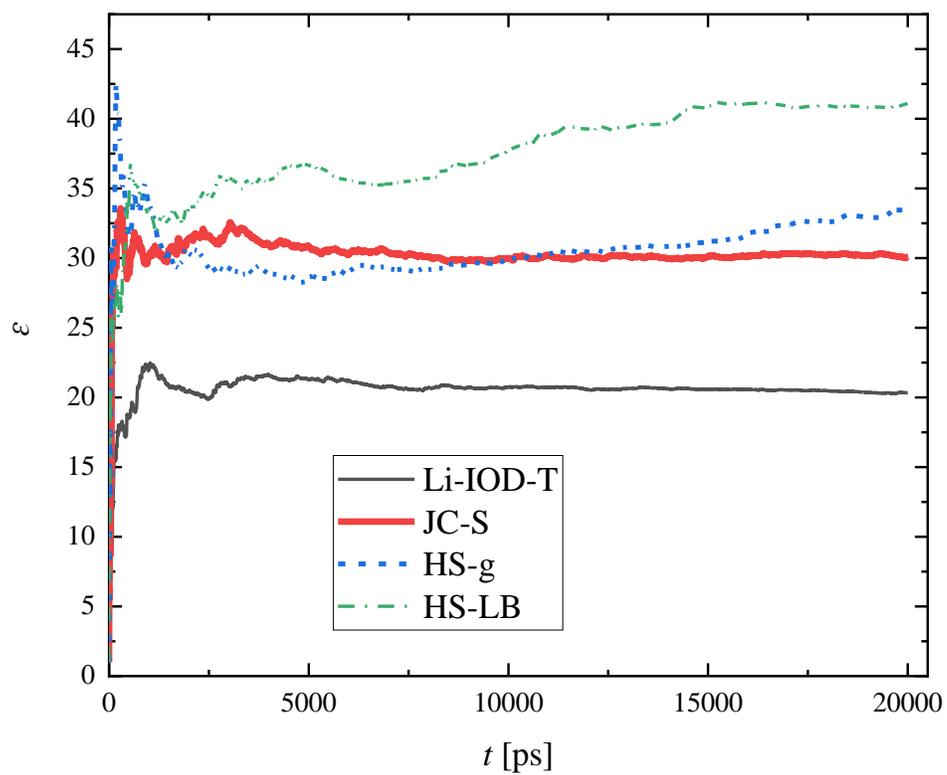

**Figure S4.** Convergence of the static dielectric constant for some selected models (Li-IOD-T, JC-S, HS-g, and HS-LB) for the 15 mol % solution. The curve is convergent for the Li-IOD-T and JC-S models, but it is still evolving at $t = 20$ ns for the HS-g (and maybe for the HS-LB) model.

**Table S2** Static dielectric constants of the investigated solutions obtained in MD simulations. Experimental values are also presented (in bold) for comparison.

|                  | 1.5 mol % | 7.5 mol % | 15 mol % |
|------------------|-----------|-----------|----------|
| **experimental[S8]** | **69.9** | **50.4** | **40.3** |
| DS               | 64.2      | 38.7      | 28.3     |
| JC-S             | 61.0      | 40.1      | 30.1     |
| HL-g             | 64.7      | 35.2      | 23.0     |
| HL-LB            | 59.2      | 36.3      | 24.5     |
| HM-g             | 59.6      | 36.0      | 24.7     |
| HM-LB            | 62.2      | 37.4      | 26.3     |
| HS-g             | 61.4      | 39.7      | 31.5     |
| HS-LB            | 69.0      | 51.9      | 40.2     |
| Gee              | 58.6      | 36.3      | 24.4     |
| RH               | 63.2      | 35.3      | 18.5     |
| RM               | 64.3      | 35.9      | 19.6     |
| RL               | 62.0      | 36.2      | 22.0     |
| KS-RL            | 60.0      | 38.1      | 25.4     |
| FN6              | 60.9      | 40.2      | 29.5     |
| FN9              | 61.1      | 37.5      | 24.3     |
| MP-S             | 60.6      | 34.8      | 22.2     |
| DVH              | 62.1      | 39.9      | 29.2     |
| RDVH             | 61.0      | 41.3      | 28.9     |
| Li-HFE-S         | 59.3      | 38.3      | 28.2     |
| Li-IOD-S         | 58.3      | 36.2      | 24.2     |
| JC-T3            | 78.9      | 40.4      | 25.5     |
| Li-HFE-T3        | 77.9      | 42.8      | 28.2     |
| MS               | 79.9      | 43.3      | 29.0     |
| JJ               | 43.3      | 25.0      | 15.3     |
| JC-T             | 56.9      | 32.2      | 21.0     |
| MP-T             | 57.3      | 30.8      | 20.3     |
| Li-HFE-T         | 54.4      | 34.2      | 24.3     |
| Li-IOD-T         | 54.0      | 33.4      | 20.6     |
| KS-Li-IOD-T      | 57.0      | 35.5      | 23.7     |
| Madrid           | 48.6      | 29.4      | 15.5     |

The ratio of the static dielectric constant of the solution to that of pure water was also calculated to eliminate the effect of the water model and to focus on the differences between the FFs of the ions. The simulated $\varepsilon/\varepsilon_{\text{water}}$ values were compared with the same ratio for the experimental values. (The simulated values of the static dielectric constant of the pure water are: 50, 58, 66, 68, and 94 for TIP4P, TIP4P/2005, TIP4P-Ew, SPC/E, and TIP3P water models, respectively, while the experimental one is 78.5 (see e. g. [S9]).) The result of this comparison is shown in Fig. S5. The deviations of the simulated and experimental normalized dielectric permittivity are smaller: they are below 10 % for the 1.5 mol % solution (except for the HS-LB model, which is 14 %), 6 – 33 % for the 7.5 mol % salt solution, and 10 – 48 % for the 15 mol % sample. The difference averaged over the three investigated concentrations is a useful characteristic of the performance of the FFs; its value was 6 – 28 % for the different models. According to this indicator, the 5 best models are HS-g, JC-S, RDVH, FN6, and DVH; the 5 worst performing models in this test are JC-T3, Li-HFE-T3, MS, RH, and MP-T models.

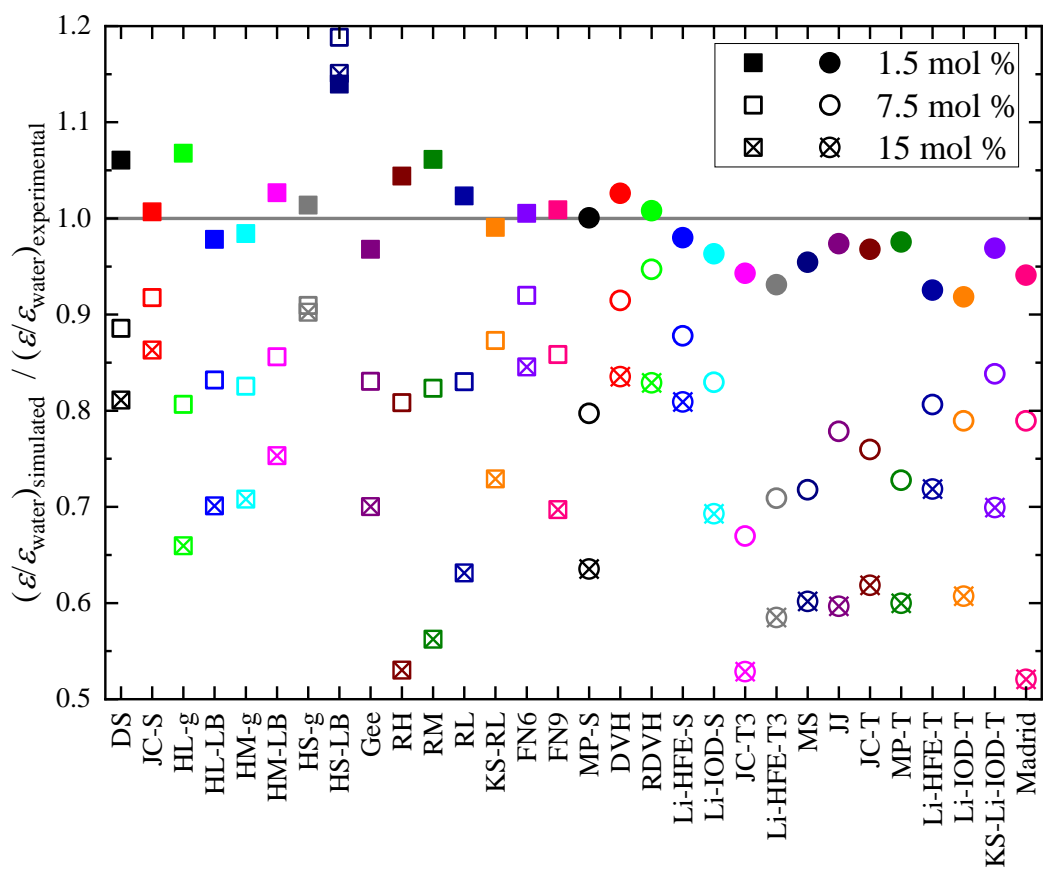

**Figure S5.** Ratio of the simulated and experimental static dielectric constants normalized with the (simulated and experimental) $\varepsilon$ values of pure water.

*4.3 Self-diffusion coefficients*

The self-diffusion coefficients were calculated from the mean-square displacement (MSD) using the Einstein-relationship (Eq. S2):

$$6D_A t = \lim_{t \to \infty} \langle \|r_i(t) - r_i(0)\|^2 \rangle_{i \in A} \qquad \text{(Eq. S2)}$$

Here $A$ refers to either $Cs^+$, $Cl^-$ ions, or water molecules. The full trajectories were used by restarting the MSD-calculation every 10 ps. The 'gmx msd' program of the GROMACS package was used for the calculations. The linear part of the MSD-$t$ curve was fitted to determine $D_A$. The MSD-$t$ curves of the water molecules were linear for the full 0 – 20 ns region; they were fitted in the 2 – 18 ns part. However, the curves obtained by calculating the displacement of the ions have non-linear segments in the high $t$ regime for several FFs, mostly at lower salt concentrations. Some of these curves already deviate from linear at 6 ns, others are linear up to 16 or 18 ns; only their linear segments were used to determine $D_A$.

The obtained self-diffusion coefficients are presented in Figs. 3, S6, S7, and in Tables S3-S5. Available experimental values are also shown; unfortunately, as far as we know, there is no experimental $D_{Cl^-}$ value for the highest salt concentration.

The simulated self-diffusion values can depend on the system sizes [S10]; the effect of the finite system size can be estimated by Eq. S3:

$$D = D_{PBC} + \frac{\xi k_B T}{6\pi\eta L}. \qquad \text{(Eq. S3)}$$

Here $D$ is the diffusion coefficient in the infinite size limit, $D_{PBC}$ is the diffusion coefficient calculated from a simulation with periodic boundary conditions. For a cubic simulation box, $\xi=2.837297$, $L$ is the box length, and $\eta$ is the viscosity. The contributions of the finite size effect, calculated with the experimental viscosity values taken from Refs. [S11, S12] are 1.54, 1.49 and 1.13 × $10^{-10}$ $m^2/s$ for 1.5 mol %, 7.5 mol % and 15 mol % solution, respectively. After subtracting these corrections from the experimental results one can get "reduced experimental values", which represent the target that an ideal FF using periodic boundary conditions should be able to reproduce. This reduced value is shown as dotted line in Figs. 3, S6, S7.

The simulated water self-diffusion coefficients are close to the experimental values for the 1.5 mol % solution, except for the models with TIP3P and TIP4P water models, which give higher self-diffusion values for pure water as well (5.49 and 3.89 × $10^{-9}$ $m^2/s$, compare to the experimental 2.3 × $10^{-9}$ $m^2/s$ value). As the salt concentration increases, the simulated self-diffusion values decrease in contrast with the experimental results: the latter has a maximum at around 5 – 6 mol %. For the 7.5 mol % solution, only the JJ, JC-T3, MS, and Li-HFE-T3 models give values higher than the experimental results; the models with SPC/E or TIP4P-Ew water show 8 – 45 % lower values. At the highest salt concentration, each FFs give self-diffusion values lower than the experiments. The deviation averaged over the three investigated concentrations is the smallest (the best) for the KS-Li-IOD-T and the Li-HFE-T models. The charge scaled KS models perform significantly better than the original models (17 % for KS-Li-IOD-T, 26 % for Li-IOD-T, and 24 % for KS-RL and 29 % RL). Water molecules move more slowly in the Reif-Hünenberger models (RL, RM, RH) than in the other models, $D_{water}$ obtained by the RH model is especially low ($D_{RH} < D_{RM} < D_{RL}$ for all concentrations).

**Table S3** Water self-diffusion coefficients of the investigated solutions obtained in MD simulations (in $10^{-9}$ m$^2$/s). (The simulated values shown are not corrected to finite size effect!) Experimental values are also presented (in bold) for comparison.

|  | 1.5 mol % | 7.5 mol % | 15 mol % |
|---|---|---|---|
| **experimental[S13]** | **2.43** | **2.49** | **2.16** |
| DS | 2.62 | 1.93 | 1.13 |
| JC-S | 2.43 | 1.90 | 1.20 |
| HL-g | 2.47 | 1.90 | 1.14 |
| HL-LB | 2.55 | 1.97 | 1.16 |
| HM-g | 2.48 | 1.93 | 1.22 |
| HM-LB | 2.58 | 1.93 | 1.20 |
| HS-g | 2.59 | 1.95 | 1.28 |
| HS-LB | 2.59 | 1.87 | 1.25 |
| Gee | 2.49 | 1.92 | 1.20 |
| RH | 2.32 | 1.21 | 0.29 |
| RM | 2.43 | 1.51 | 0.58 |
| RL | 2.51 | 1.72 | 0.90 |
| KS-RL | 2.54 | 1.89 | 1.12 |
| FN6 | 2.52 | 1.94 | 1.21 |
| FN9 | 2.55 | 1.94 | 1.15 |
| MP-S | 2.53 | 1.79 | 0.97 |
| DVH | 2.55 | 2.07 | 1.16 |
| RDVH | 2.60 | 2.05 | 0.95 |
| Li-HFE-S | 2.59 | 2.06 | 1.16 |
| Li-IOD-S | 2.54 | 1.80 | 1.01 |
| JC-T3 | 5.18 | 3.39 | 1.65 |
| Li-HFE-T3 | 5.15 | 3.62 | 1.73 |
| MS | 5.15 | 3.55 | 1.93 |
| JJ | 3.53 | 2.55 | 1.18 |
| JC-T | 2.43 | 1.73 | 1.08 |
| MP-T | 2.41 | 1.75 | 0.97 |
| Li-HFE-T | 2.48 | 2.04 | 1.19 |
| Li-IOD-T | 2.50 | 1.80 | 1.02 |
| KS-Li-IOD-T | 2.51 | 2.13 | 1.32 |
| Madrid | 2.09 | 1.64 | 1.06 |

The simulated cesium ion self-diffusion coefficients are mainly close to the experimental one (within 15 %) for the 1.5 mol % solution, except for the models with TIP3P or TIP4P water, which models give higher values and the HS-LB model, in which both $D_{Cs+}$ and $D_{Cl-}$ are very low, meaning that the ions do not move in this model. As the salt concentration increases, the self-diffusion coefficients decrease for simulation and experiments alike, and the decrease is higher for the simulated values. The deviation is mostly 15 – 50 % for the 7.5 mol % solution and 30 – 70 % for the 15 mol % sample. According to the averaged deviation, the best models are the KS-Li-IOD-T, Gee, FN9, and HL-g models; the worst performing models are HS-LB, RH, Li-HFE-T3, RM, and MS.

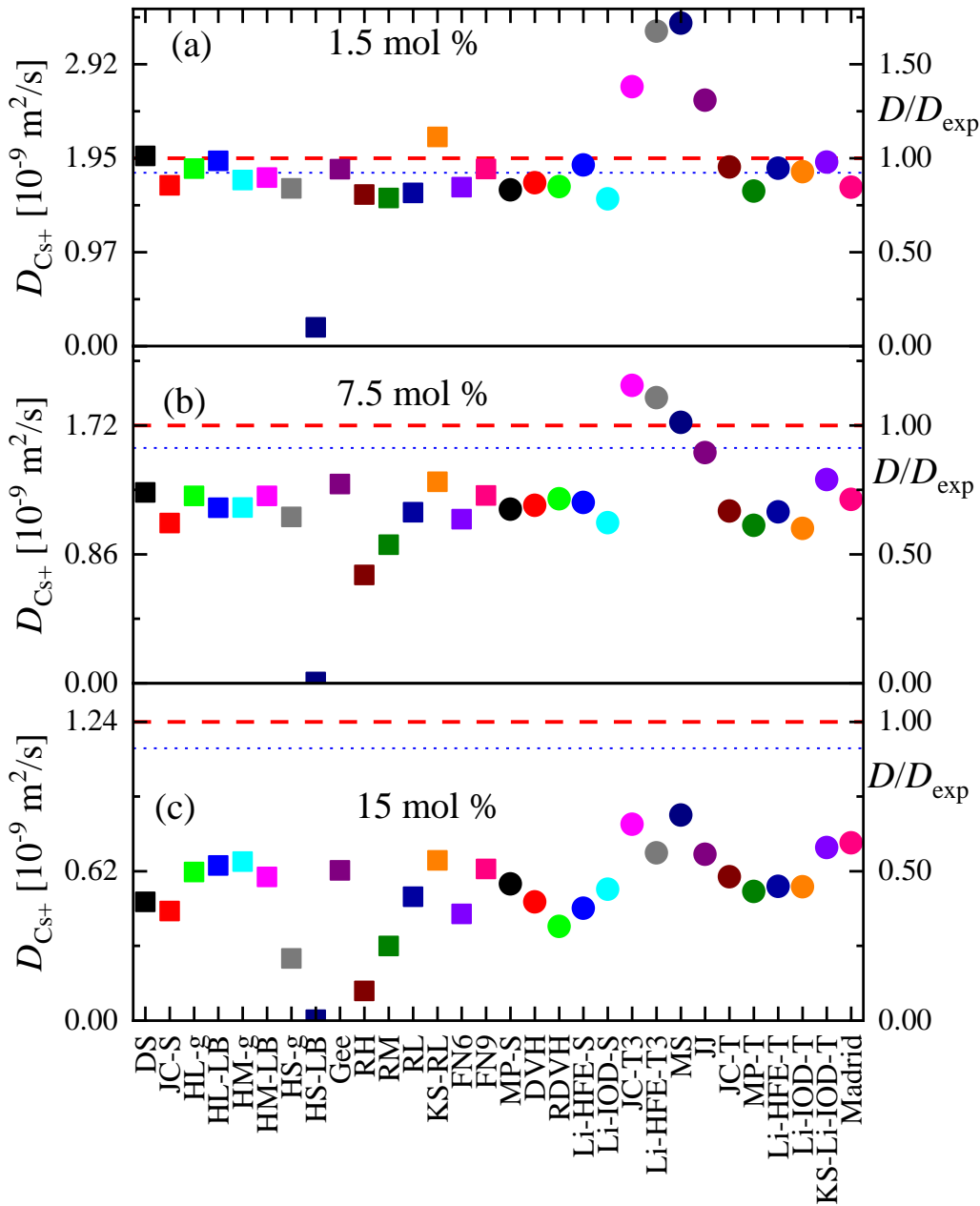

**Figure S6.** Self-diffusion coefficients of $Cs^+$ ions in aqueous CsCl solutions obtained in MD simulations. The experimental values are marked by dashed lines. Experimental $D_{Cs+}$ data (1.95, 1.72, and 1.24 × $10^{-9}$ $m^2$/s for 1.5 mol %, 7.5 mol %, and 15 mol % solutions, respectively) were estimated taking into account the data in Refs. [S14-S16]. The experimental value corrected by the finite size effect is represented by dotted lines.

**Table S4** Cs$^+$ self-diffusion coefficients of the investigated solutions obtained in MD simulations (in 10$^{-9}$ m$^2$/s) (The simulated values shown are not corrected to finite size effect!) Experimental values are also presented (in bold) for comparison.

|  | 1.5 mol % | 7.5 mol % | 15 mol % |
| --- | --- | --- | --- |
| **experimental[S14-S16]** | **1.95** | **1.72** | **1.24** |
| DS | 1.97 | 1.27 | 0.49 |
| JC-S | 1.67 | 1.07 | 0.45 |
| HL-g | 1.84 | 1.25 | 0.62 |
| HL-LB | 1.92 | 1.17 | 0.64 |
| HM-g | 1.72 | 1.17 | 0.66 |
| HM-LB | 1.75 | 1.25 | 0.60 |
| HS-g | 1.64 | 1.11 | 0.26 |
| HS-LB | 0.19 | 0.01 | 0.00 |
| Gee | 1.84 | 1.33 | 0.62 |
| RH | 1.57 | 0.72 | 0.12 |
| RM | 1.54 | 0.92 | 0.31 |
| RL | 1.59 | 1.14 | 0.51 |
| KS-RL | 2.17 | 1.34 | 0.66 |
| FN6 | 1.65 | 1.09 | 0.44 |
| FN9 | 1.84 | 1.25 | 0.63 |
| MP-S | 1.62 | 1.16 | 0.57 |
| DVH | 1.70 | 1.19 | 0.49 |
| RDVH | 1.66 | 1.23 | 0.39 |
| Li-HFE-S | 1.88 | 1.21 | 0.47 |
| Li-IOD-S | 1.53 | 1.07 | 0.54 |
| JC-T3 | 2.69 | 1.99 | 0.82 |
| Li-HFE-T3 | 3.27 | 1.90 | 0.70 |
| MS | 3.35 | 1.74 | 0.85 |
| JJ | 2.55 | 1.54 | 0.69 |
| JC-T | 1.86 | 1.15 | 0.60 |
| MP-T | 1.61 | 1.05 | 0.54 |
| Li-HFE-T | 1.85 | 1.14 | 0.56 |
| Li-IOD-T | 1.81 | 1.03 | 0.56 |
| KS-Li-IOD-T | 1.91 | 1.36 | 0.72 |
| Madrid | 1.65 | 1.23 | 0.74 |

Concerning the self-diffusion coefficients of the chloride ion, the tendencies are similar to that found for the cesium ion. As the available experimental data are insufficient, only a qualitative comparison can be made. The chloride ions in the HS-LB model do not move. In the 15 mol % solutions, the self-diffusion coefficients of chloride ions are very low for the RH and HS-g models. The mobility of ions is highest for the TIP3P models, and the charge scaling results in higher self-diffusion values of the chloride ions, similarly to the cesium ions.

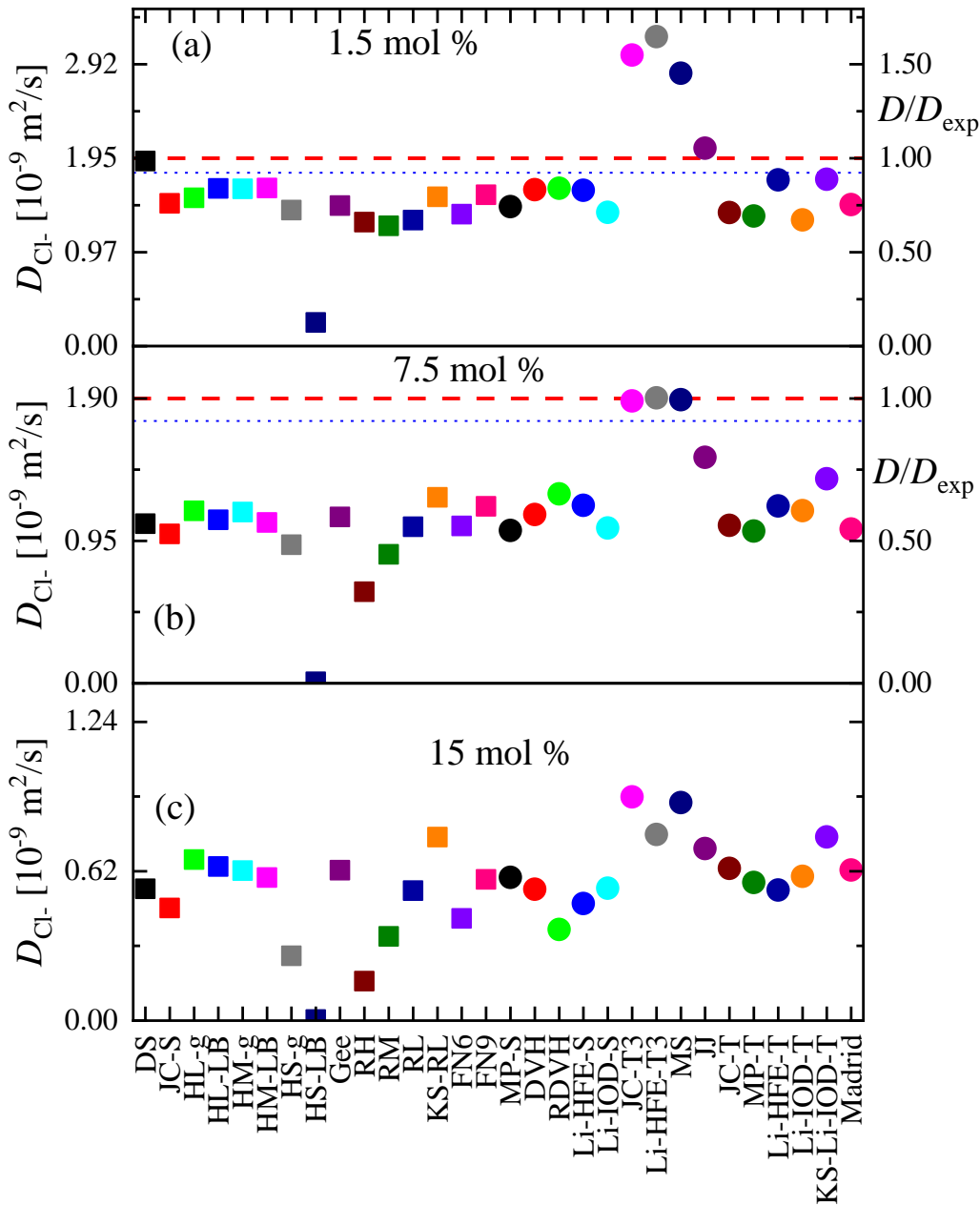

**Figure S7.** Self-diffusion coefficients of Cl$^-$ ions in aqueous CsCl solutions obtained in MD simulations. The experimental values are marked by dashed lines. Experimental $D_{Cl^-}$ data (1.95 and 1.90× 10$^{-9}$ m$^2$/s for 1.5 mol % and 7.5mol % solutions, respectively) were estimated taking into account the data in Refs. [S14, S15]. The available experimental data concerning $D_{Cl^-}$ is incomplete and uncertain at high concentrations. The experimental value corrected by the finite size effect is represented by dotted lines.

**Table S5** Cl⁻ self-diffusion coefficients of the investigated solutions obtained in MD simulations (in $10^{-9}$ m$^2$/s) (The simulated values shown are not corrected to finite size effect!) Experimental values are also presented (in bold) for comparison.

|  | 1.5 mol % | 7.5 mol % | 15 mol % |
|---|---|---|---|
| **experimental[S14,S15]** | **1.95** | **1.9** | -- |
| DS | 1.92 | 1.06 | 0.55 |
| JC-S | 1.48 | 1.00 | 0.47 |
| HL-g | 1.54 | 1.15 | 0.67 |
| HL-LB | 1.64 | 1.09 | 0.64 |
| HM-g | 1.63 | 1.14 | 0.62 |
| HM-LB | 1.64 | 1.07 | 0.59 |
| HS-g | 1.41 | 0.92 | 0.27 |
| HS-LB | 0.25 | 0.01 | 0.00 |
| Gee | 1.46 | 1.11 | 0.62 |
| RH | 1.29 | 0.61 | 0.16 |
| RM | 1.25 | 0.86 | 0.35 |
| RL | 1.31 | 1.04 | 0.54 |
| KS-RL | 1.55 | 1.24 | 0.76 |
| FN6 | 1.37 | 1.05 | 0.42 |
| FN9 | 1.57 | 1.18 | 0.59 |
| MP-S | 1.45 | 1.02 | 0.59 |
| DVH | 1.62 | 1.13 | 0.55 |
| RDVH | 1.64 | 1.26 | 0.38 |
| Li-HFE-S | 1.62 | 1.19 | 0.49 |
| Li-IOD-S | 1.39 | 1.04 | 0.55 |
| JC-T3 | 3.02 | 1.88 | 0.93 |
| Li-HFE-T3 | 3.21 | 1.90 | 0.77 |
| MS | 2.83 | 1.89 | 0.90 |
| JJ | 2.05 | 1.51 | 0.71 |
| JC-T | 1.39 | 1.05 | 0.63 |
| MP-T | 1.35 | 1.02 | 0.57 |
| Li-HFE-T | 1.72 | 1.18 | 0.54 |
| Li-IOD-T | 1.31 | 1.15 | 0.60 |
| KS-Li-IOD-T | 1.73 | 1.36 | 0.76 |
| Madrid | 1.47 | 1.03 | 0.62 |

**5. Neutron and X-ray total scattering structure factors obtained for the 15 mol % solution**

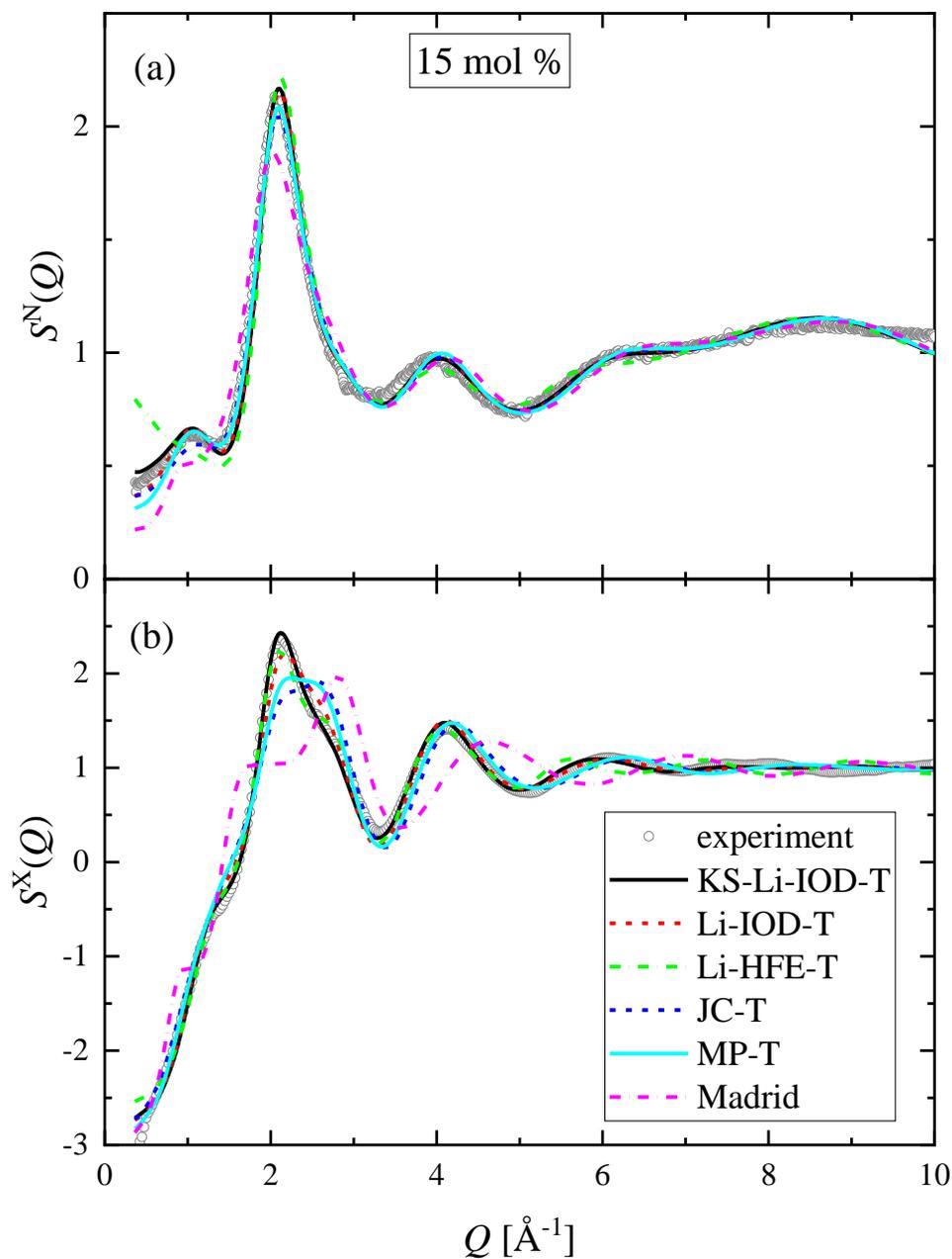

**Figure S8.** (a) Neutron and (b) X-ray total structure factors of the 15 mol % solutions obtained by simulations (lines) compared with the experimental curve (symbols) from Ref. [S3]. The simulated curves were obtained using the force fields: KS-Li-IOD-T, Li-IOD-T, Li-HFE-T, JC-T, MP-T, and Madrid.

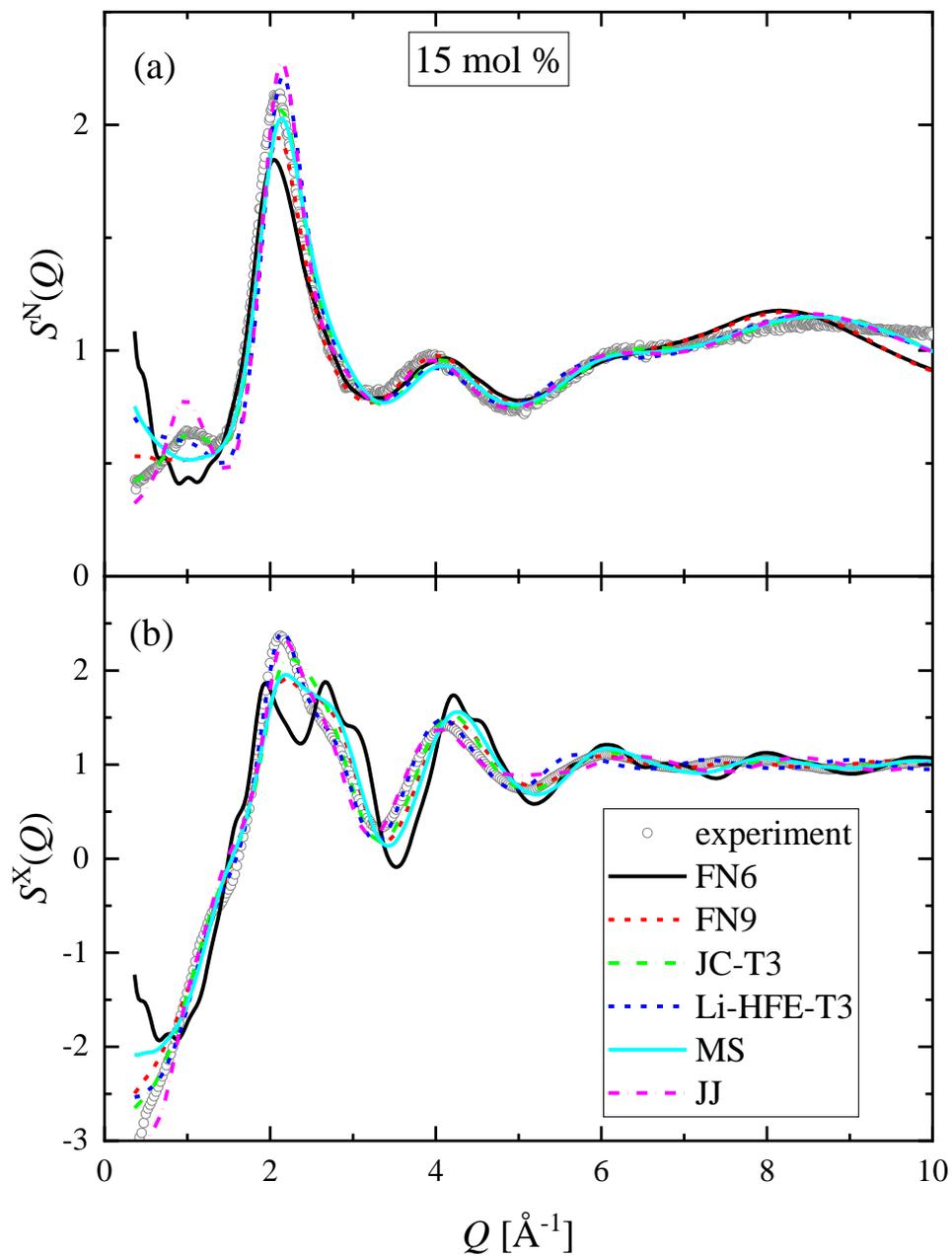

**Figure S9.** (a) Neutron and (b) X-ray total structure factors of the 15 mol % solutions obtained by simulations (lines) compared with the experimental curve (symbols) from Ref. [S3]. The simulated curves were obtained using the force fields: FN6, FN9, JC-T3, Li-HFE-T3, MS, and JJ.

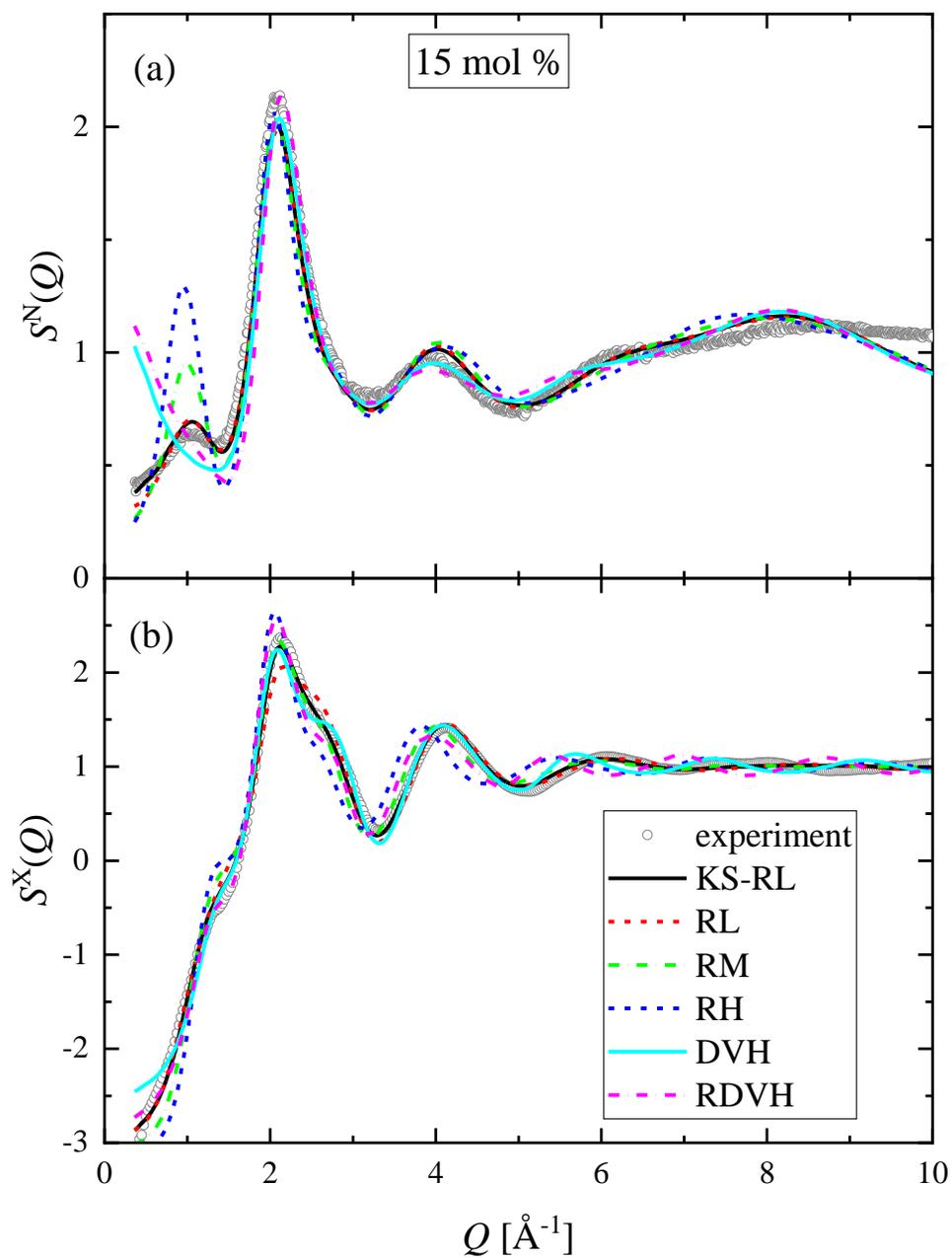

**Figure S10.** (a) Neutron and (b) X-ray total structure factors of the 15 mol % solutions obtained by simulations (lines) compared with the experimental curve (symbols) from Ref. [S3]. The simulated curves were obtained using the force fields: KS-RL, RL, RM, RH, DVH, and RDVH.

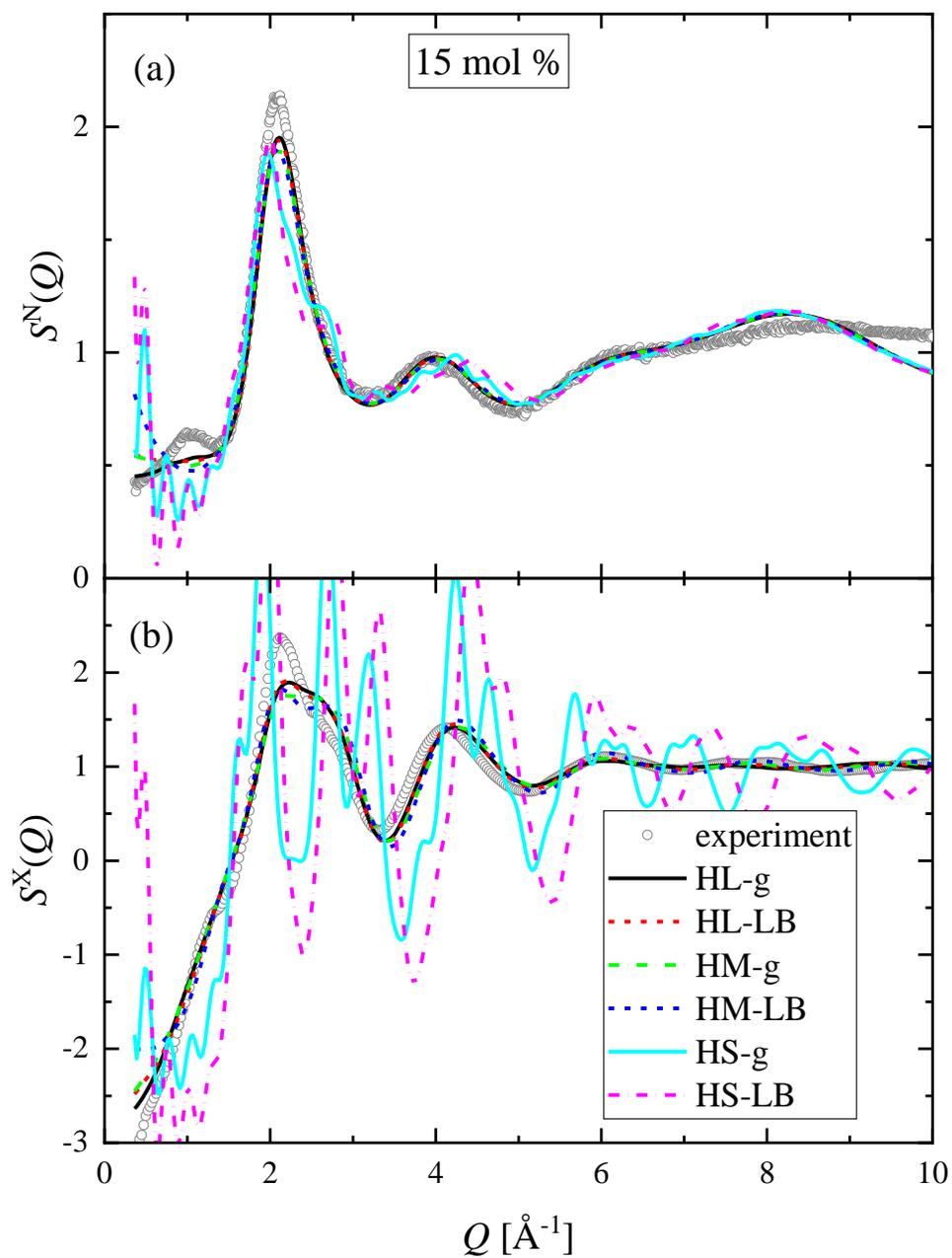

**Figure S11.** (a) Neutron and (b) X-ray total structure factors of the 15 mol % solutions obtained by simulations (lines) compared with the experimental curve (symbols) from Ref. [S3]. The simulated curves were obtained using the force fields: HL-g, HL-LB, HM-g, HM-LB, HS-g, and HS-LB.

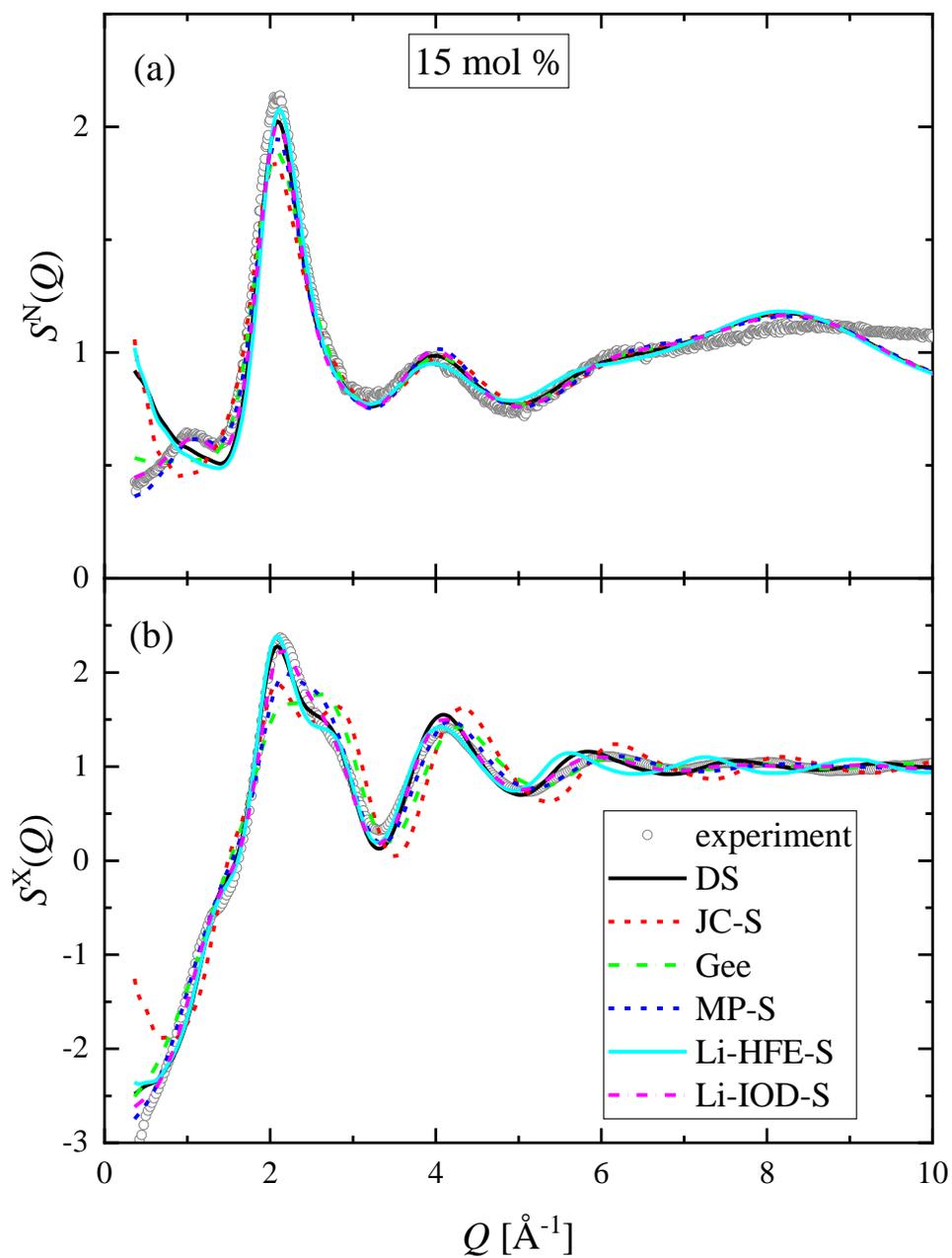

**Figure S12.** (a) Neutron and (b) X-ray total structure factors of the 15 mol % solutions obtained by simulations (lines) compared with the experimental curve (symbols) from Ref. [S3]. The simulated curves were obtained using the force fields: DS, JC-S, Gee, MP-S, Li-HFE-S, and Li-IOD-S.

# 6. Structure of CsCl solutions: Partial radial distribution functions, bond angle distributions, coordination numbers

**Table S6** Position of the first minimum of the Cl$^-$-H PRDF and the $N_{ClH}$ coordination numbers (calculated up to the first minimum) obtained in MD simulations. (The best models are in bold.)

|  | $r_{min}$ [Å] | | | $N_{ClH}$ | | |
|---|---|---|---|---|---|---|
|  | 1.5 mol % | 7.5 mol % | 15 mol % | 1.5 mol % | 7.5 mol % | 15 mol % |
| DS | 3 | 3 | 3 | 6.46 | 5.35 | 4.32 |
| JC-S | 2.96 | 2.98 | 2.94 | 6.25 | 4.84 | 3.71 |
| HL-g | 3 | 3 | 3 | 6.66 | 5.89 | 5 |
| HL-LB | 3 | 3 | 3 | 6.62 | 5.74 | 4.76 |
| HM-g | 3 | 3 | 3 | 6.62 | 5.7 | 4.74 |
| HM-LB | 3 | 3 | 3 | 6.53 | 5.39 | 4.31 |
| HS-g | 3 | 3 | 3 | 6.38 | 4.87 | 2.69 |
| HS-LB | 3 | 3 | 3 | 1.79 | 1.25 | 1.18 |
| Gee | 3 | 3 | 3 | 6.64 | 5.74 | 4.81 |
| RH | 2.74 | 2.8 | 2.74 | 6.74 | 6.44 | 5.91 |
| RM | 2.86 | 2.86 | 2.84 | 6.74 | 6.25 | 5.65 |
| **RL** | **2.92** | **2.94** | **2.94** | **6.69** | **6.07** | **5.37** |
| **KS-RL** | **2.94** | **2.94** | **2.9** | **6.37** | **5.71** | **4.91** |
| FN6 | 3.04 | 3.04 | 3.02 | 6.48 | 5.06 | 3.66 |
| FN9 | 3.04 | 3.04 | 3.02 | 6.68 | 5.79 | 4.79 |
| MP-S | 2.94 | 2.96 | 2.94 | 6.59 | 5.9 | 5.11 |
| DVH | 3.16 | 3.12 | 3.04 | 6.72 | 5.52 | 4.3 |
| RDVH | 3.24 | 3.18 | 3.1 | 6.83 | 5.61 | 4.4 |
| Li-HFE-S | 3.16 | 3.1 | 3.06 | 6.83 | 5.57 | 4.4 |
| **Li-IOD-S** | **2.98** | **3** | **2.96** | **6.74** | **5.88** | **5** |
| JC-T3 | 2.9 | 2.92 | 2.94 | 6.71 | 5.86 | 5.04 |
| Li-HFE-T3 | 3.06 | 3.06 | 3.04 | 6.99 | 5.87 | 4.85 |
| MS | 2.98 | 2.98 | 2.96 | 6.71 | 5.43 | 4.37 |
| JJ | 3.04 | 3.04 | 3 | 6.97 | 6.52 | 5.78 |
| JC-T | 3 | 2.98 | 2.96 | 6.44 | 5.72 | 4.97 |
| MP-T | 2.92 | 2.94 | 2.92 | 6.4 | 5.82 | 5.15 |
| Li-HFE-T | 3.14 | 3.14 | 3.06 | 6.57 | 5.55 | 4.47 |
| **Li-IOD-T** | **2.96** | **2.96** | **2.92** | **6.55** | **5.8** | **5.02** |
| **KS-Li-IOD-T** | **2.96** | **2.96** | **2.94** | **6.15** | **5.38** | **4.58** |
| Madrid | 2.88 | 2.84 | 2.86 | 5.69 | 5.49 | 5.35 |

**Table S7** Position of the first minimum of the Cl$^-$-O PRDF and the $N_{\text{ClO}}$ coordination numbers (calculated up to the first minimum) obtained in MD simulations. (The best models are in bold.)

|  | $r_{\min}$ [Å] | | | $N_{\text{ClO}}$ | | |
| --- | --- | --- | --- | --- | --- | --- |
|  | 1.5 mol % | 7.5 mol % | 15 mol % | 1.5 mol % | 7.5 mol % | 15 mol % |
| DS | 3.9 | 3.8 | 3.7 | 6.71 | 5.57 | 4.45 |
| JC-S | 3.8 | 3.7 | 3.64 | 6.48 | 4.9 | 3.77 |
| HL-g | 3.88 | 3.84 | 3.72 | 7.01 | 6.2 | 5.15 |
| HL-LB | 3.88 | 3.84 | 3.72 | 6.98 | 6.07 | 4.93 |
| HM-g | 3.88 | 3.8 | 3.72 | 7.04 | 5.92 | 4.9 |
| HM-LB | 3.88 | 3.8 | 3.72 | 6.9 | 5.64 | 4.49 |
| HS-g | 3.88 | 3.74 | 3.7 | 6.79 | 5.01 | 2.78 |
| HS-LB | 3.64 | 3.64 | 3.62 | 1.76 | 1.25 | 1.17 |
| Gee | 3.8 | 3.8 | 3.7 | 6.8 | 5.96 | 4.92 |
| RH | 3.6 | 3.6 | 3.52 | 6.88 | 6.52 | 5.96 |
| RM | 3.68 | 3.68 | 3.58 | 6.87 | 6.37 | 5.7 |
| **RL** | **3.76** | **3.72** | **3.64** | **6.89** | **6.17** | **5.42** |
| **KS-RL** | **3.76** | **3.72** | **3.64** | **6.58** | **5.86** | **5.06** |
| FN6 | 3.88 | 3.78 | 3.7 | 6.83 | 5.25 | 3.77 |
| FN9 | 3.88 | 3.84 | 3.74 | 6.98 | 6.07 | 5 |
| MP-S | 3.8 | 3.74 | 3.66 | 6.82 | 6 | 5.18 |
| DVH | 3.9 | 3.9 | 3.78 | 6.84 | 5.88 | 4.64 |
| RDVH | 4 | 3.96 | 3.8 | 7.11 | 6.12 | 4.77 |
| Li-HFE-S | 3.96 | 3.9 | 3.8 | 7.24 | 6.03 | 4.84 |
| **Li-IOD-S** | **3.8** | **3.76** | **3.66** | **6.98** | **6.05** | **5.07** |
| JC-T3 | 3.8 | 3.78 | 3.7 | 7.13 | 6.19 | 5.23 |
| Li-HFE-T3 | 3.94 | 3.9 | 3.84 | 7.68 | 6.48 | 5.5 |
| MS | 3.86 | 3.78 | 3.7 | 6.85 | 5.72 | 4.57 |
| JJ | 3.84 | 3.86 | 3.76 | 7.28 | 6.93 | 6.19 |
| JC-T | 3.8 | 3.74 | 3.7 | 6.63 | 5.83 | 5.13 |
| MP-T | 3.8 | 3.74 | 3.66 | 6.71 | 5.98 | 5.26 |
| Li-HFE-T | 3.94 | 3.84 | 3.78 | 7.04 | 5.79 | 4.88 |
| **Li-IOD-T** | **3.74** | **3.72** | **3.66** | **6.73** | **5.97** | **5.2** |
| **KS-Li-IOD-T** | **3.76** | **3.74** | **3.66** | **6.46** | **5.67** | **4.83** |
| Madrid | 3.6 | 3.68 | 3.66 | 5.69 | 5.69 | 5.52 |

**Table S8** Position of the first minimum of the Cs$^+$-O PRDF and the $N_{CsO}$ coordination numbers (calculated up to the first minimum) obtained in MD simulations. (The best models are in bold.)

|  | $r_{min}$ [Å] | | | $N_{CsO}$ | | |
|---|---|---|---|---|---|---|
|  | 1.5 mol % | 7.5 mol % | 15 mol % | 1.5 mol % | 7.5 mol % | 15 mol % |
| DS | 4 | 4 | 4 | 8.35 | 7.06 | 6.22 |
| JC-S | 3.7 | 3.7 | 3.7 | 6.9 | 5.34 | 4.31 |
| HL-g | 3.8 | 3.8 | 3.84 | 7.87 | 6.88 | 6.19 |
| HL-LB | 3.8 | 3.8 | 3.84 | 7.84 | 6.72 | 5.95 |
| HM-g | 3.7 | 3.7 | 3.74 | 7.26 | 6.21 | 5.47 |
| HM-LB | 3.7 | 3.7 | 3.74 | 7.16 | 5.9 | 5.04 |
| HS-g | 3.6 | 3.6 | 3.54 | 6.24 | 4.81 | 2.68 |
| HS-LB | 3.5 | 3.5 | 3.5 | 1.74 | 1.23 | 1.2 |
| Gee | 3.7 | 3.7 | 3.74 | 7 | 6.02 | 5.39 |
| RH | 4.5 | 4.4? | 4.4? | 10.74 | 10.64 | 10.72 |
| RM | 4.3 | 4.2? | 4.2? | 9.67 | 9.08 | 8.87 |
| **RL** | **4.1** | **4** | **4.14** | **9.18** | **7.81** | **7.91** |
| **KS-RL** | **4.1** | **4.1** | **4.14** | **9.16** | **8.27** | **7.87** |
| FN6 | 3.7 | 3.7 | 3.64 | 6.95 | 5.44 | 3.99 |
| FN9 | 3.9 | 3.8 | 3.86 | 8.23 | 6.72 | 6.02 |
| MP-S | 3.9 | 3.9 | 3.9 | 8.43 | 7.43 | 6.71 |
| DVH | 3.9 | 3.9 | 3.84 | 8.05 | 6.75 | 5.62 |
| RDVH | 4 | 4 | 4.1 | 8.73 | 7.49 | 7.01 |
| Li-HFE-S | 4 | 3.9 | 3.9 | 8.64 | 6.87 | 5.93 |
| **Li-IOD-S** | **4** | **4** | **4** | **8.78** | **7.68** | **6.95** |
| JC-T3 | 4 | 4 | 3.94 | 8.88 | 7.71 | 6.69 |
| Li-HFE-T3 | 3.9 | 3.9 | 4 | 8.35 | 7.11 | 6.65 |
| MS | 3.9 | 3.8 | 3.86 | 8.02 | 6.24 | 5.48 |
| JJ | 4 | 4 | 4.1 | 8.12 | 7.85 | 8.13 |
| JC-T | 3.8 | 3.8 | 3.8 | 7.78 | 6.86 | 6.19 |
| MP-T | 3.9 | 3.9 | 3.94 | 8.27 | 7.4 | 6.98 |
| Li-HFE-T | 4 | 3.9 | 3.9 | 8.37 | 6.84 | 6.06 |
| **Li-IOD-T** | **4** | **4** | **4** | **8.6** | **7.67** | **7.09** |
| **KS-Li-IOD-T** | **4** | **4** | **4.1** | **8.56** | **7.58** | **7.4** |
| Madrid | 3.64 | 3.64 | 3.72 | 6.62 | 6.26 | 6.3 |

**Table S9** Position of the first minimum of the Cs$^+$-Cl$^-$ PRDF and the $N_{CsCl}$ coordination numbers (calculated up to the first minimum) obtained in MD simulations. (The best models are in bold.)

|  | $r_{min}$ [Å] | | | $N_{CsCl}$ | | |
| --- | --- | --- | --- | --- | --- | --- |
|  | 1.5 mol % | 7.5 mol % | 15 mol % | 1.5 mol % | 7.5 mol % | 15 mol % |
| DS | 4.5 | 4.6 | 4.56 | 0.45 | 1.55 | 2.69 |
| JC-S | 4.2 | 4.2 | 4.3 | 0.48 | 1.65 | 2.74 |
| HL-g | 4.3 | 4.4 | 4.3 | 0.26 | 1.05 | 1.99 |
| HL-LB | 4.3 | 4.4 | 4.3 | 0.29 | 1.16 | 2.15 |
| HM-g | 4.3 | 4.4 | 4.3 | 0.28 | 1.09 | 2.1 |
| HM-LB | 4.2 | 4.4 | 4.3 | 0.35 | 1.37 | 2.4 |
| HS-g | 4.1 | 4.2 | 4.2 | 0.41 | 1.57 | 3.53 |
| HS-LB | 4.3 | 4.3 | 4.2 | 4.06 | 4.53 | 4.61 |
| Gee | 4.2 | 4.3 | 4.3 | 0.28 | 1.1 | 2.04 |
| RH | 4.5 | 4.5 | 4.66 | 0.16 | 0.68 | 1.83 |
| RM | 4.5 | 4.5 | 4.48 | 0.23 | 0.87 | 1.76 |
| **RL** | **4.5** | **4.44** | **4.4** | **0.28** | **0.99** | **1.86** |
| **KS-RL** | **4.5** | **4.5** | **4.56** | **0.28** | **1.08** | **2.15** |
| FN6 | 4.2 | 4.26 | 4.3 | 0.41 | 1.57 | 2.85 |
| FN9 | 4.3 | 4.4 | 4.42 | 0.3 | 1.16 | 2.2 |
| MP-S | 4.3 | 4.36 | 4.38 | 0.26 | 1 | 1.92 |
| DVH | 4.5 | 4.6 | 4.6 | 0.4 | 1.51 | 2.74 |
| RDVH | 4.7 | 4.8 | 4.8 | 0.43 | 1.64 | 3.01 |
| Li-HFE-S | 4.6 | 4.6 | 4.6 | 0.44 | 1.59 | 2.82 |
| **Li-IOD-S** | **4.5** | **4.5** | **4.5** | **0.38** | **1.3** | **2.29** |
| JC-T3 | 4.3 | 4.4 | 4.4 | 0.26 | 1.13 | 2.07 |
| Li-HFE-T3 | 4.6 | 4.6 | 4.6 | 0.37 | 1.46 | 2.58 |
| MS | 4.3 | 4.4 | 4.4 | 0.34 | 1.43 | 2.44 |
| JJ | 4.5 | 4.6 | 4.66 | 0.18 | 0.85 | 2.04 |
| JC-T | 4.3 | 4.3 | 4.3 | 0.26 | 0.97 | 1.84 |
| MP-T | 4.2 | 4.36 | 4.38 | 0.25 | 0.95 | 1.82 |
| Li-HFE-T | 4.6 | 4.5 | 4.6 | 0.47 | 1.49 | 2.68 |
| **Li-IOD-T** | **4.5** | **4.4** | **4.44** | **0.38** | **1.19** | **2.13** |
| **KS-Li-IOD-T** | **4.5** | **4.6** | **4.6** | **0.38** | **1.35** | **2.42** |
| Madrid | 4? | 4? | 3.94? | 0.06 | 0.25? | 0.5? |

**Table S10** Position of the first minimum of the O-H PRDF and the $N_{OH}$ coordination numbers (calculated up to 2.4 Å) obtained in MD simulations. (The best models are in bold.)

|  | $r_{min}$ [Å] | | | $N_{OH}$ | | |
| --- | --- | --- | --- | --- | --- | --- |
|  | 1.5 mol % | 7.5 mol % | 15 mol % | 1.5 mol % | 7.5 mol % | 15 mol % |
| DS | 2.4 | 2.34 | 2.34 | 1.78 | 1.48 | 1.23 |
| JC-S | 2.44 | 2.42 | 2.34 | 1.78 | 1.5 | 1.29 |
| HL-g | 2.4 | 2.4 | 2.36 | 1.78 | 1.43 | 1.09 |
| HL-LB | 2.44 | 2.42 | 2.4 | 1.78 | 1.44 | 1.13 |
| HM-g | 2.4 | 2.4 | 2.38 | 1.78 | 1.43 | 1.12 |
| HM-LB | 2.4 | 2.4 | 2.34 | 1.78 | 1.46 | 1.2 |
| HS-g | 2.4 | 2.4 | 2.34 | 1.78 | 1.5 | 1.45 |
| HS-LB | 2.4 | 2.4 | 2.4 | 1.84 | 1.76 | 1.68 |
| Gee | 2.4 | 2.4 | 2.34 | 1.78 | 1.44 | 1.1 |
| RH | 2.4 | 2.4 | 2.34 | 1.78 | 1.43 | 0.97 |
| RM | 2.4 | 2.4 | 2.36 | 1.78 | 1.42 | 1 |
| **RL** | **2.4** | **2.38** | **2.3** | **1.78** | **1.43** | **1.04** |
| **KS-RL** | **2.4** | **2.4** | **2.32** | **1.79** | **1.46** | **1.12** |
| FN6 | 2.4 | 2.4 | 2.36 | 1.78 | 1.49 | 1.3 |
| FN9 | 2.4 | 2.4 | 2.34 | 1.78 | 1.44 | 1.13 |
| MP-S | 2.44 | 2.42 | 2.4 | 1.78 | 1.43 | 1.07 |
| DVH | 2.4 | 2.4 | 2.34 | 1.78 | 1.48 | 1.24 |
| RDVH | 2.4 | 2.38 | 2.3 | 1.79 | 1.49 | 1.27 |
| Li-HFE-S | 2.4 | 2.38 | 2.32 | 1.78 | 1.47 | 1.23 |
| **Li-IOD-S** | **2.44** | **2.42** | **2.38** | **1.78** | **1.44** | **1.11** |
| JC-T3 | 2.4 | 2.38 | 2.34 | 1.71 | 1.35 | 1.04 |
| Li-HFE-T3 | 2.44 | 2.42 | 2.38 | 1.71 | 1.37 | 1.11 |
| MS | 2.42 | 2.42 | 2.36 | 1.71 | 1.38 | 1.13 |
| JJ | 2.42 | 2.42 | 2.36 | 1.79 | 1.39 | 0.99 |
| JC-T | 2.44 | 2.42 | 2.38 | 1.81 | 1.49 | 1.07 |
| MP-T | 2.4 | 2.4 | 2.36 | 1.81 | 1.43 | 1.04 |
| Li-HFE-T | 2.4 | 2.4 | 2.36 | 1.81 | 1.47 | 1.19 |
| **Li-IOD-T** | **2.42** | **2.42** | **2.36** | **1.81** | **1.43** | **1.08** |
| **KS-Li-IOD-T** | **2.4** | **2.4** | **2.36** | **1.82** | **1.47** | **1.17** |
| Madrid | 2.46 | 2.44 | 2.42 | 1.79 | 1.44 | 1.99 |

**Table S11** Position of the first maximum of the O-O PRDF and the $N_{OO}$ coordination numbers (calculated up to 3.2 Å) obtained in MD simulations. (The best models are in bold.)

|  | $r_{max}$ [Å] | $N_{OH}$ 1.5mol % | $N_{OH}$ 7.5mol % | $N_{OH}$ 15 mol % |
|---|---|---|---|---|
| DS | 2.74 | 3.82 | 3.27 | 2.94 |
| JC-S | 2.75 | 3.83 | 3.18 | 2.58 |
| HL-g | 2.74 | 3.81 | 3.26 | 2.92 |
| HL-LB | 2.76 | 3.83 | 3.15 | 2.55 |
| HM-g | 2.74 | 3.81 | 3.25 | 3.25 |
| HM-LB | 2.74 | 3.81 | 3.16 | 2.66 |
| HS-g | 2.74 | 3.81 | 3.15 | 2.61 |
| HS-LB | 2.75 | 3.93 | 3.73 | 3.62 |
| Gee | 2.74 | 3.81 | 3.21 | 2.8 |
| RH | 2.74 | 3.81 | 3.17 | 2.68 |
| RM | 2.75 | 3.81 | 3.15 | 2.63 |
| **RL** | **2.73** | **3.83** | **3.16** | **2.4** |
| **KS-RL** | **2.73** | **3.81** | **3.13** | **2.41** |
| FN6 | 2.74 | 3.81 | 3.12 | 2.48 |
| FN9 | 2.74 | 3.81 | 3.13 | 2.53 |
| MP-S | 2.76 | 3.83 | 3.14 | 2.49 |
| DVH | 2.74 | 3.83 | 3.29 | 3.01 |
| RDVH | 2.74 | 3.84 | 3.36 | 3.2 |
| Li-HFE-S | 2.74 | 3.82 | 3.28 | 2.98 |
| **Li-IOD-S** | **2.76** | **3.85** | **3.28** | **2.93** |
| JC-T3 | 2.74 | 3.81 | 3.17 | 2.64 |
| Li-HFE-T3 | 2.76 | 3.83 | 3.17 | 2.59 |
| MS | 2.77 | 3.78 | 3.12 | 2.59 |
| JJ | 2.77 | 3.8 | 3.21 | 2.85 |
| JC-T | 2.76 | 3.85 | 3.26 | 2.78 |
| MP-T | 2.74 | 3.81 | 3.25 | 2.98 |
| Li-HFE-T | 2.74 | 3.81 | 3.17 | 2.68 |
| **Li-IOD-T** | **2.78** | **3.79** | **3.18** | **2.78** |
| **KS-Li-IOD-T** | **2.74** | **3.82** | **3.2** | **2.66** |
| Madrid | 2.78 | 3.84 | 3.11 | 2.25 |

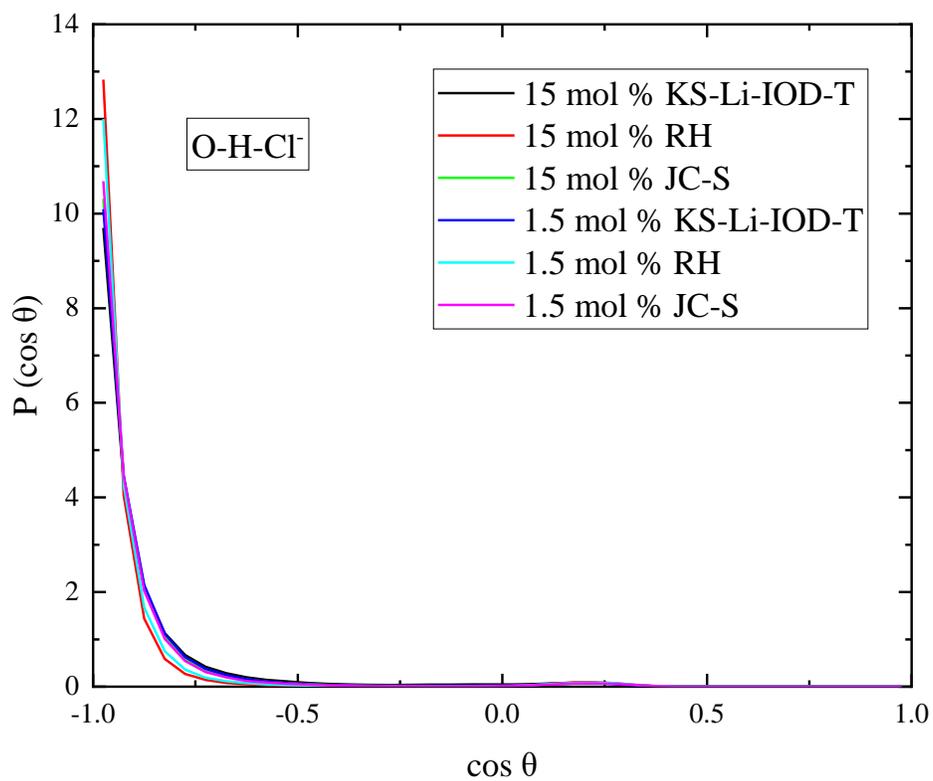

**Figure S13.** Cosine distributions of the O-H-Cl⁻ angle, obtained by MD simulations using FFs KS-Li-IOD-T, RH, and JC-S.

*PRDFs not shown in the main text*

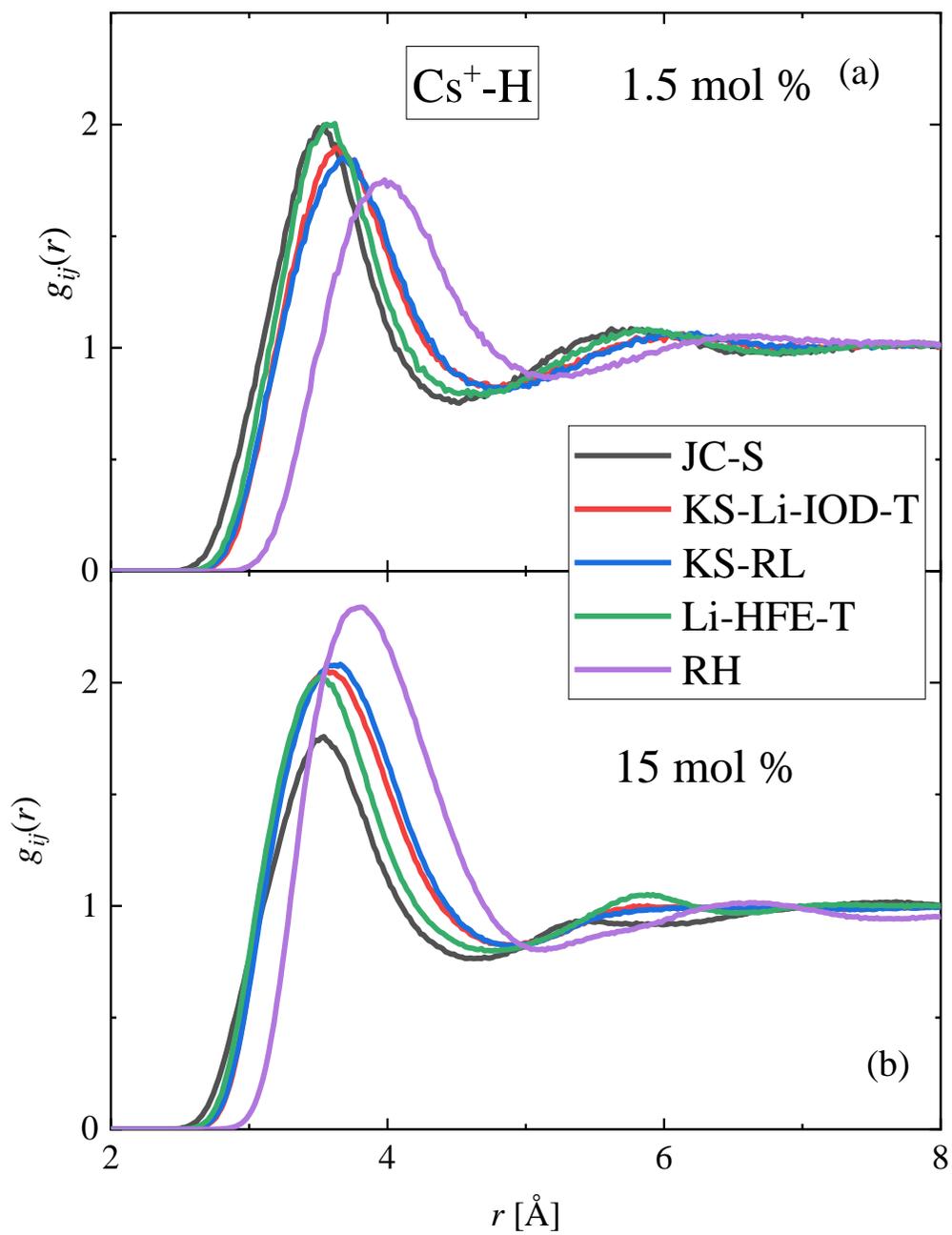

**Figure S14.** $Cs^+$ - H partial radial distribution functions, as calculated from MD simulations with different interatomic potential models. The curves are shown for (a) 1.5 mol % and (b) 15 mol % solutions.

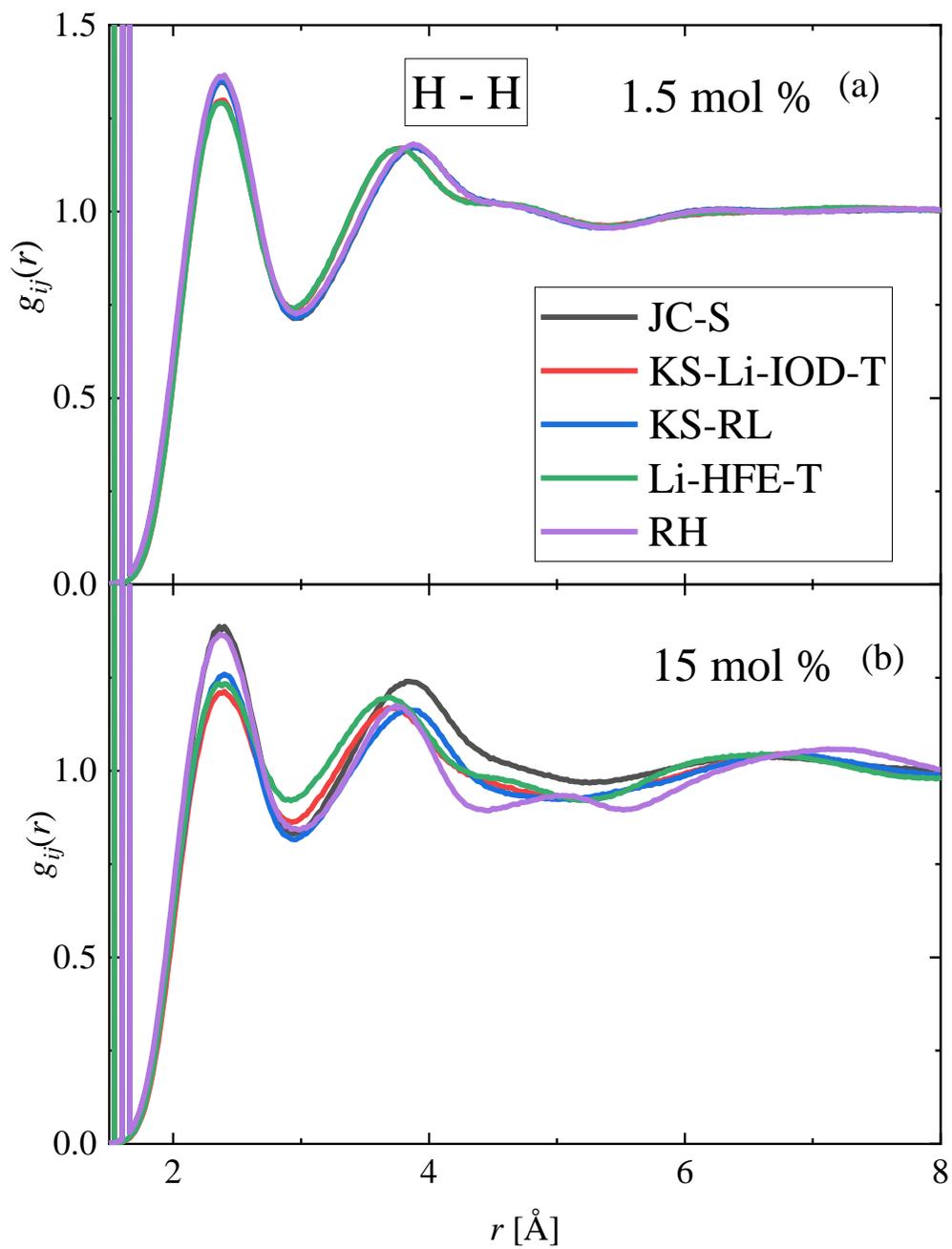

**Figure S15.** H – H partial radial distribution functions, as calculated from MD simulations with different interatomic potential models. The curves are shown for (a) 1.5 mol % and (b) 15 mol % solutions.

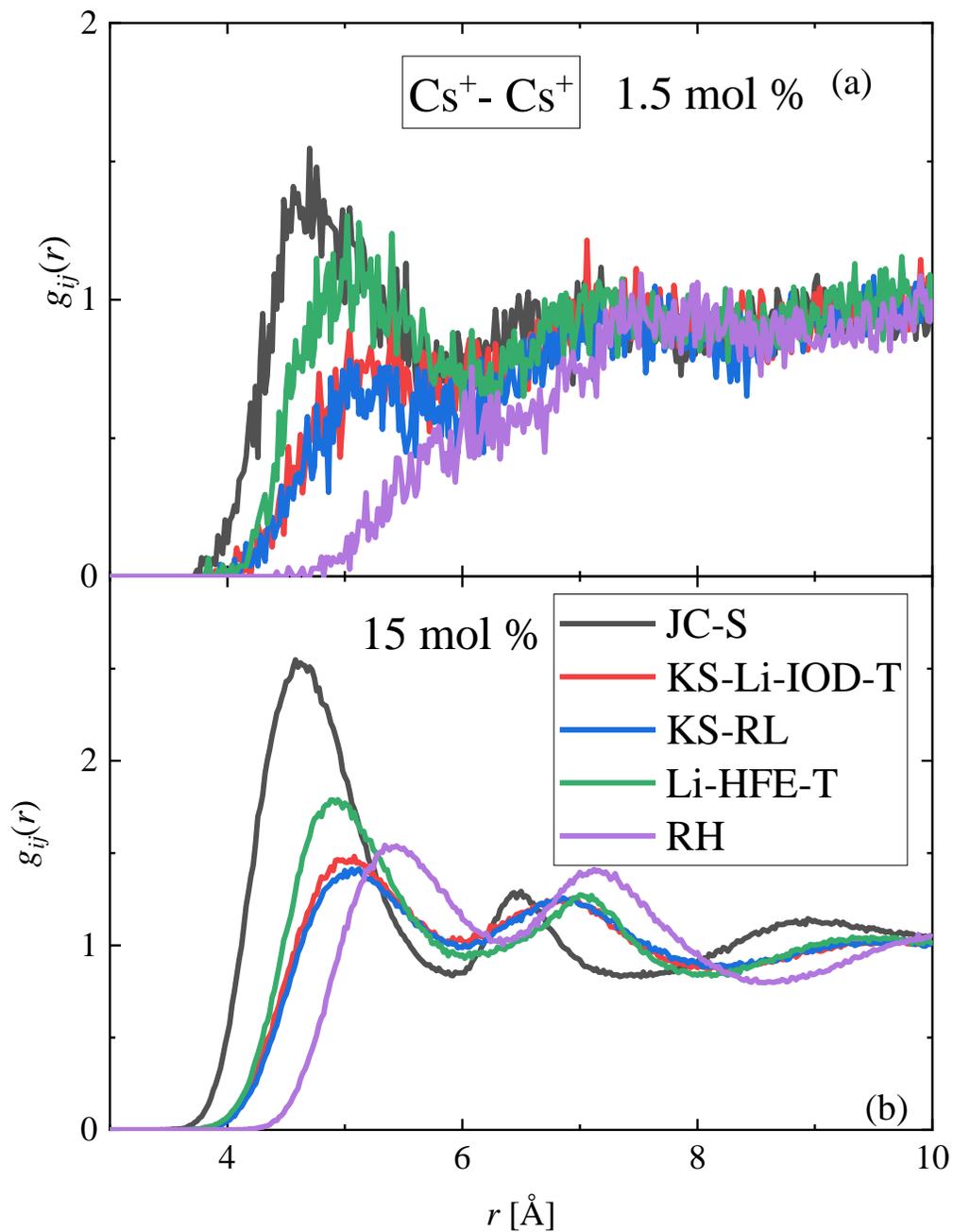

**Figure S16.** $Cs^+$ - $Cs^+$ partial radial distribution functions, as calculated from MD simulations with different interatomic potential models. The curves are shown for (a) 1.5 mol % and (b) 15 mol % solutions.

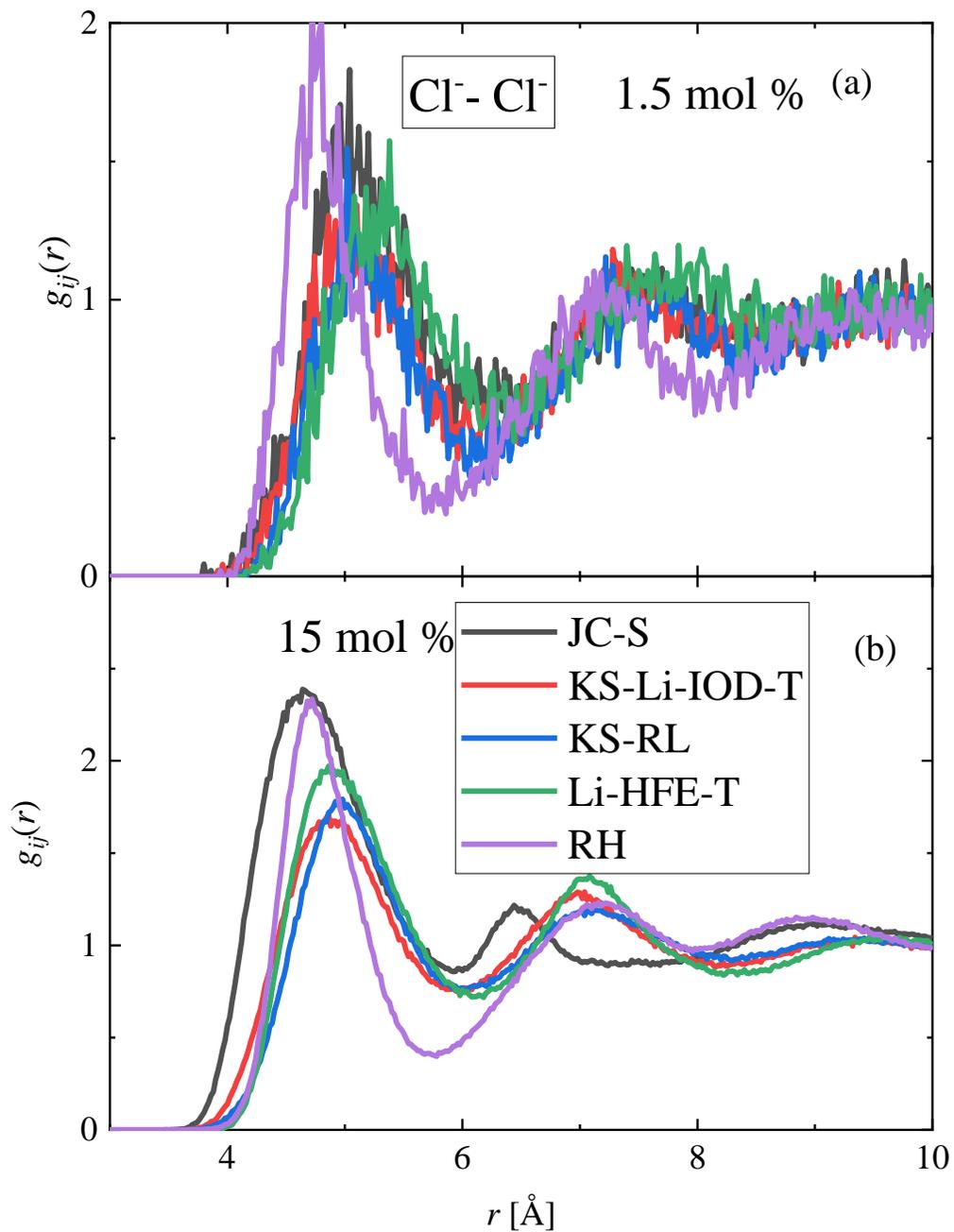

**Figure S17.** $Cl^-$ - $Cl^-$ partial radial distribution functions, as calculated from MD simulations with different interatomic potential models. The curves are shown for (a) 1.5 mol % and (b) 15 mol % solutions.

## 7. Searching for the structural origin of special features in the total scattering structure factors

In Fig. 4 and 5 (in the main text) the main features of the total scattering structure factors can be seen, the experimental ones together with those obtained from simulations using some selected potential model. In the $S^N(Q)$ curve of the 15 mol % solution a peak can be seen about at 1 Å$^{-1}$, which is often called as pre-peak, or first sharp diffraction peak. The presence of this peak is often associated with the medium range order in the system. As it can be seen in Fig. 4 (in the main text), the presence and height of this peak in the simulated curves highly depends on the applied potential model. This can help us to investigate the structural origin of this peak. The amplitude of $S(Q)$ at a given $Q$ value is determined as a weighted sum of the partial $S_{ij}(Q)$ curves, which are calculated from the PRDFs (see Eqs. 2, 3 in the main text) The low $Q$ part of the $S^N(Q)$ curve and the neutron weighted partial structure factors ($w_{ij}*S_{ij}(Q)$) of some selected FFs are shown in Fig. S18.

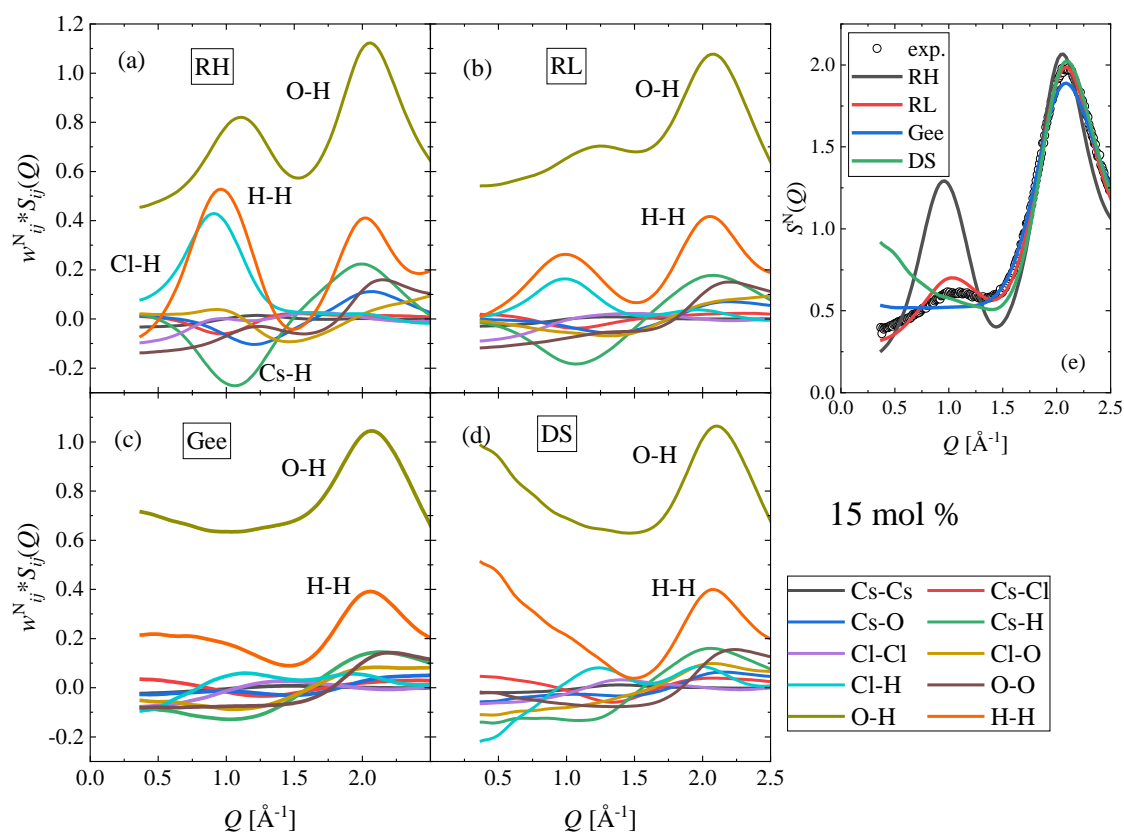

**Figure S18.** (a)-(d) Neutron weighted partial structure factors obtained in simulations using the (a) RH, (b) RL, (c) Gee, and (d) DS force fields. The weighted sum of the partial $S_{ij}(Q)$ curves (the $S^N(Q)$ curve) is shown in (e) together with the experimental curve.

The highest contribution at 1 Å$^{-1}$ comes from the O-H, H-H, Cl$^-$-H partials (and the negative contribution of the Cs$^+$-H PRDF can also be significant). (As it was mentioned in Section 2 in the main text, in the diffraction experiments of Ref. [S3] the H atoms were replaced by D, in the calculation of $w_{ij}^N$ the weight of D was taken into account accordingly.) The values of the weighted partial structure factors at the pre-peak position are compared at Fig. S19. The main difference at $Q = 1$ Å$^{-1}$ originates from the differences in the Cl$^-$-H, H-H and O-H partial structure factors, thus the Cl$^-$-H, H-H and O-H PRDFs.

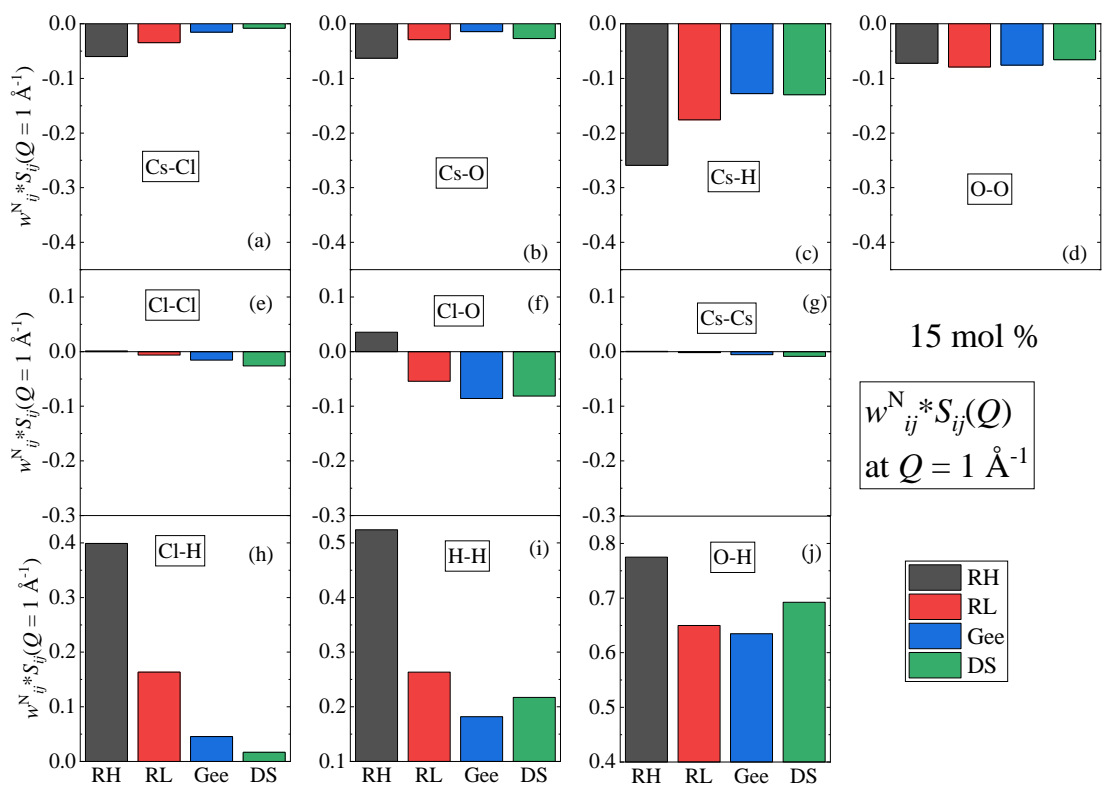

**Figure S19.** Values of the weighted partial structure factors at $Q = 1$ Å$^{-1}$ for the selected force field. Note: the *y*-scales, although starting from different values, span the same range to make the comparison easier.

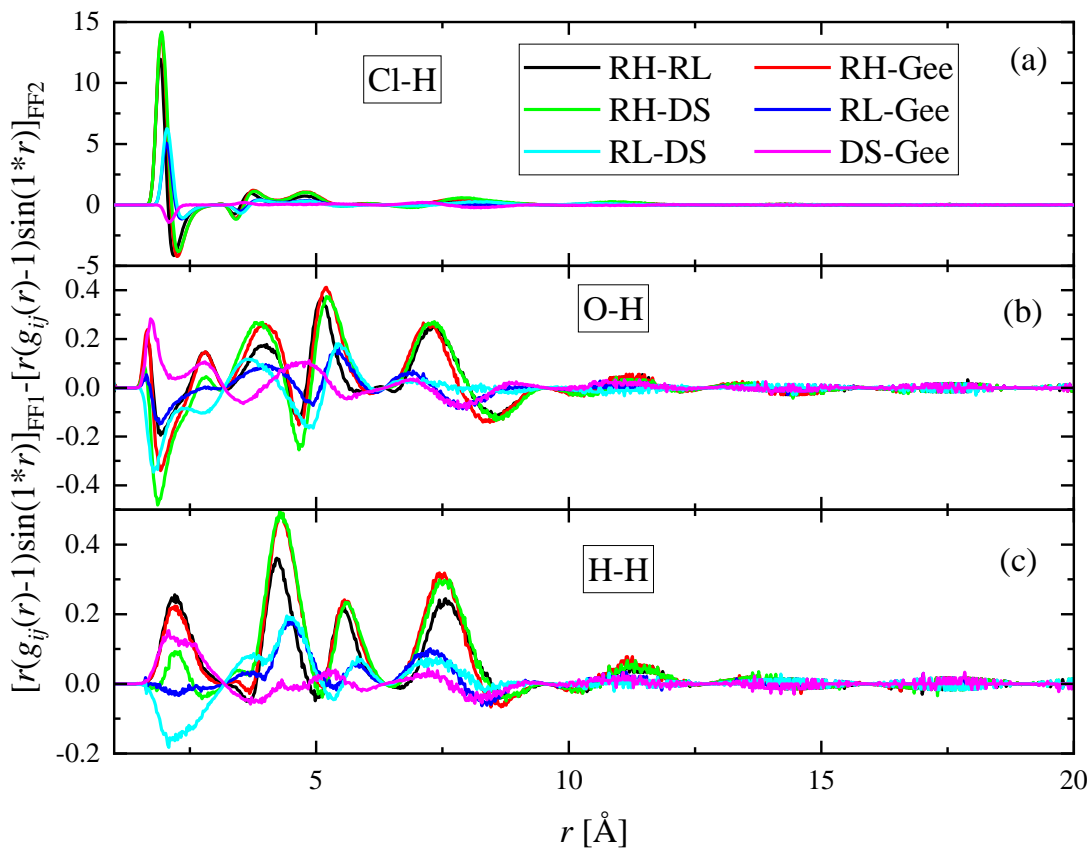

**Figure S20.** Differences in the $r*(g_{ij}(r)-1)*\sin(Qr)$ functions of the selected force fields at $Q = 1$ Å$^{-1}$, for (a) Cl$^-$-H, (b) O-H and (c) H-H partials.

According to Eq. 3 (of the main text), $S_{ij}(Q)$ is calculated as the integral of the $r*(g(r)-1)*\sin(Qr)$ function. To determine which part of the curves give significantly different contribution to the integral for different FFs, the curves were calculated at $Q = 1$ Å$^{-1}$ for the Cl$^-$-H, H-H and O-H pairs, the differences of the curves obtained by the selected FFs are shown in Fig. S20. Concerning the Cl$^-$-H partial, the main differences between the FFs are in the region around 2 Å, it originates from the position and height of the first peak of the $g_{ClH}(r)$ curves. Thus the height of the pre-peak is partly caused by the (high) number and positions of water molecules around chloride ions. In the H-H partials the differences are present in more (four) regions between 2 and 8 Å (around 2.25, 4.2, 5.5 and 7.5 Å), while in the O-H partials the main deviations can be found in the 3.5 – 7.5 Å region (around 3.9, 5.2 and 7.3 Å). These distances cannot be assigned simply to one structural motif. E. g. the first intermolecular peak of the H-H PRDFs is the result not only the H-H distances in H-bonded water molecules, but also the H-H pairs between water molecules hydrating the same chloride ion. Thus the number of H-bonded water molecules, the $r_{ClH}$ bond lengths as well as the number of hydrating water molecules can affect the shape of the H-H PRDF and cause the pre-peak of $S^N(Q)$.

Another interesting part of the simulated $S^N(Q)$ curves is below the pre-peak region. As $Q \to 0$, some of the calculated $S^N(Q)$ functions begin to increase in contrast to the experimental curve (e. g. for the MS, Li-HFE-S, Li-HFE-T, Li-HFE-T3, DVH, RDVH, JC-S, and FN6 models). This behavior was observed previously in Ref. [S3] also, in that study the DS parameters were applied in the MD simulation. To investigate the origin of this discrepancies a comparison, similar to that detailed in the previous paragraph, was performed at $Q = 0.5$ Å$^{-1}$.

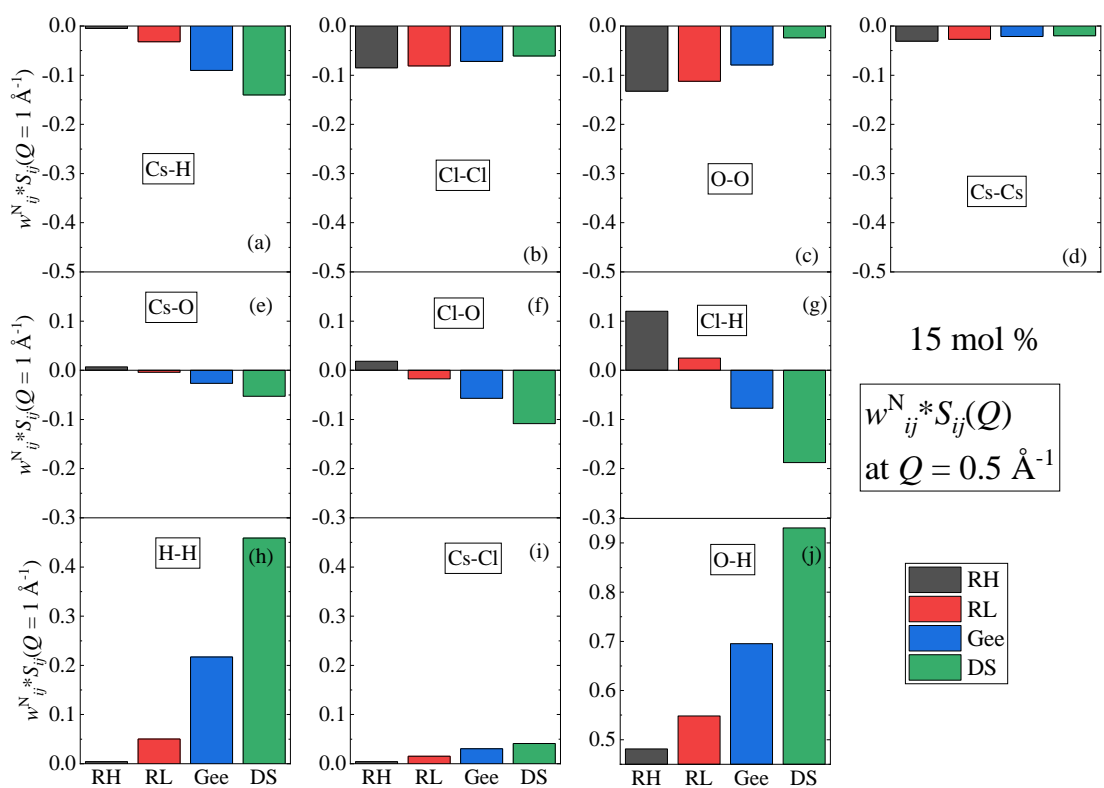

**Figure S21.** Values of the weighted partial structure factors at $Q = 0.5$ Å$^{-1}$ for the selected force field. Note: the y-scales, although starting from different values, span the same range to make the comparison easier.

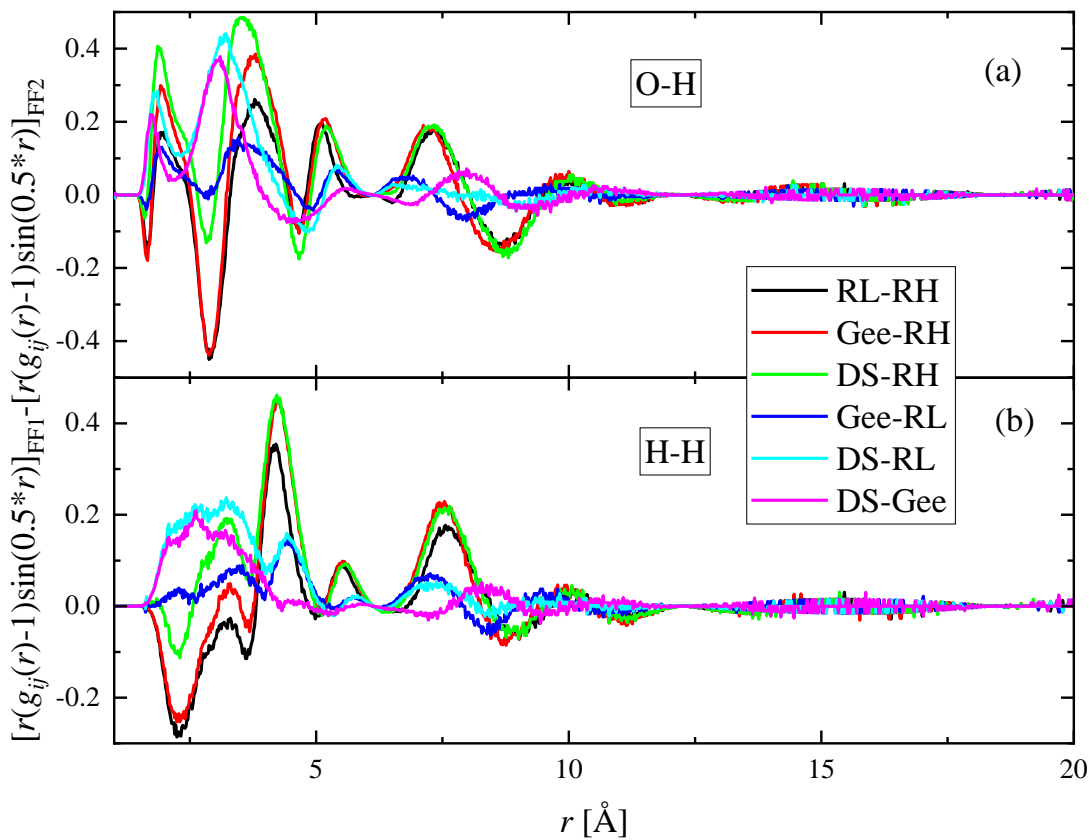

**Figure S22.** Differences in the $r*(g_{ij}(r)-1)*\sin(Qr)$ functions of the selected force fields at $Q = 0.5$ Å$^{-1}$, for (a) O-H and (b) H-H partials.

Here the main differences between the FFs are in the H-H and O-H partial structure factors, whose effect is partly compensated by the Cl$^-$-H, Cs$^+$-H and Cl$^-$-O partials (Fig. S21). In the $r$-space, several regions contribute to the difference in the 2 – 8 Å regime, see Fig. S22. Let's choose the DS and Gee FFs, which are significantly different at $Q = 0.5$ Å$^{-1}$, but are similar at higher $Q$ values. Comparison of the DS and Gee FFs shows, that the main difference between these two FFs is in the 2 – 4 Å region both in H-H and O-H PRDFs. As the Cl$^-$-H bond distance is similar in these models, the difference is mostly caused by the different number of H-bonded water molecules: the $N_{OH}$ coordination number is 1.23 for DS and 1.1 for Gee model. This difference seems to be moderate, however the shape of the $S^N(Q)$ curve is sensitive to the $N_{OH}$ coordination number. A comparison of the $N_{OH}$ coordination numbers and $S^N(Q = 0.5$ Å$^{-1})$ is shown in Fig. S23. The increase in the $S^N(Q)$ function at low $Q$ values coincides with higher number of H-bonded water molecules. A similar relation can be observed between the $S^N(Q = 0.5$ Å$^{-1})$ and the $N_{CsCl}$ coordination number. The clustering of the ions goes together with the decrease of ion-water pairs (the hydration shells of ions are connected for clustered ions, they create a common hydrate shell), while the water molecules can establish H-bonds between themselves. This phenomenon leads to the increase in the structure factor at low Q values.

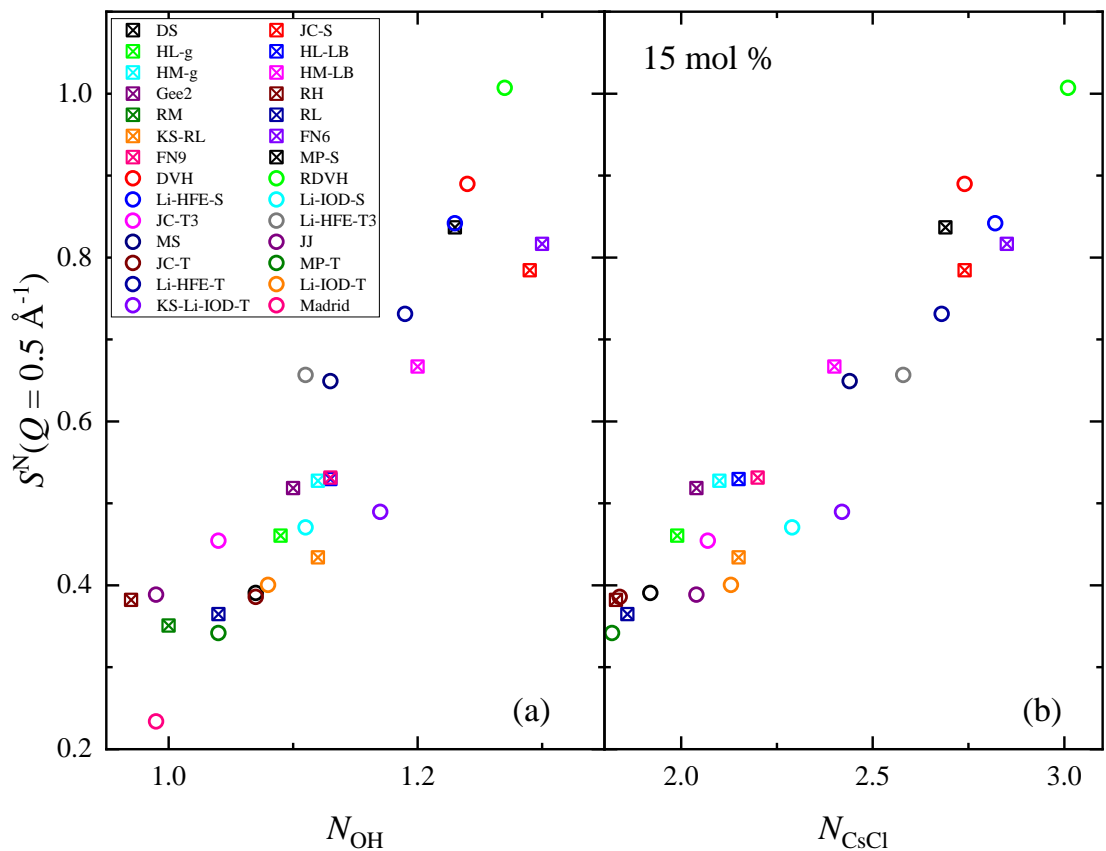

**Figure S23.** Neutron total structure factors ($S^N(Q)$) values at $Q = 0.5$ Å$^{-1}$ as a function of the (a) $N_{OH}$ and (b) $N_{CsCl}$ coordination numbers obtained in simulations using different force fields. (Note: models, where ion precipitation occurred under the solubility limit (HS-g and HS-LB), are not shown.)

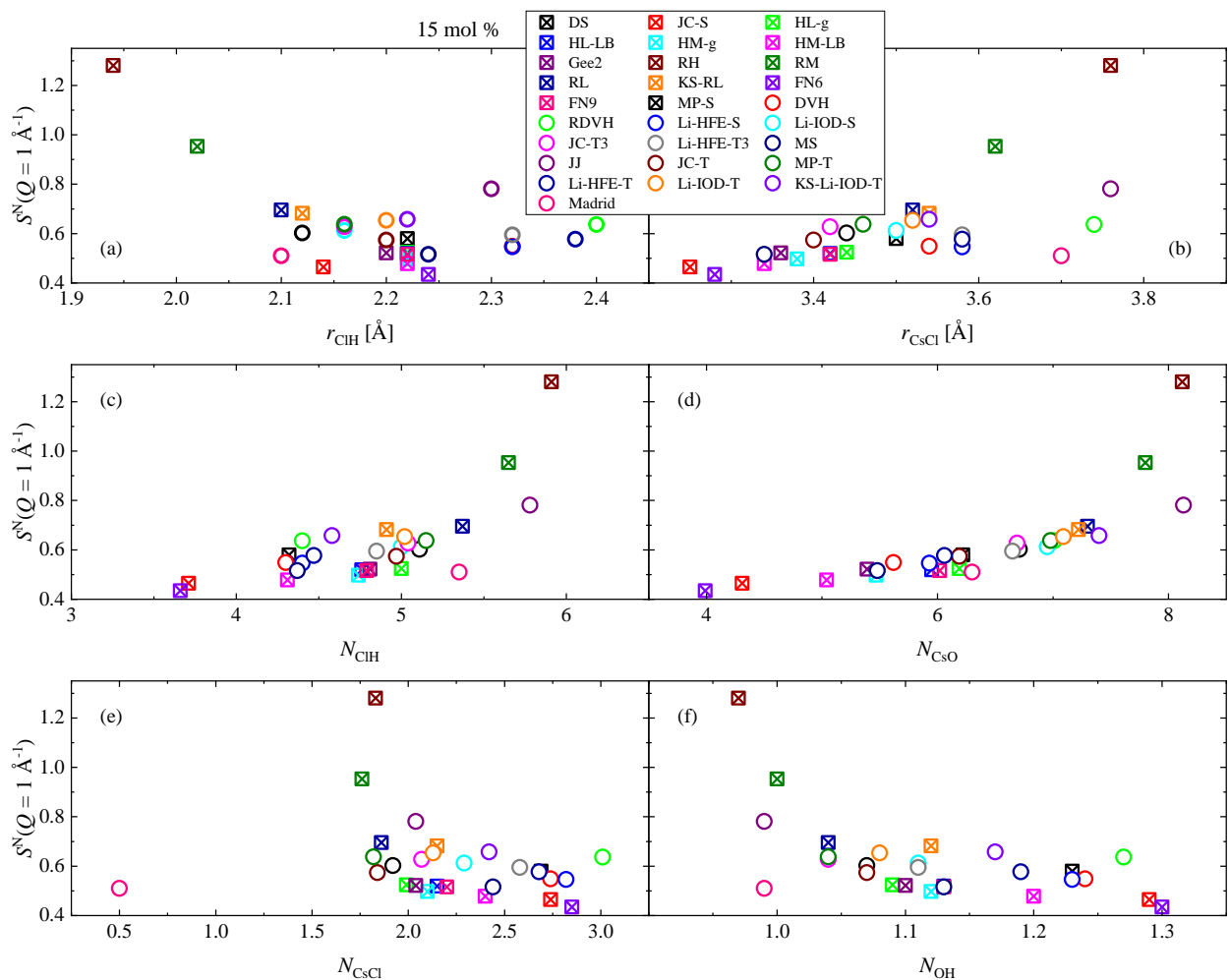

**Figure S24.** Neutron total structure factors ($S^N(Q)$) values at $Q = 1$ Å$^{-1}$ as a function of the (a) $r_{ClH}$ and (b) $r_{CsCl}$ bond distances, (c) $N_{ClH}$, (d) $N_{CsO}$, (e) $N_{CsCl}$, and (f) $N_{OH}$ coordination numbers obtained in simulations using different force fields. (Note: models, where ion precipitation occurred under the solubility limit (HS-g and HS-LB), are not shown.)

## 8. Hydrogen bonded network: ring-type distributions of wrong models

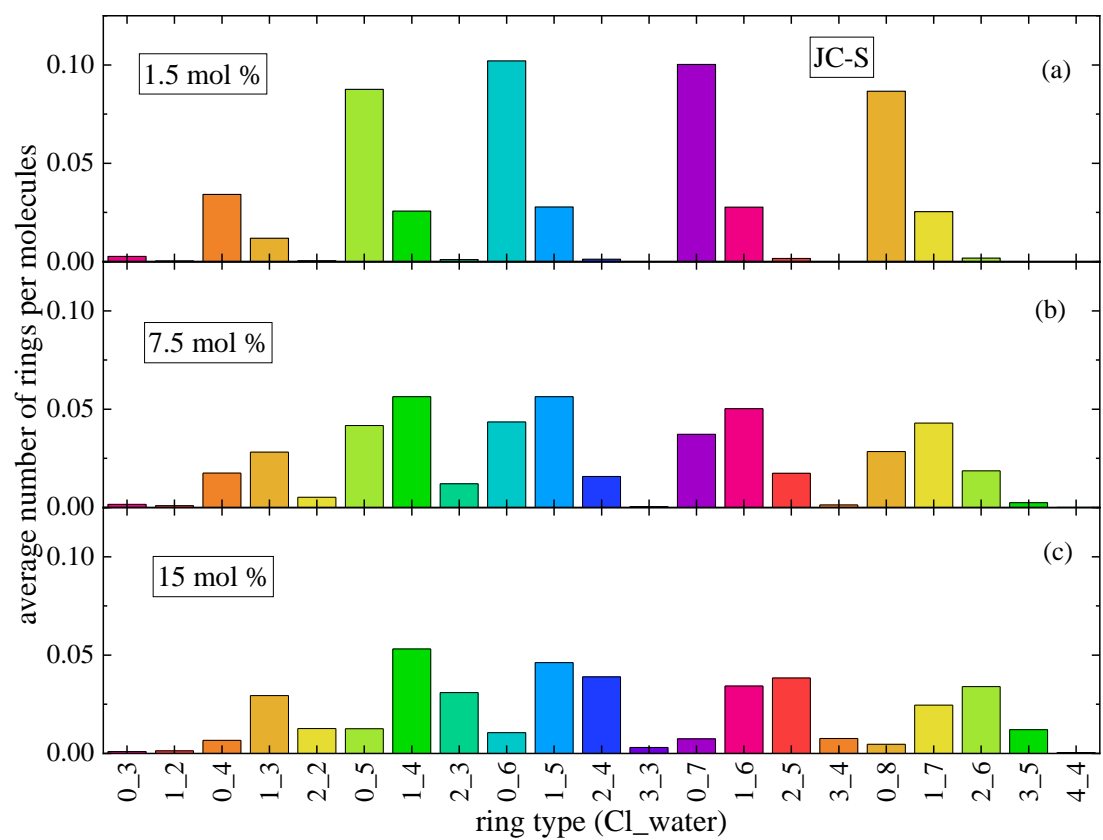

**Figure S25.** Distribution of different types of rings (rings contain water molecules and Cl$^-$ ions) normalized by the number of molecules (water + Cl$^-$ ions) in the configurations at different salt concentrations obtained from the JC-S model.

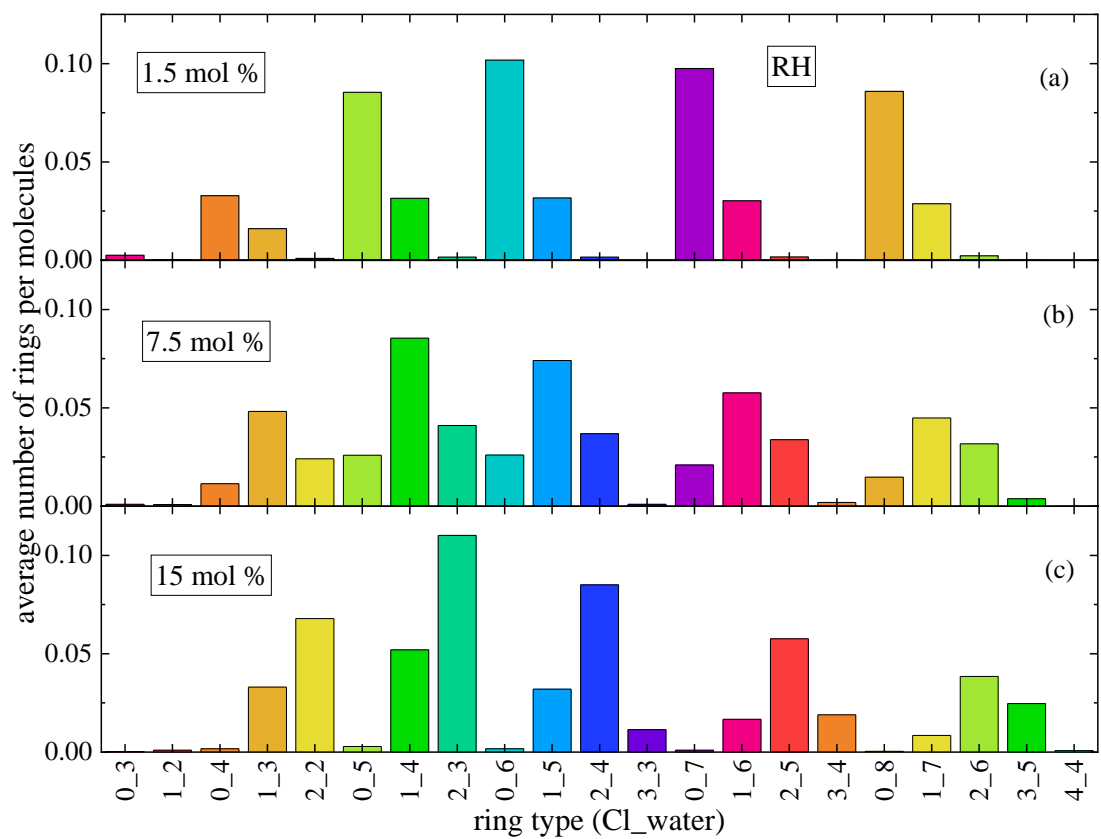

**Figure S26.** Distribution of different types of rings (rings contain water molecules and Cl[-] ions) normalized by the number of molecules (water + Cl[-] ions) in the configurations at different salt concentrations obtained from the RH model.

# References

[S1]  M. Parrinello, A. Rahman, Polymorphic transitions in single crystals: A new molecular dynamics method, J. Appl. Phys. 52 (1981) 7182–7190. doi:10.1063/1.328693.

[S2]  S. Nosé, M.L. Klein, Constant pressure molecular dynamics for molecular systems, Mol. Phys. 50 (1983) 1055–1076. doi:10.1080/00268978300102851.

[S3]  V. Mile, L. Pusztai, H. Dominguez, O. Pizio, Understanding the structure of aqueous cesium chloride solutions by combining diffraction experiments, molecular dynamics simulations, and reverse Monte Carlo modeling, J. Phys. Chem. B 113 (2009) 10760–10769. doi:10.1021/jp900092g.

[S4]  E.W. Washburn, C.J. West, eds., International Critical Tables of Numerical Data, Physics, Chemistry and Technology, McGraw-Hill Book Company, New York, volume III (1928), p97 doi:10.17226/20230.

[S5]  P. Novotny, O. Sohnel, Densities of binary aqueous solutions of 306 inorganic substances, J. Chem. Eng. Data 33 (1988) 49–55. doi:10.1021/je00051a018.

[S6]  O. Gereben, L. Pusztai, On the accurate calculation of the dielectric constant from molecular dynamics simulations: The case of SPC/E and SWM4-DP water, Chem. Phys. Lett. 507 (2011) 80–83. doi:10.1016/j.cplett.2011.02.064.

[S7]  I. Pethes, A comparison of classical interatomic potentials applied to highly concentrated aqueous lithium chloride solutions, J. Mol. Liq. 242 (2017) 845–858. doi:10.1016/j.molliq.2017.07.076.

[S8]  A.C. Tikanen, W.R. Fawcett, Application of the mean spherical approximation and ion association to describe the activity coefficients of aqueous 1:1 electrolytes, J. Electroanal. Chem. 439 (1997) 107–113. doi:10.1016/S0022-0728(97)00376-8.

[S9]  C. Vega, J.L.F. Abascal, Simulating water with rigid non-polarizable models: a general perspective, Phys. Chem. Chem. Phys. 13 (2011) 19663. doi:10.1039/c1cp22168j.

[S10] I.-C. Yeh, G. Hummer, System-size dependence of diffusion coefficients and viscosities from molecular dynamics simulations with periodic boundary conditions, J. Phys. Chem. B 108 (2004) 15873–15879. doi:10.1021/jp0477147.

[S11] T. Nakai, S. Sawamura, Y. Taniguchi, Effect of pressure on the viscosity of aqueous cesium chloride solution at 25°C, J. Mol. Liq. 65–66 (1995) 365–368. doi:10.1016/0167-7322(95)00832-4.

[S12] P.A. Lyons, J.F. Riley, Diffusion coefficients for aqueous solutions of calcium chloride and cesium chloride at 25°, J. Am. Chem. Soc. 76 (1954) 5216–5220. doi:10.1021/ja01649a081.

[S13] K.J. Müller, H.G. Hertz, A Parameter as an indicator for water−water association in solutions of strong electrolytes, J. Phys. Chem. 100 (1996) 1256–1265. doi:10.1021/jp951303w.

[S14] J. Anderson, R. Paterson, Application of irreversible thermodynamics to isotopic diffusion. Part 1.—Isotope–isotope coupling coefficients for ions and water in concentrated aqueous solutions of alkali metal chlorides at 298.16 K, J. Chem. Soc. Faraday Trans. 1 Phys. Chem. Condens. Phases 71 (1975) 1335. doi:10.1039/f19757101335.

[S15] H.G. Hertz, R. Mills, The effect of structure on self-diffusion in concentrated electrolytes : relationship between the water and ionic self-diffusion coefficients, J. Chim. Phys. 73 (1976) 499–508. doi:10.1051/jcp/1976730499.

[S16] B.M. Braun, H. Weingaertner, Accurate self-diffusion coefficients of lithium(1+), sodium(1+), and cesium(1+) ions in aqueous alkali metal halide solutions from NMR spin-echo experiments, J. Phys. Chem. 92 (1988) 1342–1346. doi:10.1021/j100316a065.